\documentclass[prc,floatfix,%
 reprint,
superscriptaddress,
 amsmath,amssymb,
 aps,
]{revtex4-2}

\usepackage{soul}
\usepackage{graphicx}
\usepackage{dcolumn}
\usepackage{bm}
\usepackage[mathscr]{euscript}
\usepackage{xcolor}
\usepackage[dvipsnames]{xcolor}
\usepackage[normalem]{ulem}
\usepackage{float}

\usepackage{etoolbox}

\makeatletter
\patchcmd{\bibsection}
  {\onecolumngrid@push}
  {\onecolumngrid@push\vspace*{28pt}}
  {}{}
\makeatother

\definecolor{amethyst}{rgb}{0.75, 0.58, 0.89}

\newcommand{\vd}[1]{\textcolor{amethyst}{{\textbf{#1}}} }

\begin{document}


\title{Many-body effects on dense matter with hyperons at finite temperature} 

\author{Rafael B\'{a}n Jacobsen}
\affiliation{Departamento de F\'{\i}sica, Universidade Federal de Santa Maria, 97105-900 Santa Maria, RS, Brazil}%
\affiliation{Center for Nuclear Research, Department of Physics, Kent State University, Kent, OH 44242 USA}

\author{Ricardo Luciano Sonego Farias}%
\affiliation{Departamento de F\'{\i}sica, Universidade Federal de Santa Maria, 97105-900 Santa Maria, RS, Brazil}
\affiliation{Center for Nuclear Research, Department of Physics, Kent State University, Kent, OH 44242 USA}%


\author{Veronica Dexheimer}
\affiliation{Center for Nuclear Research, Department of Physics, Kent State University, Kent, OH 44242 USA}


\date{\today}

\begin{abstract}
In this work, we present the first extension of the Many-Body Forces (MBF) Model to finite temperature. The MBF Model describes nuclear matter in a relativistic quantum hadrodynamics formalism that takes many-body forces into account through a field dependence of the nuclear interaction coupling constants. Assuming nuclear matter to be charge neutral, beta-equilibrated, and populated by the baryon octet, electrons, muons, and neutrinos, we explore the parameter of the model that controls many-body forces, three different hyperon coupling schemes (also introduced here for the first time in the MBF Model), and temperature effects to describe basic properties of nuclear matter, including the speed of sound, compressibility, and adiabatic index. We also investigate the mass-radius relation of neutron and protoneutron stars by solving the Tolman-Oppenheimer-Volkoff equations at zero and finite temperature, including scenarios with fixed entropy per baryon and lepton fraction. We find that different hyperon coupling schemes soften the equation of state to different degrees, requiring different degrees of 
many-body force effects to reproduce observations, although they are never zero. Multidimensional tables with our results are available in the CompOSE repository, ready to be used in astrophysical simulations.

\end{abstract}

\maketitle


\section{\label{sec:intro}Introduction}

Understanding strongly interacting matter under extreme conditions remains one of the central problems of modern nuclear astrophysics. Quantum chromodynamics (QCD) is the fundamental theory of the strong interaction and, in principle, governs the behavior of dense nuclear matter~\cite{Gell-Mann:1964ewy,Zweig:1964jf}. QCD calculations, in most cases, involve the use of high-performance computers, mainly for the study of lattice QCD~\cite{Creutz:1983njd}. However, first-principles calculations remain severely limited at high baryon density because of the sign problem in lattice QCD~\cite{Philipsen:2008zz,deForcrand:2009zkb}. Moreover, perturbative QCD (PQCD)~\cite{Ghiglieri:2020dpq} becomes applicable only in asymptotic regions of the phase diagram~\cite{Haque:2014rua,Gorda:2021znl}.

Nonetheless, at the energy scales relevant for compact stars, hadronic degrees of freedom provide an adequate effective description for strongly interacting matter. In this context, in 1974, Walecka~\cite{Walecka:1974qa}, following the work of Johnson and Teller~\cite{PhysRev.98.783} and Duerr~\cite{PhysRev.103.469}, proposed an effective theory that describes nuclear matter through the exchange of scalar and vector mesons, within a covariant mean-field framework. The Walecka model can be regarded as a \emph{relativistic mean-field theory (RMF)}, based on Yukawa's formulation~\cite{Yukawa:1935xg}. Although successful in reproducing several nuclear properties, this original RMF formulation requires extensions to improve the description of saturation and effective masses~\cite{book:Walecka}.

Several extensions were subsequently proposed, including nonlinear mesonic interactions, such as the \emph{Nonlinear Boguta-Bodmer Model}~\cite{Boguta:1977xi}, and derivative couplings, like the \emph{Zimanyi and Moszkowski Model}, or simply \emph{ZM Model}~\cite{PhysRevC.42.1416}. Another model that follows the same path is the \emph{Many-Body Forces Model} (MBF Model)~\cite{Taurines:2000xz, PhysRevC.63.065801}. From a formal point of view, the special feature of this model consists in a generalization of the previously developed effective models, considering a parameterizable derivative coupling that simulates many-body forces. The MBF Model is the one to be adopted in this work and its complete formalism is presented in Sec.~\ref{sec:MBFformalism}.  

Aiming to constrain the behavior of the EoS of nuclear matter at high densities, the interior of a neutron star (NS) offers a unique meeting point between astrophysics and nuclear physics, because the macroscopic properties of these objects depend directly on the microscopic behavior of strongly interacting matter. Neutron stars (NSs) are formed during the core-collapse supernovae of massive stars ($8$ M$_{\rm Sun}$ $\lesssim M \lesssim$ $25$ M$_{\rm Sun}$), when the stellar core can no longer be supported against gravitational collapse. Owing to the conservation of magnetic flux and angular momentum during the collapse, the remnant has extremely strong magnetic fields and rapid rotation~\cite{1992ApJ...392L...9D, 1996ApJ...473..322T, Peng:2007uu}. The resulting proto-neutron star (PNS) is initially hot, with temperatures of tens of MeV, and subsequently cools through photon and neutrino emission, eventually evolving into a cold NS~\cite{Dexheimer:2008ax}. Typical NSs have masses of $1$--$2$ M$_{\rm Sun}$, radii of $10$--$13$ km, and central densities that can exceed several times the nuclear saturation density~\cite{doi:10.1146/annurev-nucl-102419-124827}. NSs and PNSs therefore constitute natural laboratories for the study of strongly interacting matter at high densities (or chemical potentials) and relatively low temperatures ($\mu \gtrsim 900~\rm{MeV}$, $T \lesssim 30~\rm{MeV}$).

Effective hadronic models are particularly suitable for describing compact-star matter because the enormous number of particles involved justifies the thermodynamic limit commonly assumed in nuclear-matter calculations. Moreover, the covariant formulation of quantum hadrodynamics naturally preserves relativistic consistency. Finally, the high densities found in NS interiors favor the validity of the mean-field approximation employed in RMF models.

Despite their name, NSs are not composed exclusively of neutrons. Depending on density, temperature, and the underlying microscopic description, several degrees of freedom may become energetically favorable in their interior. These are the possible compositions mostly explored in the literature, as well as some of the pioneering works on such descriptions: purely hadronic matter~\cite{1971NuPhA.175..225B}, pion- and kaon-condensed phases~\cite{PhysRevLett.30.1340,Kaplan:1986yq}, hyperonic matter~\cite{1959ApJ...130..884C,1960SvA.....4..187A}, hybrid configurations containing deconfined quarks~\cite{Ivanenko:1965dg}, and even self-bound quark stars~\cite{Itoh:1970uw}.

\begin{figure}
\includegraphics[width=0.45\textwidth]{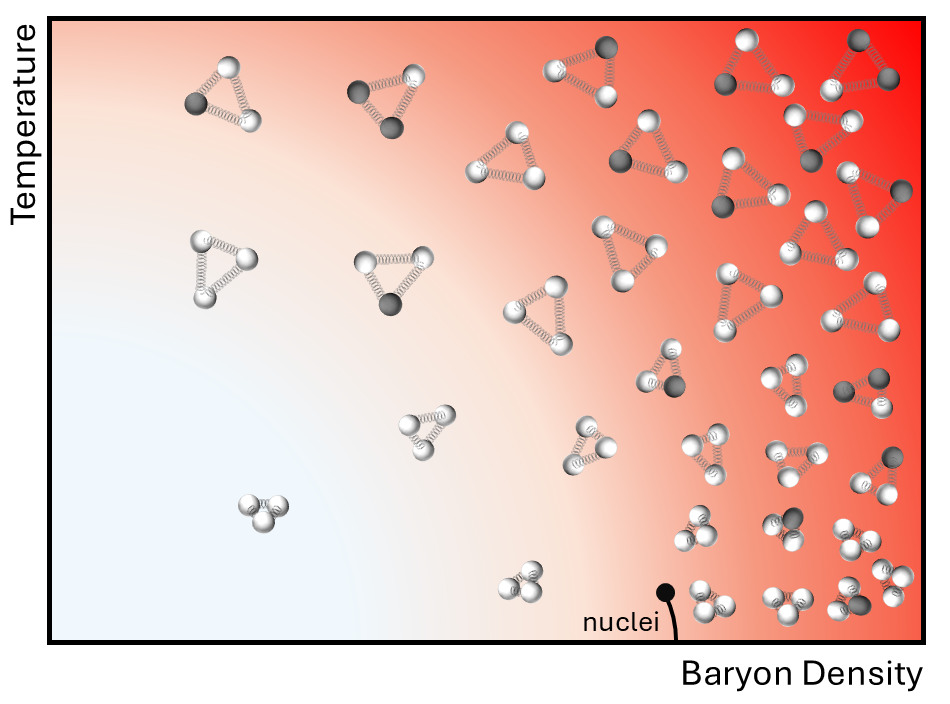}
\caption{\label{PhaseDiagram} Schematic representation of the low/high-baryon-density and low/medium-temperature sector of the QCD phase diagram, where strongly interacting matter is dominated by baryonic degrees of freedom. Light quarks ($u$ and $d$) are shown in white, such that ordinary nucleons are depicted as bound triplets of three white circles. The onset of strangeness is illustrated through the appearance of strange quarks ($s$), represented in gray; hyperons are therefore indicated as composite states containing one or two gray circles in combination with the light-quark constituents. The background red gradient points toward increasing energy scale.}
\end{figure}

In the present work, we focus on purely nucleonic and hyperonic matter, corresponding to the hadronic sector illustrated in Fig.~\ref{PhaseDiagram}. It schematically illustrates the low-temperature, high-density region of the QCD phase diagram relevant for neutron stars. In this regime, matter is predominantly hadronic and composed of baryons built from light quarks. As density and, to a lesser extent, temperature increase, the appearance of strange baryons becomes energetically favorable, providing an intermediate stage between purely nucleonic matter and possible deconfined phases at still higher densities~\cite{Fukushima_2010}. The present work is restricted to this hadronic sector and investigates the role of hyperonic degrees of freedom under finite-temperature conditions.

Regardless of their microscopic composition, NSs and PNSs are assumed to satisfy baryon-number conservation, electric charge neutrality, and chemical equilibrium.

Baryon-number conservation is expressed as
\begin{equation}
\label{conservdensbar}
\sum_{j=b} n_{B_j} - n_B = 0\, ,
\end{equation}
where $n_{B_j}=\mathscr{B}_j n_j$, with 
$\mathscr{B}_{j}$ denoting the baryon number ($1$ for baryons and $0$ for leptons) and $n_j$ the number density of species $j$, and $n_B$ is the total baryon (number) density.

Compact stars must also be electrically neutral. The enormous disparity between gravitational and electromagnetic forces implies the well-known constraint $Z_{\rm net}/A \sim 10^{-36}$~\cite{livro:Glendenning}. Charge neutrality is therefore imposed through
\begin{equation}
\label{eqdecarga}
\sum_{j=l} \mathscr{Q}_{j} \, n_{l_j} + \sum_{j=b} \mathscr{Q}_{j} \, n_{B_j} = 0 \, , 
\end{equation}
where $\mathscr{Q}_{j}$ is the electric charge, and $n_{l_j}$ is the lepton (number) density of species $j$.

Chemical equilibrium is imposed through the chemical potentials of the particle species participating in the relevant reactions. In equilibrium, the chemical potentials satisfy relations determined by the underlying weak and strong interaction processes, ensuring thermodynamic stability. As a consequence, particle abundances adjust according to the available reaction channels. Several reactions may occur during the evolution from a hot PNS to a cold NS.
At sufficiently high densities, the nucleon Fermi energy may exceed the rest mass of hyperons, making the appearance of strange baryons energetically favorable. In addition to hyperon formation and decay processes, beta decay and inverse beta decay play a central role in establishing beta equilibrium in dense matter. The corresponding equilibrium conditions can be expressed in terms of a small set of chemical potentials associated with the conserved charges of the system:
\begin{equation}
\label{generalchemicaleq}
\mu_j = \mathscr{B}_{j} \ \mu_B + \mathscr{Q}_{j} (\mu_q + \mu_l) + \mathfrak{L}_{j} \ \mu_l\, .
\end{equation}
Here, $\mu_j$ denotes the chemical potential of species $j$, while $\mu_B$, $\mu_q$, and $\mu_l$ are the baryon, charge, and lepton chemical potentials, respectively. Eq.~(\ref{generalchemicaleq}) applies to both baryons and leptons and fully specifies the conditions for beta equilibrium. The same expression is valid for both NSs and PNSs, although $\mu_l$ is nonzero only in neutrino-trapped matter, where the lepton fraction is conserved. If strangeness were conserved, as in the early stages of a heavy-ion collision, an additional strange chemical potential, $\mu_{\mathscr{S}}$, would be required.

Recent observational advances have considerably improved the constraints on the properties of neutron stars. Measurements from NASA's NICER mission have provided reliable estimates of stellar radii~\cite{Riley:2019yda, Miller:2019cac}, while observations by the LIGO and Virgo collaborations have constrained their tidal deformabilities \cite{LIGOScientific:2017vwq} and maximum masses~\cite{Rezzolla:2017aly}. As a result, many models of dense matter that are inconsistent with these observations have been significantly constrained or ruled out. At the same time, progress in the modeling of core-collapse supernovae and neutron-star mergers has highlighted the importance of finite-temperature effects in dense matter. Traditional approaches based on \emph{ad hoc} thermal corrections are often inadequate because they do not allow the composition of matter to evolve consistently with temperature (see Fig.~\ref{PhaseDiagram}), potentially leading to significant inaccuracies. In this context, the present work aims to provide a consistent description of dense matter in compact stars beyond the zero-temperature approximation with an adjustable parameter that controls the effects of many-body forces.

\section{\label{sec:MBFformalism}The Many-Body Forces (MBF) Model}

\subsection{\label{sec:MBFgeneral}General formalism}

Inspired by extensions of the ZM Model~\cite{Koepf:1991frx,Delfino:1995ea}, the MBF Model proposed by Taurines \emph{et al.}~\cite{Taurines:2000xz,PhysRevC.63.065801} introduces meson-field-dependent couplings that effectively incorporate many-body interactions. Accordingly, it became known as the \emph{Many-Body Forces} (MBF) Model. The model generalizes both the minimal coupling of the Walecka Model~\cite{Walecka:1974qa} and the derivative coupling of the ZM Model~\cite{PhysRevC.42.1416}, allowing for a continuous family of effective interactions controlled by its parameterization. The MBF framework has been successfully applied to the study of nuclear compressibility~\cite{Dexheimer:2007mt}, hyperonic matter~\cite{Gomes:2014aka}, deconfinement transitions~\cite{Jacobsen:2010zz}, and strongly magnetized neutron stars~\cite{Gomes:2017zkc}. The most complete version of the MBF Model~\cite{Gomes:2014aka} contains eight baryon fields ($\psi_b$), six meson fields ($\sigma$, $\omega$, $\boldsymbol{\varrho}$, $\boldsymbol{\delta}$, $\phi$, and $\sigma^*$), and free lepton fields ($\psi_l$). The $\boldsymbol{\delta}$ meson improves the description of isospin-asymmetric matter, whereas the strange mesons ($\sigma^*$ and $\phi$) play an important role in hyperon interactions.

Using natural units ($\hbar=c=k_B=1$), the MBF Lagrangian density is given by
\begin{equation}\begin{split}\label{lagrangianaMBF}
\mathcal{L}&= \underset{b}{\sum}\bar{\psi}_{b}\left[\gamma_{\mu}\left(i\partial^{\mu}-g_{\omega b \xi}^{*}\omega^{\mu} -g_{\phi b \kappa}^{*}\phi^{\mu}
-\frac{1}{2}g_{\varrho b \eta}^{*}\boldsymbol{\tau}\cdot\boldsymbol{\varrho}^{\mu}\right) \right.
\\& \left. - \left(m_b-g_{\sigma b \zeta}^{*}\sigma-g_{\sigma^* b \zeta}^{*}\sigma^*
-\frac{1}{2}g_{\delta b \zeta }^{*}\boldsymbol{\tau}.\boldsymbol{\delta}\right) \right]\psi_{b}
\\& +\left(\frac{1}{2}\partial_{\mu}\sigma\partial^{\mu}\sigma-m_{\sigma}^{2}\sigma^{2}\right)
+\left(\frac{1}{2}\partial_{\mu}\sigma^*\partial^{\mu}\sigma^*-m_{\sigma^*}^{2}\sigma^{* 2}\right)
\\&+\left(-\frac{1}{16 \pi}\omega_{\mu\nu}\omega^{\mu\nu}+ \frac{1}{8 \pi} m_{\omega}^{2}\omega_{\mu}\omega^{\mu}\right)
\\&+\left(-\frac{1}{16 \pi} \phi_{\mu\nu}\phi^{\mu\nu}+ \frac{1}{8 \pi}m_{\phi}^{2}\phi_{\mu}\phi^{\mu}\right)
\\&+\left(-\frac{1}{16 \pi} \boldsymbol{\varrho}_{\mu\nu}\cdot\boldsymbol{\varrho}^{\mu\nu}+\frac{1}{8 \pi}m_{\varrho}^{2}\boldsymbol{\varrho}_{\mu}\cdot\boldsymbol{\varrho}^{\mu}\right)
\\&+\left(\frac{1}{2}\partial_{\mu}\boldsymbol{\delta.}\partial^{\mu}\boldsymbol{\delta}-m_{\delta}^{2}\boldsymbol{\delta}^{2}\right)
+\underset{l}{\sum}\bar{\psi}_{l}\gamma_{\mu}\left(i\partial^{\mu}-m_{l}\right)\psi_{l}\, , 
\end{split}\end{equation}
where 
$\omega_{\mu\nu}=\partial_{\mu}\omega_{\nu}-\partial_{\nu}\omega_{\mu}$,
$\boldsymbol{\varrho}_{\mu\nu}=\partial_{\mu}\boldsymbol{\varrho}_{\nu}-\partial_{\nu}\boldsymbol{\varrho}_{\mu}$ and
$\phi_{\mu\nu}=\partial_{\mu}\phi_{\nu}-\partial_{\nu}\phi_{\mu}$,
and the operators $\boldsymbol{\tau}=(\tau_1,\,\tau_2,\,\tau_3)$ denote the Pauli isospin matrices. Moreover, the subscripts $b$ and $l$ label, respectively, the baryon octet ($n$, $p$, $\Lambda^0$, $\Sigma^-$, $\Sigma^0$, $\Sigma^+$, $\Xi^-$, $\Xi^0$) and lepton ($e^-$, $\mu^-$, \vd{$\nu$}) degrees of freedom. The meson-baryon coupling is introduced by the coupling constants $g^*$ present in the first term of Eq.~(\ref{lagrangianaMBF}).
\begin{table}
\caption{\label{particles}%
Properties of baryons and leptons included in the MBF Model. The first column presents the different particles and $I^3$, $\mathscr{B}$, $\mathfrak{L}$, $\mathscr{Q}$ and $\mathscr{S}$ stand for the isospin projection in the $z$ direction, baryon number, lepton number, electric charge (in units of the elementary charge) and strangeness, respectively. The masses follow the values provided by the \emph{Particle Data Group} (\emph{PDG})~\cite{ParticleDataGroup:2024cfk}, rounded to four significant digits. 
}
\begin{center} 
\begin{ruledtabular}
\begin{tabular}{cccccccccc} 
particle & mass $(\mathrm{\rm{MeV}})$ & $I^{3}$ & $\mathscr{B}$ & $\mathfrak{L}$ & $\mathscr{Q}$ & $\mathscr{S}$ \tabularnewline
\hline
$p$ & $939.6$ & $1/2$ & $1$ & $0$ & $+1$ & $0$  \tabularnewline
$n$ & $938.3$ & $-1/2$ & $1$ & $0$ & $0$ & $0$ \tabularnewline
$\Lambda^{0}$ & $1116$ & $0$ & $1$ & $0$ & $0$ & $-1$ \tabularnewline
$\Sigma^{+}$ & $1189$ & $+1$ & $1$ & $0$ & $+1$ & $-1$ \tabularnewline
$\Sigma^{0}$ & $1193$ & $0$ & $1$ & $0$ & $0$ & $-1$ \tabularnewline
$\Sigma^{-}$ & $1197$ & $-1$ & $1$  & $0$ & $-1$ & $-1$ \tabularnewline
$\Xi^{0}$ & $1315$ & $+1/2$ & $1$ & $0$ & $0$ & $-2$ \tabularnewline
$\Xi^{-}$ & $1322$ & $-1/2$ & $1$ & $0$ & $-1$ & $-2$ \tabularnewline
$e^{-}$ & $0.5110$ & $0$ & $0$ & $1$ & $-1$ & $0$  \tabularnewline
$\mu^{-}$ & $105.7$ & $0$ & $0$ & $1$ & $-1$ & $0$ \tabularnewline
\end{tabular}
\end{ruledtabular}
\par\end{center}
\end{table}
The properties of all the particles included in the formalism are shown in Table \ref{particles}, and the meson fields are listed in Table \ref{mesons}.

\begin{table}
\caption{\label{mesons}%
Meson fields in the MBF Model. Particle nomenclature follows that of the \emph{PDG}~\cite{ParticleDataGroup:2024cfk}. The masses of the five heavier mesons shown in this table follow the \emph{PDG} average or estimate values, truncated in the units place. Since \emph{PDG} does not provide an average or estimate value for the mass of $\sigma$, we adopt the upper value in the range provided by this same reference, which is in accordance with~\cite{Pelaez:2015qba}. 
}
\begin{center} 
\begin{ruledtabular}
\resizebox{1.0\linewidth}{!}{
\begin{tabular}{ccccc} 
meson & particle & classification & coupling & mass \\ 
& & & constant & $(\mathrm{\rm{MeV}})$  \\
\hline 
$\sigma $ &  $f_0(500) $ & scalar-isoscalar & $g_{\sigma_b}$ & $550$ \tabularnewline  
$\boldsymbol{\delta}$ & $a_0 (980)$& scalar-isovector &$g_{\delta_b}$ & $980$ \tabularnewline 
$\omega_\mu $ & $\omega (782) $& vector-isoscalar &$g_{\omega_b}$ & $783$ \tabularnewline 
$\boldsymbol{\varrho}_{\mu}$  & $\varrho (770)$ & vector-isovector & $g_{\varrho_b}$& $775$ \tabularnewline
$\sigma^\ast $ & $f_0 (980)$ & scalar-isoscalar &$g_{\sigma^*_b}$ & $990$ \tabularnewline  
$\phi_\mu $ & $\phi (1020) $ & vector-isoscalar & $g_{\phi_b}$ & $1019$ \tabularnewline
\end{tabular}}
\end{ruledtabular}
\end{center}
\end{table}

The general definition of the meson-baryon couplings is:
\begin{equation}\label{acoplamentosMBF}
\begin{split}
& g_{\sigma b \zeta}^{*}\equiv \Pi_{\zeta b} g_{\sigma b},
\quad g_{\sigma^{*} b \zeta}^{*}\equiv \Pi_{\zeta b} g_{\sigma^* b},
\quad g_{\delta b \zeta}^{*}\equiv \Pi_{\zeta b} g_{\delta b}, \\ &
g_{\omega b \xi}^{*}\equiv \Pi_{\xi b} g_{\omega b},
\quad g_{\varrho b \kappa}^{*}\equiv \Pi_{\kappa b} g_{\varrho b},
\quad g_{\phi b \eta}^{*}\equiv \Pi_{\eta b} g_{\phi b},
\end{split}
\end{equation}
where the parametric term $\Pi_{\lambda b}$  introduces the nonlinear contributions that define the MBF Model:
\begin{equation}\label{mstar}
\Pi_{\lambda b}\equiv \left(1+\frac{g_{\sigma b}\sigma+g_{\sigma^* b}\sigma^*
+\frac{1}{2}g_{\delta b}\boldsymbol{\tau.\delta}}{\lambda\,m_{b}}\right)^{-\lambda},
\end{equation}
for $\lambda= \xi,\,\kappa,\,\eta,\,\zeta$. 

In agreement with the Lagrangian density (\ref{lagrangianaMBF}), the effective mass of baryons $m_{b}^{*}$ is defined as:
\begin{equation}
 \label{meffMBF}
 m_{b}^{*} = m_{b} - \Pi_{\zeta b} \left( g_{\sigma b}\sigma+g_{\sigma^* b}\sigma^* + \frac{1}{2} g_{\delta b } \boldsymbol{\tau.\delta} \right)\, .
\end{equation}

In expression (\ref{mstar}), $\lambda$ is an adjustable parameter that can assume different values $\xi$, $\kappa$, $\eta$ or $\zeta$ to regulate the intensity of such coupling for $\omega$, $\varrho$, $\phi$ or the scalar mesons, respectively. From expressions (\ref{acoplamentosMBF}) and (\ref{mstar}) it can be seen that the coupling constants of the model depend on the scalar fields and, since these fields vary as a function of density according to their equations of motion, the couplings have an indirect density dependence. In short, the MBF Model presents \emph{adjustable meson-nucleon couplings}. This type of adjustable coupling provides a great freedom of choice, since we can vary the indices $\xi$, $\kappa$, $\eta$ and $\zeta$ continuously to obtain different values for the relevant physical quantities.

The range of relevant values for these parameters is not as large as one might initially guess. Due to the general form of the couplings, presented in (\ref{mstar}), there is a rapid convergence to an exponential form of coupling. In order to verify this fact, it suffices to let the adjustable parameters tend to infinity ($\lambda \to \infty$):
\begin{equation}
\begin{split}
\label{exponential coupling} \lim_{\lambda \to \infty} \Pi_{\lambda b}& =
\lim_{\lambda \to \infty} \left( 1+\frac{g_{\sigma b}\sigma+g_{\sigma^* b}\sigma^*
+\frac{1}{2}g_{\delta b}\boldsymbol{\tau.\delta}}{\lambda\,m_{b}} \right)^{-\lambda} \\& = \lim_{\lambda \to \infty} \left( 1+\frac{x}{\lambda} \right)^{-\lambda} = e^{-x} ,
\end{split}
\end{equation}
where we have defined 
\begin{equation}
x \equiv (g_{\sigma b}\sigma+g_{\sigma^* b}\sigma^*
+\frac{1}{2}g_{\delta b}\boldsymbol{\tau.\delta})/m_b\, . 
\end{equation}
Given that sufficiently small values for $\lambda$ can be chosen, it is possible to carry out the explicit binomial expansion of the parametric term $\Pi_{\lambda b}$ in the following way:
\begin{equation}
\begin{split}
    \Pi_{\lambda b}  =  & 1  - \left( \frac{g_{\sigma b}\sigma+g_{\sigma^* b}\sigma^*
+\frac{1}{2}g_{\delta b}\boldsymbol{\tau.\delta}}{m_{b}}\right) \\& + \frac{(\lambda+1)}{2! \lambda} {\left( \frac{g_{\sigma b}\sigma+g_{\sigma^* b}\sigma^*
+\frac{1}{2}g_{\delta b}\boldsymbol{\tau.\delta}}{m_{b}} \right)}^2 \\ &
    - \frac{(\lambda^2 +3\lambda +2)}{3! \lambda^2} {\left( \frac{g_{\sigma b}\sigma+g_{\sigma^* b}\sigma^*
+\frac{1}{2}g_{\delta b}\boldsymbol{\tau.\delta}}{m_{b}}\right)}^3 + ...
    \end{split}
\end{equation}

The crossed terms, of order higher than one, are interpreted as many-body contributions. Important to emphasize, however, that $\lambda$ does not need to be small, and different $\lambda$ values can be seen as producing \emph{different models}, although we conservatively refer to these as \emph{different parameterizations} or \emph{different versions}. 

The dependence of the many-body interactions is introduced in the meson-baryon couplings through the parameters $\xi$, $\kappa$, $\eta$ or $\zeta$, which will define the character of the many-body contribution to each meson field. Thus, in the MBF formalism, the parameter space to be investigated is expanded and the many-body forces influence can be easily controlled in an effective manner. To reduce the number of parameters to be studied, we restrict ourselves to the standard parameterizations of the MBF Model, as summarized in Table \ref{classesMBF}.

\begin{table}  
\caption{Nomenclature for the standard parameterizations of the MBF Model. S: scalar version; V: vector version; SVIS: scalar-vector-isoscalar version; SVIV: scalar-vector-isovector version.} 
\begin{center} \label{classesMBF}
\begin{ruledtabular}
\begin{tabular}{ccccc} 
model & $\zeta$ & $\xi$ & $\kappa$ & $\eta$  \\
\hline
S &  $\neq 0 $ & $= 0$ & $=0$ & $=0$ \tabularnewline  
V & $\neq 0 $ & $\neq 0 $ &$=0$ &$=0$\tabularnewline 
SVIS & $\neq 0 $& $\neq 0 $ & $\neq 0 $ &$=0$ \tabularnewline 
SVIV  & $\neq 0 $& $\neq 0 $ & $\neq 0 $ & $\neq 0 $ \tabularnewline 
\end{tabular}
\end{ruledtabular}
\end{center}
\end{table}

Most of the literature on the MBF Model has focused on the scalar (S) and scalar-vector-isoscalar (SVIS) parameterizations, and in the latter case, the constraint $\zeta=\xi=\kappa$ has commonly been adopted~\cite{Taurines:2000xz, PhysRevC.63.065801}. However, the majority of works have focused on the scalar version (S) of the MBF Model~\cite{Gomes:2014aka}, since it has presented more consistent results for both the properties of nuclear matter and the masses and radii of NSs. For this reason, in order to allow a comparison of the effects of including temperature in the MBF Model (something unprecedented until now) with most of the results already obtained for this model, the present work will henceforth explore only the S version of the MBF Model.

Therefore, from this point on, adopting the scalar version (S) of the MBF Model, the equations are controlled by a single adjustable parameter ($\zeta$). In this scenario, the general definition of the meson-baryon couplings is:
\begin{equation}\label{MBFE couplings}
\begin{split}
& g_{\sigma b \zeta}^{*}\equiv \Pi_{\zeta b} g_{\sigma b} \, ,
\quad g_{\sigma^{*} b \zeta}^{*}\equiv \Pi_{\zeta b} g_{\sigma^* b} \, ,
\quad g_{\delta b \zeta}^{*}\equiv \Pi_{\zeta b} g_{\delta b} \, , \\ &
g_{\omega b \xi}^{*}\equiv g_{\omega b} \, ,
\quad \, \, \, \, \, \,  \quad g_{\varrho b \kappa}^{*}\equiv g_{\varrho b} \, , \quad \, \, \, \, \,
\quad g_{\phi b \eta}^{*}\equiv g_{\phi b} \, .
\end{split}
\end{equation}

Because the meson-baryon couplings depend on scalar meson fields, which themselves depend on the thermodynamic state of the system through the self-consistent field equations, the MBF Model is implicitly density dependent. In this sense, it shares several advantages of density-dependent RMF models, while preserving the variational structure of conventional RMF theories.

This point deserves some clarification, since it is directly related to thermodynamic consistency. In standard density-dependent RMF models, the meson-baryon couplings are prescribed as explicit functions of the baryon density, $g_i=g_i(n_B)$. As a consequence, the variation of the energy density with respect to the baryon fields generates additional terms proportional to $\partial g_i/\partial n_B$. These are the rearrangement contributions, which must be included in the baryon self-energy in order to preserve thermodynamic consistency, including the Gibbs–Duhem relation and the Hugenholtz–Van Hove theorem~\cite{Fuchs:1995as,Typel:1999yq,Frohaug:2025okz,Huang:2022kej,gxyr-v5h7}.

The situation is different in the MBF Model. Here, the effective couplings are not prescribed as explicit functions of $n_B$, but rather as functions of the scalar meson fields,
\begin{equation}
g_i^* = g_i^*(\sigma,\delta,\sigma^*) \, .
\end{equation}
The density dependence of the model is therefore only implicit, arising through the self-consistent solutions of the meson field equations. Equivalently, one may write schematically
\begin{equation}
\frac{d g_i^*}{d n_B}
=
\frac{\partial g_i^*}{\partial \sigma}\frac{d\sigma}{dn_B}
+
\frac{\partial g_i^*}{\partial \delta}\frac{d\delta}{dn_B}
+
\frac{\partial g_i^*}{\partial \sigma^*}\frac{d\sigma^*}{dn_B}\, ,
\end{equation}
showing that the apparent density dependence is mediated entirely by dynamical meson fields. When the Lagrangian is varied, the derivatives of $g_i^*$ with respect to these fields enter the scalar meson equations of motion directly. Thus, the medium dependence of the interaction is already accounted for at the level of the Euler-Lagrange equations.

For this reason, no independent rearrangement contribution appears in the baryon equation of motion. In other words, the MBF Model does not omit the rearrangement terms required in explicitly density-dependent RMF theories; rather, such terms are not generated in the first place, because the couplings are field-dependent instead of density-dependent. The thermodynamic consistency of the model is therefore maintained by the same variational structure that defines the field equations and the energy-momentum tensor.

\subsection{\label{sec:equationsMFT}Equations of motion and mean-field approximation}

The equations of motion for the meson and baryon fields can be obtained by the Euler-Lagrange equations from the Lagrangian density of the model (\ref{lagrangianaMBF}), as follows:
\begin{equation}
\label{eqsigmaMBF}
\begin{split}
(\partial_{\mu}\partial^{\mu}+m_ {\sigma}^2)\sigma
& = \sum_{b}\bar{\psi}_b \Bigg[ g_{\sigma b} (\Pi_{\zeta b}) - \frac{ g_{\sigma b}}{m_b} {(\Pi_{\zeta b})}^{\frac{\zeta + 1}{\zeta}} \\& \times
\left( g_{\sigma b} + \frac{1}{2} g_{\delta b} \boldsymbol{\tau.\delta} + g_{\sigma^* b} \sigma^* \right) \Bigg] \psi_b\, ,
\end{split}
\end{equation}

\begin{equation}
\label{eqomegaMBF}
-\partial^{\nu}\omega_{\mu\nu}+ m_{\omega} ^2\omega_{\mu}=
\sum_{b} g_{\omega b}
\bar{\psi}_b\gamma_{\mu}\psi_b\, ,
\end{equation}

\begin{equation}
\label{eqrhoMBF}
-\partial^{\nu}\boldsymbol{\varrho}_{\mu\nu}+ m_{\varrho} ^2
\boldsymbol{\varrho}_{\mu}= \frac{1}{2}
\sum_{b} g_{\varrho b}
\bar{\psi}_b\gamma_{\mu} \boldsymbol{\tau} \psi_b\, ,
\end{equation}

\begin{equation}
\label{eqdeltaMBF}
\begin{split}
(\partial_{\mu}\partial^{\mu}+m_ {\sigma}^2)\boldsymbol{\delta}
& =\sum_{b}\bar{\psi}_b \Bigg[ g_{\sigma b} (\Pi_{\zeta b}) - \frac{ g_{\sigma b}}{m_b} {(\Pi_{\zeta b})}^{\frac{\zeta + 1}{\zeta}} \\& \times \left( g_{\sigma b} + \frac{1}{2} g_{\delta b} \boldsymbol{\tau.\delta} + g_{\sigma^* b} \sigma^* \right) \Bigg] \frac{\boldsymbol{\tau}}{2} \psi_b\, ,
\end{split}
\end{equation}

\begin{equation}
\label{eqsigmaasMBF}
\begin{split}
(\partial_{\mu}\partial^{\mu}+m_ {\sigma}^2)\sigma^*
&=\sum_{b}\bar{\psi}_b \Bigg[ g_{\sigma^* b} (\Pi_{\zeta b}) - \frac{ g_{\sigma^* b}}{m_b} {(\Pi_{\zeta b})}^{\frac{\zeta + 1}{\zeta}} \\& \times \left( g_{\sigma b} + \frac{1}{2} g_{\delta b} \boldsymbol{\tau.\delta} + g_{\sigma^* b} \sigma^* \right) \Bigg] \psi_b\, ,
\end{split}
\end{equation}

\begin{equation}
\label{eqphiMBF}
-\partial^{\nu}\phi_{\mu\nu}+ m_{\omega} ^2\phi_{\mu}=
\sum_{b} g_{\phi b}
\bar{\psi}_b\gamma_{\mu}\psi_b\, .
\end{equation}

Eqs.~(\ref{eqsigmaMBF})--(\ref{eqphiMBF}) correspond to Klein-Gordon-type equations for the scalar mesons ($\sigma$, $\delta$, $\sigma^*$) and Proca-type equations for the vector mesons ($\omega$, $\varrho$, $\phi$), with source terms determined by the baryonic currents and densities.

For the baryon fields $\psi_b$,
\begin{equation}
\label{eqbarionsMBF}
\begin{split}
\sum_b & \Bigg[\gamma_\mu \Bigg( i \partial^\mu - g_{\omega b} \omega^\mu - \frac{1}{2} g_{\varrho b} \boldsymbol{\tau}.\boldsymbol{\varrho}^{\mu} - g_{\phi b} \phi^{\mu} \Bigg)  \\& - \Bigg( m_b - g_{\sigma b}^* \sigma-g_{\sigma^* b}^* \sigma^* - \frac{1}{2} g_{\delta b}^* \boldsymbol{\tau.\delta} \Bigg) \Bigg] \psi_b = 0 \, ,
\end{split}
\end{equation}
which corresponds to the Dirac equation in the presence of the meson fields. The scalar mesons generate the effective baryon mass $m_b^*$, according to (\ref{meffMBF}), while the vector mesons shift the kinetic term. As a consequence, the effective baryon chemical potential becomes
\begin{equation}
\label{mueffMBF}
\mu_b^*  = \mu_b - g_{\omega b} \gamma_\mu \omega^\mu - \frac{1}{2}  g_{\varrho b} \gamma_\mu \boldsymbol{\tau}.\boldsymbol{\varrho}^{\mu} - g_{\phi b} \gamma_\mu \phi^{\mu}\, ,
\end{equation}
where $\mu_b$ is the baryon chemical potential defined in Eq.~(\ref{generalchemicaleq}).

Finally, for the lepton fields, 
\begin{equation}
\label{eqleptonsMBF}
\sum_l ( i \gamma_\mu \partial^{\mu}-m_l )
\psi_l =0\, ,
\end{equation}
which corresponds to the free Dirac equation for leptons of mass $m_l$, reflecting the absence of meson-lepton couplings.

As in most RMF calculations for dense matter, Coulomb interactions can be safely neglected explicitly, and taken into account only through the charge-neutrality condition given by (\ref{eqdecarga}), what is called in literature a “Coulomb-less” model description
\cite{Hempel:2013tfa}
{\color{blue}.}

The nonlinear meson-field equations (\ref{eqsigmaMBF})--(\ref{eqphiMBF}) are solved within the mean-field approximation. This approximation, widely employed in relativistic many-body theory
\cite{osti_5938050}, replaces the meson fields by their expectation values, an approach that becomes increasingly accurate at high densities where field fluctuations are relatively small. The meson fields are rewritten as:
\begin{equation}
\begin{split}
& \sigma \rightarrow \langle \sigma \rangle = \sigma_0 , \\&
\sigma^* \rightarrow \langle \sigma^* \rangle = \sigma_0^* , \\&
 \omega^\mu \rightarrow \langle \omega^\mu \rangle = \delta_0^\mu \omega_0 ,  \\& 
\phi^\mu \rightarrow \langle \phi^\mu \rangle = \delta_0^\mu \phi_0 , \\&
 \boldsymbol{\varrho}^{\mu} \rightarrow \langle \boldsymbol{\varrho}^{\mu} \rangle = \delta_0^\mu \varrho_{03} , \\&
 \boldsymbol{\delta} \rightarrow \langle \boldsymbol{\delta} \rangle = \delta_0^3\, ,
\end{split}
\end{equation}
where $\delta_0^\mu$ is the Kroenecker delta function and $\sigma_0$, $\sigma_0^*$, $\omega_0$, $\phi_0$, $\varrho_{03}$, and $\delta_{0}^3$ denote the classical expectation values of the meson fields. It is important to highlight that, in these equivalences, the $z$ axis was chosen as the quantization axis in isospin space, so that:
\begin{equation}
\begin{split}
& \frac{1}{2} \boldsymbol{\tau}.\boldsymbol{\varrho}^{\mu} \rightarrow \frac{1}{2} \langle \boldsymbol{\tau}.\boldsymbol{\varrho}^{\mu} \rangle = \frac{1}{2} \tau^i \delta_i^\mu \varrho_{03} = I^3 \delta_0^\mu \varrho_{03} = I^3 \varrho_{0}^3\, , \\ &
\frac{1}{2} \boldsymbol{\tau}.\boldsymbol{\delta} \rightarrow \frac{1}{2} \langle\boldsymbol{\tau}.\boldsymbol{\delta} \rangle = \frac{1}{2} \tau^i \delta_{i3} \delta_{0}^3 = I^3 \delta_{0}^3\, ,
\end{split}
\end{equation}
where $I^3$ corresponds to the isospin projection of each baryon in the $z$ direction, whose values are found in Table \ref{particles}.

Applying the mean-field approximation to Eqs.~(\ref{eqsigmaMBF}), (\ref{eqomegaMBF}), (\ref{eqrhoMBF}), (\ref{eqdeltaMBF}), (\ref{eqsigmaasMBF}) and (\ref{eqphiMBF}), we obtain:
\begin{equation}\begin{split}\label{tcm}
\sigma_{0}&=\frac{1}{m_{\sigma}^{2}}\underset{b}{\sum}
\Bigg[g_{\sigma b}\left(\Pi_{\zeta b}\right) 
\\ &  -\frac{g_{\sigma b}}{m_b}
\left(\Pi_{\zeta b}\right)^{\frac{\zeta+1}{\zeta}}
\left(g_{\sigma b}\sigma_{0} + g_{\delta b}\delta_{0}^{3}\,I^{3b}+g_{\sigma^* b}\sigma_{0}^*\right)\Bigg] n_{s_b}, \\
\omega_{0}&=\frac{1}{m_{\omega}^{2}}\underset{b}{\sum}g_{\omega b} \,  n_{B_b}, \\
\varrho_{0}^{3}&=\frac{1}{m_{\varrho}^{2}}\underset{b}{\sum}g_{\varrho b}\, I^{3b}\,n_{B_b},\\
\delta_{0}^{3}&=\frac{1}{m_{\delta}^{2}}\underset{b}{\sum}
\Bigg[g_{\delta b}\left(\Pi_{\zeta b}\right) 
\\ & -\frac{g_{\delta b}}{m_b}
\left(\Pi_{\zeta b}\right)^{\frac{\zeta+1}{\zeta}}
\left(g_{\sigma b}\sigma_{0} + g_{\delta b}\delta_{0}^{3}\,I^{3b}+g_{\sigma^* b}\sigma_{0}^*\right)\Bigg]I^{3b}\,n_{s_b},\\
\sigma^*_{0}&=\frac{1}{m_{\sigma^*}^{2}}\underset{b}{\sum}
\Bigg[g_{\sigma^* b}\left(\Pi_{\zeta b}\right) 
\\ & -\frac{g_{\sigma^* b}}{m_b}
\left(\Pi_{\zeta b}\right)^{\frac{\zeta+1}{\zeta}}
\left(g_{\sigma b}\sigma_{0} + g_{\delta b}\delta_{0}^{3}\,I^{3b}+g_{\sigma^* b}\sigma_{0}^*\right)\Bigg] n_{s_b}, \\
\phi_{0}&=\frac{1}{m_{\phi}^{2}}\underset{b}{\sum}g_{\phi b} \, n_{B_b}\, .
\end{split}\end{equation}
In the above equations, the scalar density and baryon density are calculated, respectively, as $n_{s_b} = \langle \bar{\psi_b} \psi_b \rangle$ and $n_{B_b} = \langle \psi_b^{\dagger} \psi_b \rangle $ and, in accordance with (\ref{mstar}), 
\begin{equation}\label{mstarTCM}
\Pi_{\zeta b} = \left(1+\frac{g_{\sigma b}\sigma_0+g_{\sigma^* b}\sigma_0^*
+g_{\delta b} I^{3b} \delta_0^3}{\zeta\,m_{b}}\right)^{-\zeta}\, .
\end{equation}

To calculate $n_{s_b}$ and $n_{B_b}$, the equation of motion for the baryon fields $\psi_b$ is also considered in the mean field approximation (for mesons only), from (\ref{eqbarionsMBF}):
\begin{equation}
\label{eqbarionsMBFtcm}
\begin{split}
\sum_b ( i \gamma_\mu \partial^\mu & - g_{\omega b} \gamma_0 \omega_0 -  g_{\varrho b} \gamma_0 I^{3b} \varrho_{0}^3  \\& -  g_{\phi b} \gamma_0 \phi_0 - m_b^* ) \psi_b = 0\, . 
\end{split}
\end{equation}
In the mean-field approximation, according to (\ref{meffMBF}) and (\ref{mueffMBF}), the effective mass and effective chemical potential of each baryon species can be rewritten respectively as:
\begin{equation}
\label{meffMBFtcm}
m_{b}^{*} = m_{b} - \Pi_{\zeta b} \left( g_{\sigma b}\sigma_0+g_{\sigma^* b}\sigma_0^* + g_{\delta b } I^{3b} \delta_0^3 \right)\, ,
\end{equation}
\begin{equation}
\label{mueffMBFtcm}
\mu_b^* = \mu_b - g_{\omega b} \omega_0 - g_{\varrho b} I^{3b} \varrho_0^3 - g_{\phi b} \phi_0\, .
\end{equation}
Clearly, the lepton equation of motion, given by (\ref{eqleptonsMBF}), does not undergo any modification due to the application of the mean field approximation, since it does not depend on the meson fields.

\subsection{\label{sec:EoSMBF}EoS, densities and entropy at finite temperature}

To determine the baryon density $n_{B_b} = \langle\psi_b^{\dagger} \psi_b\rangle $ and the scalar density $n_{s_b} = \langle\bar{\psi_b} \psi_b\rangle$, it is necessary to consider the solutions to the modified Dirac equation (\ref{eqbarionsMBFtcm}) that describes the fields $\psi_b$. Since the mean-field approximation maintains the form of the equation and since, in this approximation, the scalar and vector mesons are represented by classical fields (i.e., constant functions in spacetime), the solutions will have the same form obtained for the free particle case. In short, they will be typical stationary plane wave solutions, with a displacement in the mass term, due to the scalar fields, according to (\ref{meffMBFtcm}), and also a displacement in the energy eigenvalues, due to the vector fields, according to (\ref{mueffMBFtcm}). Then, from the expressions for $\psi_b$ and $\psi_b^{\dagger}$ and the anti-commutation relations between the creation and annihilation operators of baryons and antibaryons, $n_{B_b}$ and $n_{s_b} $ can easily be obtained.

Regarding the calculations of energy density and pressure to construct the equation of state (EoS), both quantities can be obtained from the components of the energy-momentum tensor. In quantum field theory, the energy-momentum tensor is defined as
\begin{equation}
\label{tem} T_{\mu\nu} = -\mathcal{L}g_{\mu\nu}+
\frac{\partial\mathcal{L}}{\partial (\partial^{\mu}\psi)}
\partial_{\nu} \psi\, ,
\end{equation}
where $g_{\mu\nu}$ is the Minkowsky metric tensor. If the field $\psi$ satisfies a Dirac-like equation, such as equations (\ref{eqbarionsMBF}) and (\ref{eqleptonsMBF}), the first term in (\ref{tem}) vanishes, leaving:
\begin{equation}
\label{tensorcomponents}
T_{\mu\nu} = i\bar{\psi}\gamma^{\mu}\partial_{\nu}\psi\, . \end{equation}
Assuming and ideal fluid, one can
identify
\begin{equation}
\label{Tperfectfluid} \langle T_{\mu\nu}\rangle = (\varepsilon + p)\, u_{\mu} u_{\nu} - p g_{\mu\nu}\, ,
\end{equation}
as being the structure of the energy-momentum tensor and where the four-velocity, for a fluid at rest, is described by $u^{\mu} = (1,\bf{0})$, and satisfies ${u_{\mu}}^2 = 1$. The energy density $\varepsilon$ and the pressure $p$ of the system are thus determined by the expected value of the energy-momentum tensor. \begin{equation}
\label{t00e} \varepsilon = \langle T_{00}\rangle\, ,
\end{equation}
\begin{equation}
\label{tijp} p = \frac{1}{3} (\langle T_{xx}\rangle+ \langle T_{yy}\rangle + \langle T_{zz} \rangle )\, ,
\end{equation}
where the energy-momentum tensor assumes the symmetric and isotropic form. \begin{equation}
\label{} T_{\mu\nu} = \left(\begin{array}{cccc}
\varepsilon&0&0&0\\0&p&0&0\\0&0&p&0\\0&0&0&p
\end{array}\right)\, .
\end{equation}
Using (\ref{tensorcomponents}) and considering the standard form of Dirac matrices, Eqs.~(\ref{t00e}) and (\ref{tijp}) become
\begin{equation}
\label{EnergCampoPsi}\varepsilon = i\bar{\psi}\gamma^{0}\partial_{0}\psi \, ,
\end{equation}
\begin{equation}
\label{PressCampoPsi}
p= -\frac{i}{3}\bar{\psi}\vec{\gamma} \cdot \vec{\nabla} \psi \, .
\end{equation}

In the finite temperature case ($T \neq 0$), the integrals over the Fermi momenta (involved in the calculations of $n_{B}$, $n_s$, $\varepsilon$ and $p$) must extend to infinity, and both particles and antiparticles must be included in the statistical description. Therefore, the results for all these physical quantities can be summarized as follows:
\begin{equation}
\label{rhoBMBF} n_{B} = \sum_{j=b} n_{B_j} = \frac{1}{\pi^2} 
\sum_{j=b} \int_0^{\infty}dk_j
\ {k_j}^2 \ (f_{j+} + f_{j-})\, ,
\end{equation}
\begin{equation}
\label{rhosMBF}
n_{s} = \sum_{j=b} n_{s_j} = \sum_{j=b} \frac{m_j^*}{\pi^2} 
 \int_0^{\infty}dk_j
\ \frac{{k_j}^2}{E_j} \ (f_{j+} + f_{j-})\, ,
\end{equation}
\begin{equation}
\label{EnergiaMBF}
\begin{split}
\varepsilon & = \frac{m_{\sigma}^2 \sigma_0^2}{2}  + \frac{m_{\delta}^2 {\delta_0^3}^2}{2} + \frac{m_{\sigma^*}^2 {\sigma_0^*}^2 }{2} + \frac{m_{\omega}^2 \omega_0^2}{2} \\& + \frac{m_{\varrho}^2 {\varrho_0^3}^2}{2}  + \frac{m_{\phi}^2 \phi_0^2}{2}  +  \sum_{j=b} \varepsilon_{j} + \sum_{j=l} \varepsilon_{j}\, ,
\end{split}
\end{equation}
\begin{equation}
\label{pressoaparMBF}
\begin{split}
p & = -\frac{m_{\sigma}^2 \sigma_0^2}{2}  - \frac{m_{\delta}^2 {\delta_0^3}^2}{2} -\frac{m_{\sigma^*}^2 {\sigma_0^*}^2 }{2} + \frac{m_{\omega}^2 \omega_0^2}{2} \\& + \frac{m_{\varrho}^2 {\varrho_0^3}^2}{2}  + \frac{m_{\phi}^2 \phi_0^2}{2}  +  \sum_{j=b} p_{j} + \sum_{j=l} p_{j}\, ,
\end{split}
\end{equation}
where
\begin{equation}
\label{EnergiaGasFermiTocteto}
\varepsilon_j = \frac{1}{\pi^2}\int_0^{\infty}dk_j
\ {k_j}^2 E_j \ (f_{j+} + f_{j-})\, ,  
\end{equation}
\begin{equation}
\label{PressaoGasFermiTocteto}
p_j=  \frac{1}{\pi^2}\int_0^{\infty}dk_j\
 \frac{{k_j}^4}{E_j} \ (f_{j+} + f_{j-})  ,
\end{equation}
being $k_j$ the momentum for each species $j$ of fermion (baryon $b$ or lepton $l$), $E_b = \sqrt{{k_b}^2+{(m_b^*)}^2}$ the energy for each species of baryon $b$; $E_l = \sqrt{{k_l}^2+{m_l}^2}$ the energy for each species of lepton $l$; additionally, $f_{b+}$ e $f_{b-}$ are the Fermi-Dirac distributions for baryons and antibaryons, respectively, and $f_{l+}$ e $f_{l-}$ are the Fermi-Dirac distributions for leptons and antileptons, respectively:
\begin{equation}
\label{fermi_distribution}
    f_{b \pm} = \Bigg[ \frac{1}{e^{{(E_b \mp \mu_b^*)}/{T}}+1} \Bigg]\, , \, \, \,  f_{l \pm} = \Bigg[ \frac{1}{e^{{(E_l \mp \mu_l)}/{T}}+1} \Bigg]\, .
\end{equation}
In the above expressions, $m_b^*$ and $\mu_b^*$ are given by Eqs.~(\ref{meffMBFtcm}) and (\ref{mueffMBFtcm}), respectively. 

It is important to highlight that, technically, Eq.~(\ref{rhoBMBF}) gives the \emph{net baryon density}, i.e., the number of particles minus the number of antiparticles, although the word ``net" is often omitted. Using the net baryon density allows the value to be conserved by particle-antiparticle creation and annihilation events.

The total entropy density reads:
\begin{equation}
\label{entropiaMBF}
s= \sum_{j=b} s_{j} + \sum_{j=l} s_{j}\, ,
\end{equation}
being
\begin{equation}
\begin{split}
    \label{EntropiaGasFermiOcteto} s_{j} & =  \frac{1}{\pi^2}\int_0^{\infty} dk_j \ {k_j}^2  \Bigg[   f_{j+} \ln \left( \frac{1}{{f_{j+}}} \right) +  f_{j-} \ln \left( \frac{1}{{f_{j-}}} \right)  \\
    &  + \left(1 - f_{j+} \right) \ln \left(  \frac{1}{1 - f_{j+}}   \right) + \left(1 - f_{j-} \right) \ln \left(  \frac{1}{1 - f_{j-}}   \right)  \Bigg] ,
\end{split}
\end{equation}
where the Fermi-Dirac distributions $f_{j+}$ and $f_{j+}$ follow the prescription of (\ref{fermi_distribution}). The above expression for $s_j$ is basically the formula for the entropy density of a free Fermi gas; however, it is important to emphasize that the interactions are implicit in the Fermi-Dirac distributions for baryons, as long as they contain both the effective chemical potential $\mu_b^*$ (which depends on the scalar meson fields) and the effective mass $m_b^*$ (which is present in the definition of the energy for baryons $E_b$ and depends on the vector meson fields). In this context, the entropy per baryon $S/A$ can be defined as:
\begin{equation}
    S/A = s/n_B\, , 
\end{equation}
where $s$ and $n_B$ are defined by expressions (\ref{entropiaMBF}) and (\ref{rhoBMBF}), respectively. 

\subsection{\label{sec:Ncouplings}Meson-nucleon couplings}

In the scalar (S) version of the MBF Model, each choice of $\zeta$ corresponds to a different parameterization of the model. Then, given a specific value for $\zeta$, the meson-nucleon coupling constants can be determined in two steps, for isospin symmetric and asymmetric matter, following the procedure described in detail in the Appendix. 

\begin{table}   
\caption{\label{quadro_acoplamentos}
Effective mass of the nucleon, compressibility, and symmetry energy, symmetry energy slope (all at saturation), and meson-nucleon coupling constants for different parameterizations of the scalar (S) version of the MBF Model. These results consider a saturation density $n_0= 0.15~\rm{fm}^{-3}$ and a binding energy per baryon $B/A = -15.75~\rm{MeV}$. The lines in bold describe the two parameterizations that will be focused on in the present work. 
}
\begin{ruledtabular}
\resizebox{1.0\linewidth}{!}{
\begin{tabular}{ccccccccc} 
$\zeta$ & $\frac{{(m_N^*)}_0}{m_N}$ & $K_0$ & $a_{sym}^0$ & $L_0$ & ${(\frac{g_{\sigma}}{m_{\sigma}})}^2$ & ${(\frac{g_{\omega}}{m_{\omega}})}^2$ & ${(\frac{g_{\varrho}}{m_{\varrho}})}^2$ & ${(\frac{g_{\delta}}{m_{\delta}})}^2$ \\ 
 & & \scriptsize(\rm{MeV}) & \scriptsize(\rm{MeV}) & \scriptsize(\rm{MeV}) & \scriptsize{($fm^2$)} & \scriptsize{($fm^2$)}  & \scriptsize{($fm^2$)}  & \scriptsize{($fm^2$)}  \\
\hline 
$0.040$ & $0.65$ & $297$ & $25$ & $76$ & $14.41$ & $8.68$ & $2.52$ & $0.38$ \\ 
$0.040$ & $0.65$ & $297$ & $25$ & $90$ & $14.41$  & $8.68$ & $6.49$ & $6.50$ \\
$\bf{0.040}$ & $\bf{0.65}$ & $\bf{297}$ & $\bf{33}$ & $\bf{100}$ & $\bf{14.41}$  & $\bf{8.68}$ & $\bf{4.69}$ & $\bf{0.38}$ \\ 
$0.040$ & $0.65$ & $297$ & $33$ & $114$ & $14.41$  & $8.68$ & $8.73$ & $6.64$ \\ \hline 
$0.049$ & $0.67$ & $272$ & $25$ & $74$ & $13.90$ & $8.08$ & $2.50$ & $0.16$ \\ 
$0.049$ & $0.67$ & $272$ & $25$ & $90$ & $13.90$ & $8.08$ & $7.15$ & $7.55$ \\
$0.049$ & $0.67$ & $272$ & $33$ & $100$ & $13.90$ & $8.08$ & $5.26$ & $1,11$ \\
$0,049$ & $0.67$ & $272$ & $33$ & $115$ & $13.90$ & $8.08$ & $9.60$ & $8.01$ \\ \hline  
$0.059$ & $0.69$ & $253$ & $25$ & $73$ & $13.34$ & $7.49$ & $2.72$ & $0.35$ \\ 
$0.059$ & $0.69$ & $253$ & $25$ & $90$ & $13.34$  & $7.49$ & $7.75$ & $8.55$ \\
$0.059$ & $0.69$ & $253$ & $33$ & $100$ & $13.34$  & $7.49$ & $5.78$ & $1.81$ \\
$0.059$ & $0.69$ & $253$ & $33$ & $115$ & $13.34$  & $7.49$ & $10.20$ & $9.03$ \\ \hline  
$0.071$ & $0.71$ & $237$ & $25$ & $72$ & $12.74$  & $6.89$ & $2.97$ & $0.61$ \\ 
$0.071$ & $0.71$ & $237$ & $25$ & $90$ & $12.74$  & $6.89$ & $8.12$ & $9.25$ \\ 
$0.071$ & $0.71$ & $237$ & $33$ & $100$ & $12.74$  & $6.89$ & $6.31$ & $2.57$ \\
$0.071$ & $0.71$ & $237$ & $33$ & $115$ & $12.74$  & $6.89$ & $10.50$ & $9.62$ \\ \hline  
$0.085$ & $0.73$ & $225$ & $25$ & $71$ & $12.13$  & $6.31$ & $3.06$ & $0.61$ \\ 
$0.085$ & $0.73$ & $225$ & $25$ & $90$ & $12.13$  & $6.31$ & $8.41$ & $9.84$ \\  
$0.085$ & $0.73$ & $225$ & $33$ & $100$ & $12.13$  & $6.31$ & $7.21$ & $3.98$ \\ 
$0.085$ & $0.73$ & $225$ & $33$ & $115$ & $12.13$ & $6.31$ & $10.71$ & $10.10$ \\ \hline 
$0.104$ & $0.75$ & $216$ & $25$ & $71$ & $11.45$ & $5.70$ & $3.41$ & $1.08$ \\
$0.104$ & $0.75$  & $216$ & $25$ & $90$ & $11.45$ & $5.70$ & $8.42$ & $10.00$ \\
$0.104$ & $0.75$  & $216$ & $33$ & $100$ & $11.45$ & $5.70$ & $7.07$ & $3.70$ \\
$0.104$ & $0.75$  & $216$ & $33$ & $115$ & $11.45$ & $5.70$ & $10.65$ & $9.78$ \\ \hline 
$0.129$ & $0.77$ & $211$ & $25$ & $73$ & $10.77$ & $5.12$ & $4.44$ & $2.79$ \\ 
$0.129$ & $0.77$  & $211$ & $25$ & $90$ & $10.77$  & $5.12$  & $9.13$ & $11.58$ \\ 
$\bf{0.129}$ & $\bf{0.77}$  & $\bf{211}$ & $\bf{33}$ & $\bf{100}$ & $\bf{10.77}$  & $\bf{5.12}$  & $\bf{7.26}$ & $\bf{4.00}$ \\
$0.129$ & $0.77$  & $211$ & $33$ & $113$ & $10.77$  & $5.12$  & $10.76$ & $10.54$  
\end{tabular}}
\end{ruledtabular}
\end{table}

Observing the data in Table \ref{quadro_acoplamentos}, it is evident that the adjustable parameter $\zeta$ functions as a control that allows fine-tuning the intensity of the many-body contributions in the MBF Model, establishing a middle ground between the predictions of Walecka and Zimanyi–Moszkowski models. Note that, as the parameter $\zeta$ increases, the effective mass of the nucleon at saturation ${(m_N^*)}_0$ also increases and the compressibility at saturation $K_0$ decreases, disclosing a spectrum between the limiting values of Walecka~\cite{book:Walecka} and Zimanyi–Moszkowski~\cite{PhysRevC.42.1416} models, which are, respectively, in their original configurations, ${(m_N^*)}_0=0.54~m_N$ with $K_0=560~\rm{MeV}$ and ${(m_N^*)}_0=0.85~ m_N$ with $K_0=225~\rm{MeV}$. Furthermore, it should be highlighted that, due to the specific form of the parametric term $\Pi_{\zeta b}$ that characterizes the MBF Model, which is explicitly dependent on the scalar meson fields (see Eq.~(\ref{mstarTCM})), an increase in the parameter $\zeta$ implies a reinforcement of the attractive components of the nuclear interaction, which is the reason for the decrease in the compressibility when $\zeta$ is increased. Without the intension to be exhaustive, Table \ref{quadro_acoplamentos} presents seven possible values for the pair $(m_N^*)_0$/$K_0$, and, for each one of those, two values for $a_{sym}^0$ and four values for $L_0$. 

\subsection{\label{sec:Hcouplings}Meson-hyperon couplings}

Hyperons are baryons containing one or more strange quarks (see Table \ref{particles}). Their large masses suppress their population near nuclear saturation density, preventing the direct calibration of meson-hyperon couplings from ordinary nuclear-matter properties, as is commonly done for nucleonic couplings. Different proposals for the meson-hyperon coupling constants have been considered in literature, and three of them will be adopted in the present work:\\

{\textbf{U - Universal Coupling:}} In this very simple proposal, the coupling constants for hyperons ($Y=\Lambda, \Sigma, \Xi$) and nucleons are considered to be identical: 
\begin{gather}
\label{Ucouplings}
g_{\sigma Y}=g_{\sigma}, \quad g_{\omega Y}=g_{\omega}, \quad g_{\varrho Y}=g_{\varrho}, \\ \nonumber g_{\delta Y}=g_{\delta}, \quad g_{\phi Y}=g_{\omega}\, . \end{gather}
Moreover, considering that the strange mesons do not couple to the nucleons, $g_{\sigma^*}=0$ and $g_{\phi}=0$. \\

{\textbf{M - Moszkowski Coupling:}} Based on the quark content of nucleons and hyperons, Moszkowski proposed~\cite{Moszkowski:1974gj} that meson-hyperon coupling constants should be corrected according to: 
\begin{gather}
\label{Mcouplings}
g_{\sigma Y}=\sqrt{\frac{2}{3}} g_{\sigma}, \quad g_{\omega Y}=\sqrt{\frac{2}{3}} g_{\omega}, \quad g_{\varrho Y}= \sqrt{\frac{2}{3}} g_{\varrho}, \\ \nonumber g_{\delta Y}= \sqrt{\frac{2}{3}} g_{\delta}, \quad g_{\phi Y}= \sqrt{\frac{2}{3}} g_{\omega}\, . \end{gather}
In this scheme, for the reasons previously stated, we also consider $g_{\sigma^*}=0$ and $g_{\phi}=0$.\\

{\textbf{SU(6) - Spin-Flavor Symmetry Coupling:}} This coupling scheme takes into account the strangeness of the particles and the isospin proportion between nucleons and hyperons~\cite{Dover:1985ba,Schaffner:1993qj}:  
\begin{gather}
\label{SU6couplings}
\frac{1}{3}g_{\omega }= \frac{1}{2}g_{\omega \Lambda}=\frac{1}{2} g_{\omega \Sigma}=  g_{\omega \Xi}, \\ \nonumber
g_{\varrho}=\frac{1}{2} g_{\varrho \Sigma}=g_{\varrho \Xi}, \quad g_{\varrho \Lambda} = 0, \\  \nonumber
-\frac{2\sqrt{2}}{3} g_{\omega}= 2 g_{\phi \Lambda}= 2 g_{\phi \Sigma}= g_{\phi \Xi}, \quad g_{\phi}=0, \\ \nonumber
g_{\delta}=\frac{1}{2} g_{\delta \Sigma}=g_{\delta \Xi}, \quad g_{\delta \Lambda} = 0\, . \end{gather}
Furthermore, we obtain the hyperon-sigma coupling associated to the attractive interaction between hyperons and nucleons by fitting the potential depths of the hyperons in nuclear matter $U_Y$, according to the prescription~\cite{Glendenning:1991es,Schaffner:1992sn}: 
\begin{equation}
\label{hyperonpotentials}
g_{\sigma Y} = \frac{g_{\omega Y}\, \omega_0(n_0) - U_Y}{\sigma_0(n_0)}\, .  
\end{equation}
Using the values presented in Ref.~\cite{Weissenborn:2011kb}, we consider
$U_{\Lambda} = -28\, \rm{MeV}$, $U_{\Sigma}= +30\, \rm{MeV}$ and $U_{\Xi}= -18\, \rm{MeV}$. 

It must be emphasized that the three coupling schemes described above should be viewed as representative scenarios that are useful for exploration, since they allow us to assess the sensitivity of the results to different assumptions about hyperon interactions.

Finally, the coupling of hyperons to the $\sigma^*$ meson must also be specified. Given the limited experimental information on hyperon-hyperon interactions and $U_{YY}$ potentials, we follow the standard MBF prescription and assume universal couplings,
$g_{\sigma^* \Lambda}=g_{\sigma^* \Sigma}=g_{\sigma^* \Xi}\equiv g_{\sigma^*Y}$.

Following \cite{Gomes:2014aka}, we consider the range $g_{\sigma^*Y}=0$--$5$. Because $\sigma^*$ is a scalar meson, increasing $g_{\sigma^*Y}$ enhances the attractive hyperon-hyperon interaction, slightly softening the EoS and reducing the maximum stellar mass~\cite{Gomes:2014aka}. The coupling strength given by $g_{\sigma^*Y}$ is also constrained by the requirement that the effective hyperon masses remain positive; once they vanish, the mean-field description ceases to be applicable~\cite{PhysRevC.63.065801,Schaffner:1995th,Ellis:1995kz}

\section{\label{sec:resultsNM}Results for nuclear matter properties at finite temperature}

As discussed in Sec.~\ref{sec:Ncouplings}, different choices of the coupling constants lead to distinct EoSs. To isolate many-body effects, we consider two limiting values of the parameter $\zeta$ that reproduce the same saturation symmetry energy $a_{sym}^0$ and slope parameter $L_0$. These parameterizations ($\zeta=0.040$ and $\zeta=0.129$) are highlighted in bold in Table \ref{quadro_acoplamentos}. They represent cases with smaller and larger many-body contributions, respectively.

Unless otherwise stated, the results shown in Secs.~\ref{sec:resultsNM} and \ref{sec:MRrelations} correspond to the limiting cases $\zeta=0.040$ and $\zeta=0.129$. In the very end of Sec.~\ref{sec:MRrelations} we expand the discussion to include different values. Similarly, we consider the limiting values $g_{\sigma^*Y}=0$ and $g_{\sigma^*Y}=5$, corresponding to the weakest and strongest hyperon-hyperon attraction explored in this work.

Furthermore, it would be possible to adopt many different configurations to study the case of finite temperature. Hence, without the intention of being exhaustive, we will present results for three paradigmatic scenarios, where entropy per baryon is fixed, following pioneering studies~\cite{Prakash:1996xs,Pons:1998mm,Bednarek:2006tw,Goussard:1997bn,Pons:2000iy} and also more recent works (see, e.g.,~\cite{Dexheimer:2012mk},~\cite{Kumar:2025dlc} and~\cite{Stone:2019blq}):
\begin{itemize}
\item $S/A=s/n_B=1$, $Y_l=0.4$ 
\item $S/A=s/n_B=2$, $\mu_{\nu}=0$ 
\item $S/A=s/n_B=0$, $\mu_{\nu}=0$ 
\end{itemize}
These correspond to three snapshots of the time evolution of a compact star in its first minutes of life. Initially, the star is relatively warm (represented by fixed entropy per particle $S/A=1$) and has a large number of trapped neutrinos, since their mean free path is much shorter than the star’s radius, which can be represented by a fixed lepton fraction $Y_l$, defined as
\begin{equation}
    \label{leptonfraction}
    Y_l=\sum_{j=l} Y_{lj}=\sum_{j=l} \frac{n_{lj}}{n_B}=\sum_{j=l} \frac{\mathfrak{L}_j \, n_{j}}{n_B} \, . 
\end{equation}
Next, in the deleptonization stage of the PNS evolution, neutrinos gradually diffuse outward and escape; as they diffuse, they deposit energy via scattering and absorption in certain layers of the star, in a process that can locally increase the temperature (the so-called “Joule heating”), here represented by the fixed entropy per particle $S/A=2$~\cite{Prakash:1996xs,1986ApJ...307..178B}. Finally, the star becomes transparent to neutrinos, cools down gradually by neutrino emission and settles into a cold NS ($\sim S/A=0$ or $T=0$). The value $Y_l\simeq0.4$ is taken from from numerical simulations of proto-neutron-star evolution~\cite{1986ApJ...307..178B,Pons:1998mm,Fischer_2010,2010PhRvL.104y1101H,PhysRevD.87.043006}.

The choice of fixed entropy per baryon $S/A$ is physically motivated by the nearly adiabatic evolution of proto-neutron stars. In this description, the temperature adjusts self-consistently with density, providing a more realistic thermal profile than calculations performed at fixed temperature.

It is important to emphasize that the finite-temperature stages considered in this work should be interpreted as idealized thermodynamic snapshots of proto-neutron-star matter. In a realistic PNS evolution, the system is not necessarily in exact beta equilibrium at all times, since neutrino diffusion, finite weak-interaction rates, convection, and hydrodynamical evolution may drive the matter temporarily out of chemical equilibrium. These effects can modify the instantaneous particle composition, in particular the onset and abundance of hyperons, whose thresholds are sensitive to the balance among baryon, electric-charge, and lepton chemical potentials~\cite{Stone:2019blq}. A fully time-dependent treatment including neutrino transport and finite weak-interaction rates is beyond the scope of the present work and is left for future applications of the finite-temperature MBF equations of state, made possible with the multidimensional tables we provide with this work.

In the following subsections, our results are consistently presented in two sets of plots:

\begin{itemize}
    \item {\bf{SET 1:}} Each panel shows the numerical results for fixed values of $\zeta$ and $g_{\sigma^*}$, displaying all three scenarios of entropy per baryon and covering four different possibilities for the inclusion of hyperons: $N$ - pure nucleonic matter, containing only neutrons, protons, electrons and muons (black lines); $U$ - universal coupling (green lines); $M$ - Moszkowski coupling (red lines); $SU(6)$ - spin-flavor symmetry coupling (blue lines). Each row (line) of plots in this set aims to enable a direct, side-by-side comparison of the changes caused in the physical quantity as a result of the change in the value of the adjustable parameter $\zeta$ characterizing many-body forces.
    \item {\bf{SET 2:}} Each panel shows the numerical results for fixed values of entropy per baryon and $g_{\sigma^*}$, displaying together the two options for $\zeta$ and covering the four different possibilities for the inclusion of hyperons. Each row (line) of plots in this set aims to provide a glimpse of the time evolution of the star, from a hot PNS into a cold NS.  
\end{itemize}

\subsection{\label{sec:EoS}Equations of state}

\begin{figure*}
\includegraphics[trim=0 0 0 0, clip, width=0.7\textwidth]{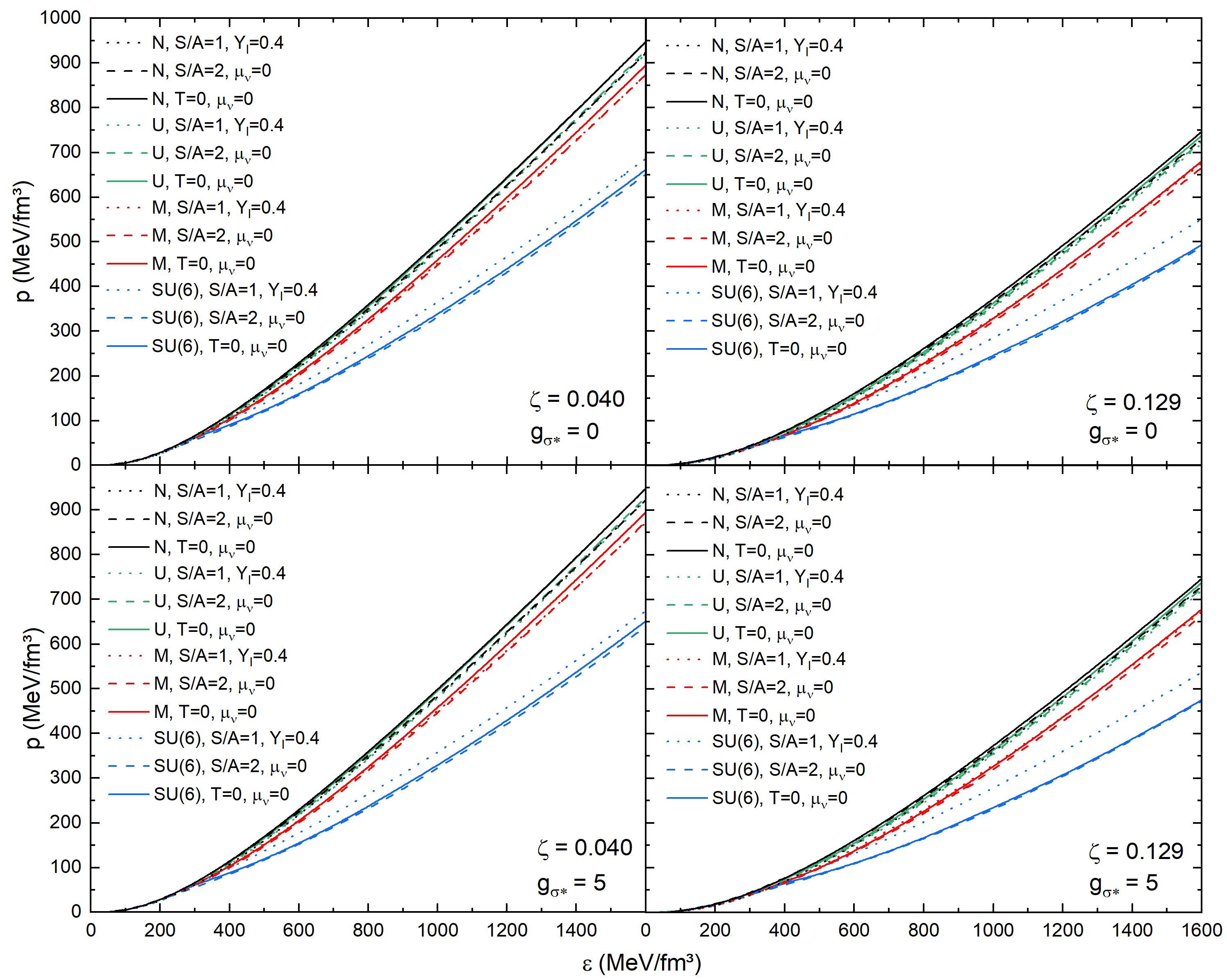}
\caption{ {\bf{SET 1 (Equations of state):}} Pressure ($p$) as a function of energy density ($\varepsilon$) for different proto-neutron star evolution snapshots, particle composition, and hyperon couplings. The different panels show different MBF (left vs. right) and strange scalar meson parameterizations (top vs. bottom).}
\label{EOSgeral}
\end{figure*}

\begin{figure*}
\includegraphics[width=1.0\textwidth]{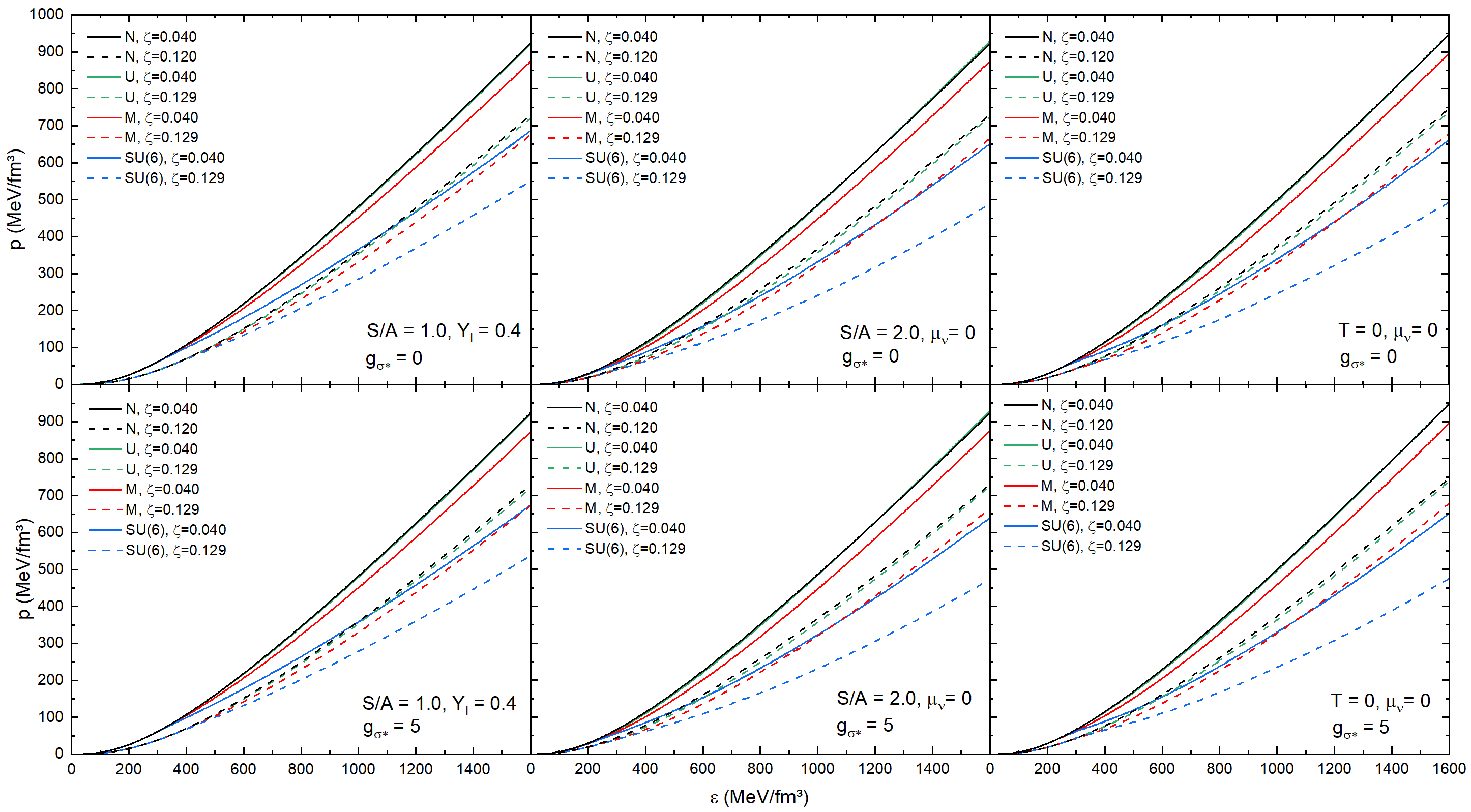}
\caption{ {\bf{SET 2 (Equations of state):}} Same as Fig. \ref{EOSgeral} but now the different panels show different evolution snapshots (left/center/right) and strange scalar meson parameterizations (top vs. bottom).}
\label{EOSsequencia}
\end{figure*}

Fig. \ref{EOSgeral} presents the equations of state (pressure vs. energy density) obtained within the MBF Model for the two parameterizations considered in this work, $\zeta=0.040$ (left panel) and $\zeta=0.129$ (right panel), at different temperatures. As expected, the $\zeta=0.040$ (weaker many-body effects) parameterization yields a stiffer EoS, with pressure increasing more rapidly with the energy density. The inclusion of the scalar strange meson has only a minor effect on the overall shape of the EoS, as can be seen by comparing the upper and lower panels of Figs. \ref{EOSgeral} and \ref{EOSsequencia}.

Fig.~\ref{EOSsequencia} illustrates the evolution of the EoS from the hot proto-neutron-star stage (left panels) to the cold neutron-star configuration (right panels), for $g_{\sigma^*}=0$ (upper panels) and $g_{\sigma^*}=5$ (lower panels). The displayed curves span the range from the softest ($\zeta=0.129$, $SU(6)$) to the stiffest ($\zeta=0.040$, nucleonic matter) EoSs considered in this work.

For purely nucleonic matter, the EoSs remain nearly unchanged throughout the evolution represented by different styles of lines in Fig.~\ref{EOSgeral} or different columns of panels in Fig.~\ref{EOSsequencia}, except for a slightly higher pressure at high densities in the $T=0$ case. For hyperonic matter, the stiffness follows the ordering $U>M>SU(6)$, all lying below the nucleonic EoS. The stronger softening observed for the $SU(6)$ coupling reflects its earlier and more abundant hyperon population, which redistributes the baryon number among additional degrees of freedom and reduces the nucleon Fermi pressure. Accordingly, the $SU(6)$ curves exhibit a pronounced change in slope at the onset of hyperons, particularly at $T=0$. These results are consistent with previous density-dependent RMF calculations \cite{Shrivastava:2025ags}.

When hyperons are included, the EoSs display a richer thermal evolution, with crossings occurring at different energy densities. The universal coupling ($U$) produces an EoS that nearly overlaps the nucleonic one. This demonstrates that hyperonic matter is not necessarily associated with a soft EoS, provided an appropriate coupling scheme is adopted.

Here, all sets of EoSs are numerically checked for thermodynamic consistency using Gibbs-Duhem relation:
\begin{equation}
    \label{gibbsduhem}
    \varepsilon + p = T  \Bigg( \sum_{j=b} s_{j} + \sum_{j=l} s_{j} \Bigg) + \sum_{j=b} n_{B_j}\, \mu_{j} + \sum_{j=l} n_{l_j}\, \mu_{j}\, ,
\end{equation}
where $\mu_j$ is given by Eq.~(\ref{generalchemicaleq}) and $n_{l_j}$ is the number density of each lepton species $j=l$, calculated through the usual integration over momentum space:
\begin{equation}
    \label{rho_leptons}
    n_{l_j} = \frac{1}{\pi^2} \int_0^{\infty}dk_j
\ {k_j}^2 \ (f_{j+} + f_{j-})\, . 
\end{equation}
The numerical verification of Eq.~(\ref{gibbsduhem}) for all EoSs considered here corroborates the discussion presented in Sec.~\ref{sec:MBFgeneral}: the finite-temperature extension of the MBF Model remains fully thermodynamically consistent over the entire parameter space investigated, without the need for rearrangement terms.

It is worth commenting on the comparatively modest influence of the strange scalar meson $\sigma^*$ throughout the present results. At first sight, one might expect that introducing an additional attractive interaction in the hyperonic sector would produce significant modifications to the equation of state and, consequently, to the global properties of neutron stars. However, the limited impact observed here is physically expected and follows naturally from the role played by the $\sigma^*$ meson in the MBF Model.

Unlike the ordinary scalar meson $\sigma$, which couples to all baryons and therefore contributes throughout the entire density range, the $\sigma^*$ meson couples exclusively to strange baryons. Consequently, its contribution vanishes identically in purely nucleonic matter and remains negligible until the first hyperons become energetically favored. Since hyperons themselves only appear at densities of a few times the nuclear saturation density, the $\sigma^*$ field acts only in the high-density core of the star, where the strange baryon fraction becomes appreciable.

Even after the onset of hyperons, the additional scalar attraction mediated by $\sigma^*$ competes with the dominant repulsive vector interactions and modifies primarily the effective masses and chemical potentials of the strange baryons. As a consequence, its most direct effects are limited to small shifts in hyperon threshold densities and relative particle abundances, while global thermodynamic quantities, such as the pressure, energy density, sound velocity, compressibility, adiabatic index, and stellar mass-radius relation, remain largely controlled by the many-body parameter $\zeta$ and by the adopted hyperon-meson coupling prescription. For this reason, in the next section, which specifically addresses particle populations, results will be presented for the cases $g_{\sigma^*Y} = 0$ and $g_{\sigma^*Y} = 5$; however, in subsequent sections, dealing with the macroscopic properties of dense matter, only the results for $g_{\sigma^*Y} = 5$ will be presented.

An equally important point is that the interval explored in the present work, $0 \le g_{\sigma^*Y} \le 5$, is itself constrained by current hypernuclear phenomenology and by the internal consistency of the model. Larger values would indeed enhance the attractive hyperon-hyperon interaction, but they would also produce excessively small hyperon effective masses and eventually drive the formalism beyond the range of validity of the mean-field approximation, as discussed previously. Therefore, the relatively small quantitative differences found here should not be interpreted as indicating that the $\sigma^*$ meson is irrelevant. Rather, they demonstrate that, within the phenomenologically acceptable range of hyperon-hyperon couplings, the $\sigma^*$ interaction acts as a secondary correction to the strange sector, whereas the overall stiffness of the EoS is primarily determined by the many-body dynamics encoded in $\zeta$ and by the hyperon coupling scheme.

Because the EoS alone does not fully characterize the properties of dense matter; therefore, additional properties of dense matter and their dependence on the MBF parameterization are examined in the following sections. 

\subsection{\label{sec:populations}Populations of particles and antiparticles}

\begin{figure*}
\includegraphics[width=0.7\textwidth]{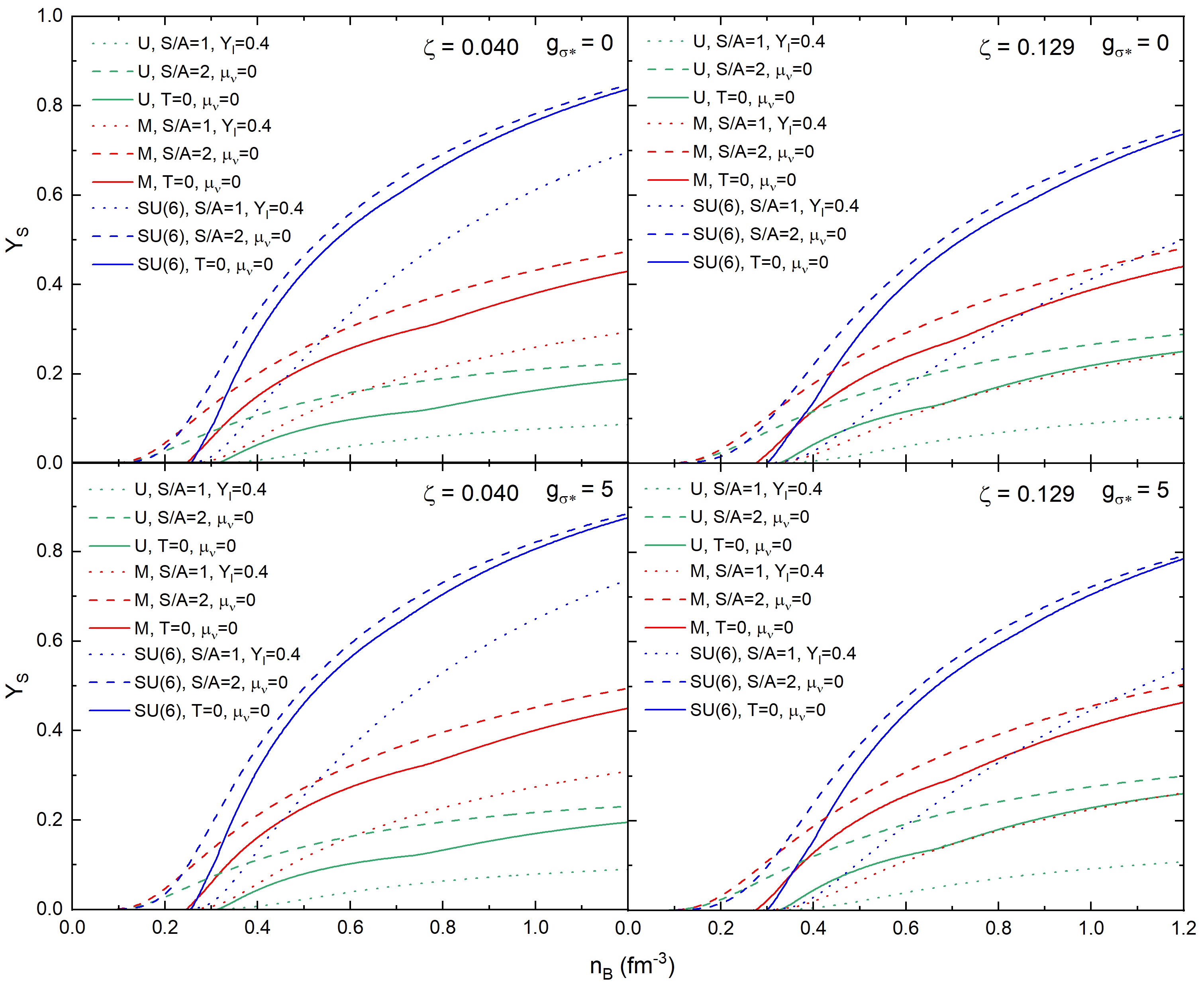}
\caption{\label{FSgeral} {\bf{SET 1 (Fraction of strangeness):}} Fraction of strangeness ($Y_S$) as a function of baryon density ($n_B$) for different proto-neutron star evolution snapshots, particle composition, and hyperon couplings. The different panels show different MBF (left vs. right) and strange scalar meson parameterizations (top vs. bottom).}
\end{figure*}

At sufficiently high densities, the appearance of hyperons becomes energetically favorable, introducing strangeness into dense matter \cite{livro:Glendenning}. Hyperons are expected to populate hadronic matter once their chemical potentials satisfy the threshold conditions \cite{Balberg:1998ug}. Eq.~(\ref{generalchemicaleq}) determines these thresholds and, consequently, the particle populations as functions of baryon density. A convenient measure of the hyperon content is the strangeness fraction,
\begin{equation}\label{strangeness}
Y_S= \sum_{j}\frac{n_{B_j}\, |{\mathscr{S}}_j|}{n_{B}}=\frac{n_{B_\Lambda}+n_{B_\Sigma}+2n_{B_\Xi}}{n_{B}}\, , 
\end{equation}
where ${\mathscr{S}}_j$ is the strangeness of these particles, according to the values given in Table \ref{particles}. 

Nonetheless, a more detailed analysis can be carried out by calculating the populations of particles (or antiparticles), i.e., the ratio $Y_j$ between the number density of a given species $j$ present in the system ($n_j$) and the total baryon density $n_B$:
\begin{equation}
    \label{fractions}
    Y_j= \frac{n_j}{n_B} \, .
\end{equation}
For baryons, $n_j \equiv n_{B j}$, which can be determined from Eq.~(\ref{rhoBMBF}); for leptons, $n_j \equiv n_{l j}$, which can be determined from Eq.~(\ref{rho_leptons}). These two equations remind us that these densities are \emph{net densities}, i.e., the result that emerges form subtracting the number of particles and the number of antiparticles of a given species $j$. In order to calculate the number density of a certain antiparticle separately, one must consider
\begin{equation}
    \label{rho_antiparticles}
    n_{j} = \frac{1}{\pi^2} \int_0^{\infty}dk_j
\ {k_j}^2 \ f_{j-} \, , 
\end{equation}
where the Fermi-distribution function for antiparticles $f_{j-}$ is defined in (\ref{fermi_distribution}). 

\begin{figure*}
\includegraphics[width=1.0\textwidth]{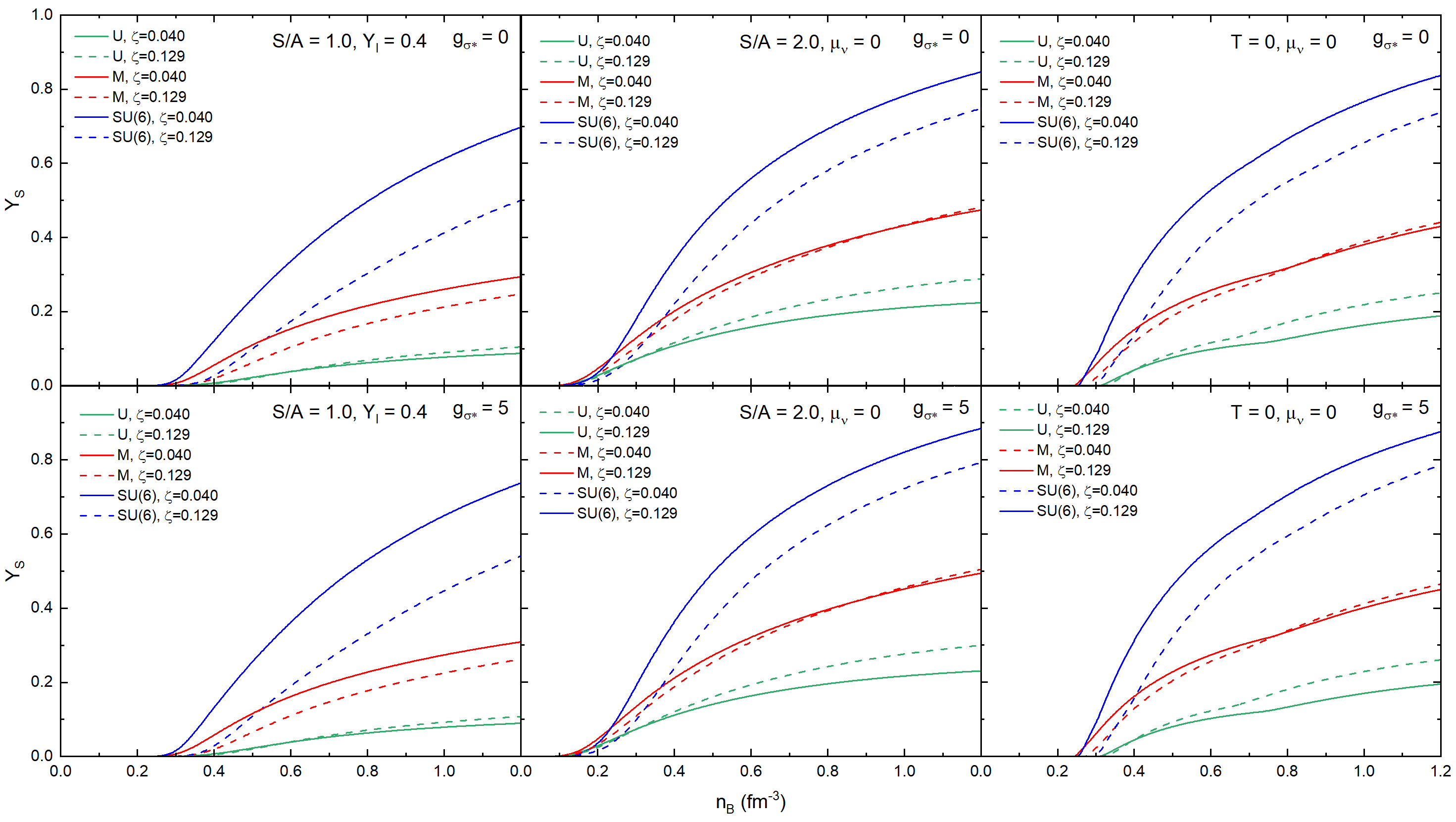}
\caption{\label{FSsequencia} {\bf{SET 2 (Fraction of strangeness):}} Same as Fig. \ref{FSgeral} but now the different panels show different evolution snapshots (left/center/right) and strange scalar meson parameterizations (top vs. bottom).}
\end{figure*}

\begin{figure*}
\includegraphics[width=1.0\textwidth]{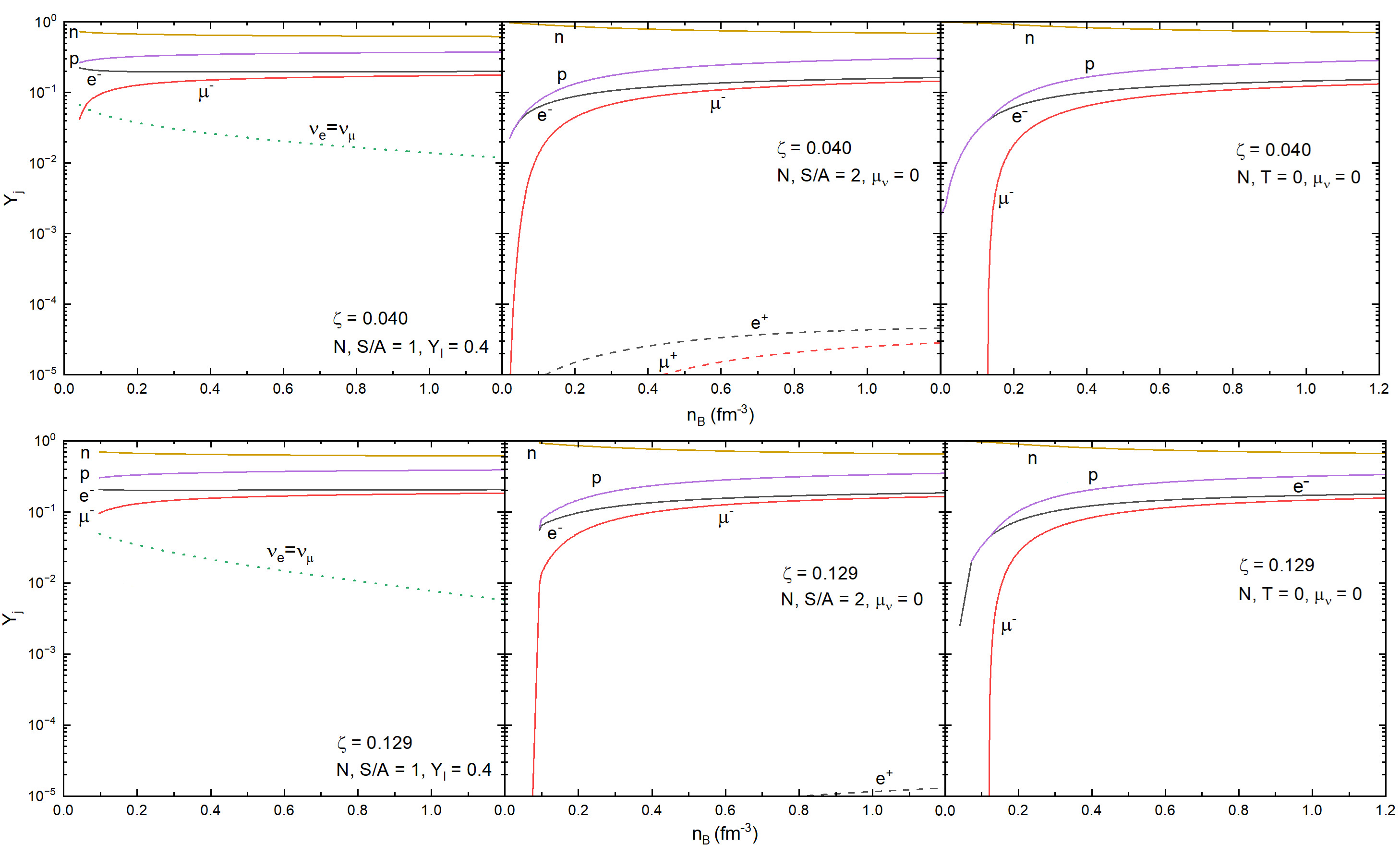}
\caption{\label{POPnucleons} {\bf{Populations of particles -- Pure nucleonic matter ($N$):}} Particle fraction ($Y_j$) as a function of baryon density ($n_B$). The different panels show different proto-neutron star evolution snapshots (left/center/right) and different MBF parameterizations (top vs. bottom). Antiparticles appear in dashed lines (center panels only).}
\end{figure*}

\begin{figure*}
\includegraphics[width=1.0\textwidth]{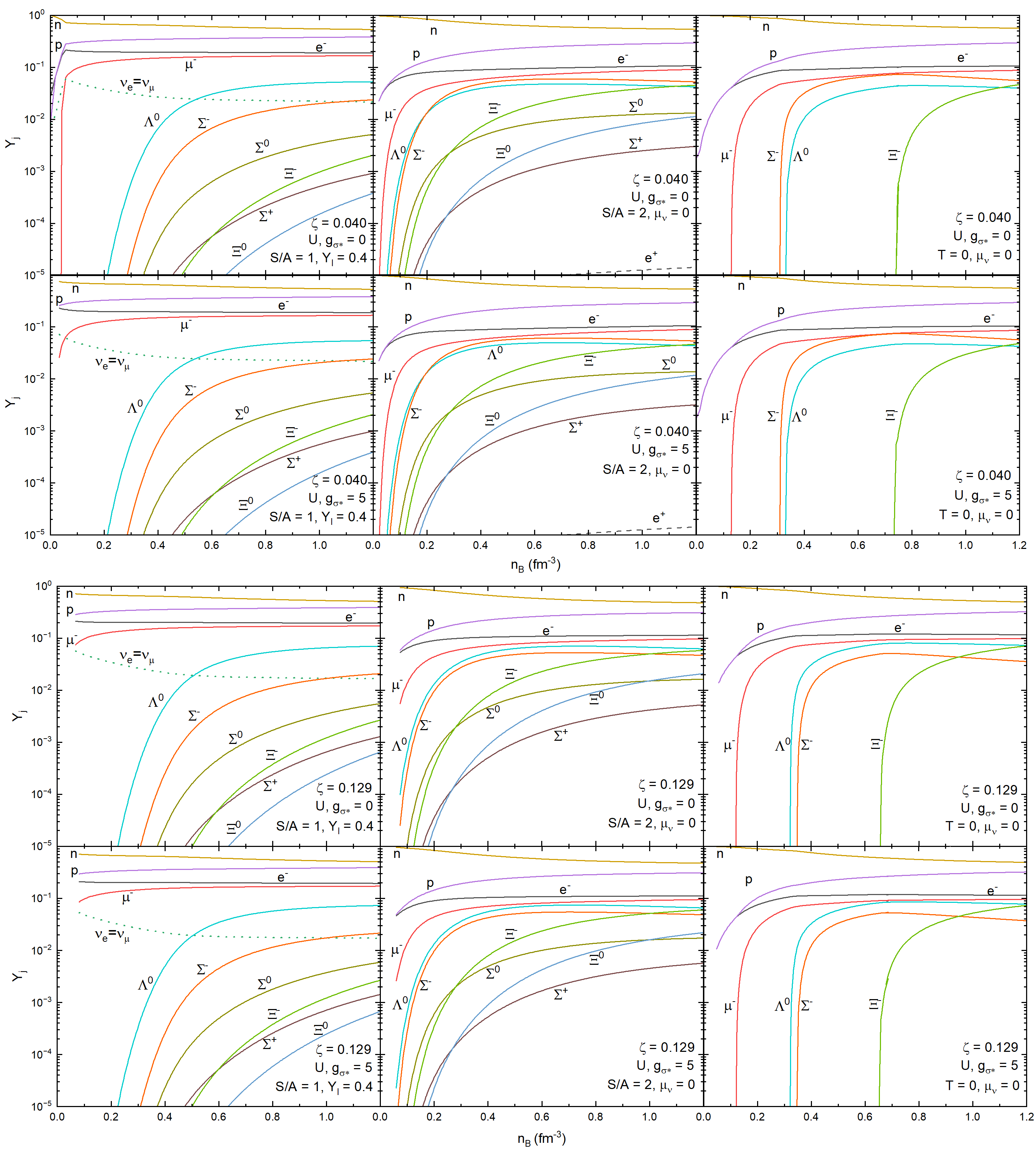}
\caption{\label{POPU} {\bf{Populations of particles -- Universal coupling ($U$):}} Same as Fig. \ref{POPnucleons} but for the universal coupling scheme for hyperons. Second and forth rows show different strange scalar meson parameterizations.}
\end{figure*}

\begin{figure*}
\includegraphics[width=1.0\textwidth]{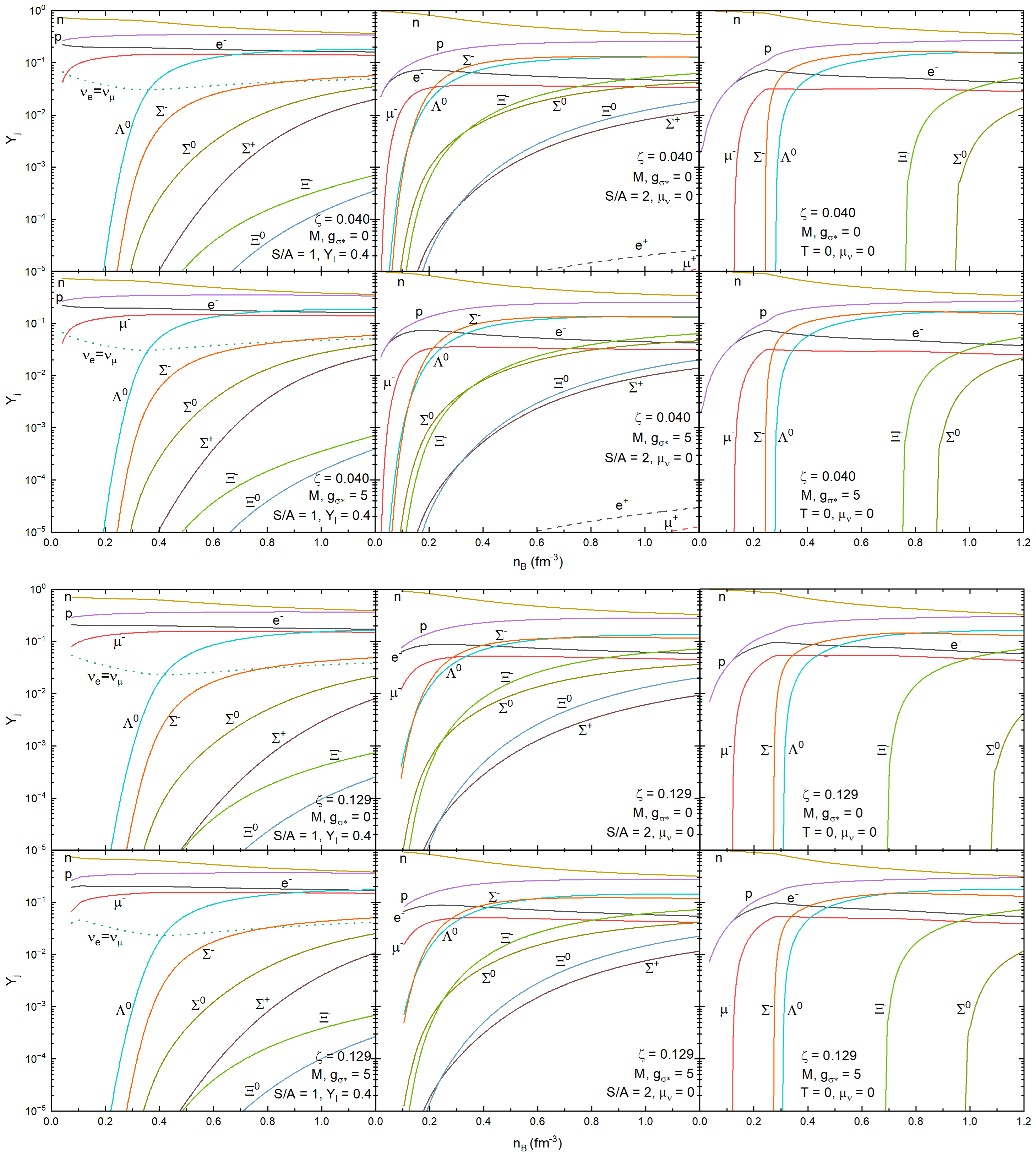}
\caption{\label{POPM} {\bf{Populations of particles -- Moszkowski coupling ($M$):}} Same as Fig. \ref{POPU} but for the Moszkowski coupling scheme for hyperons.}
\end{figure*}

\begin{figure*}
\includegraphics[width=1.0\textwidth]{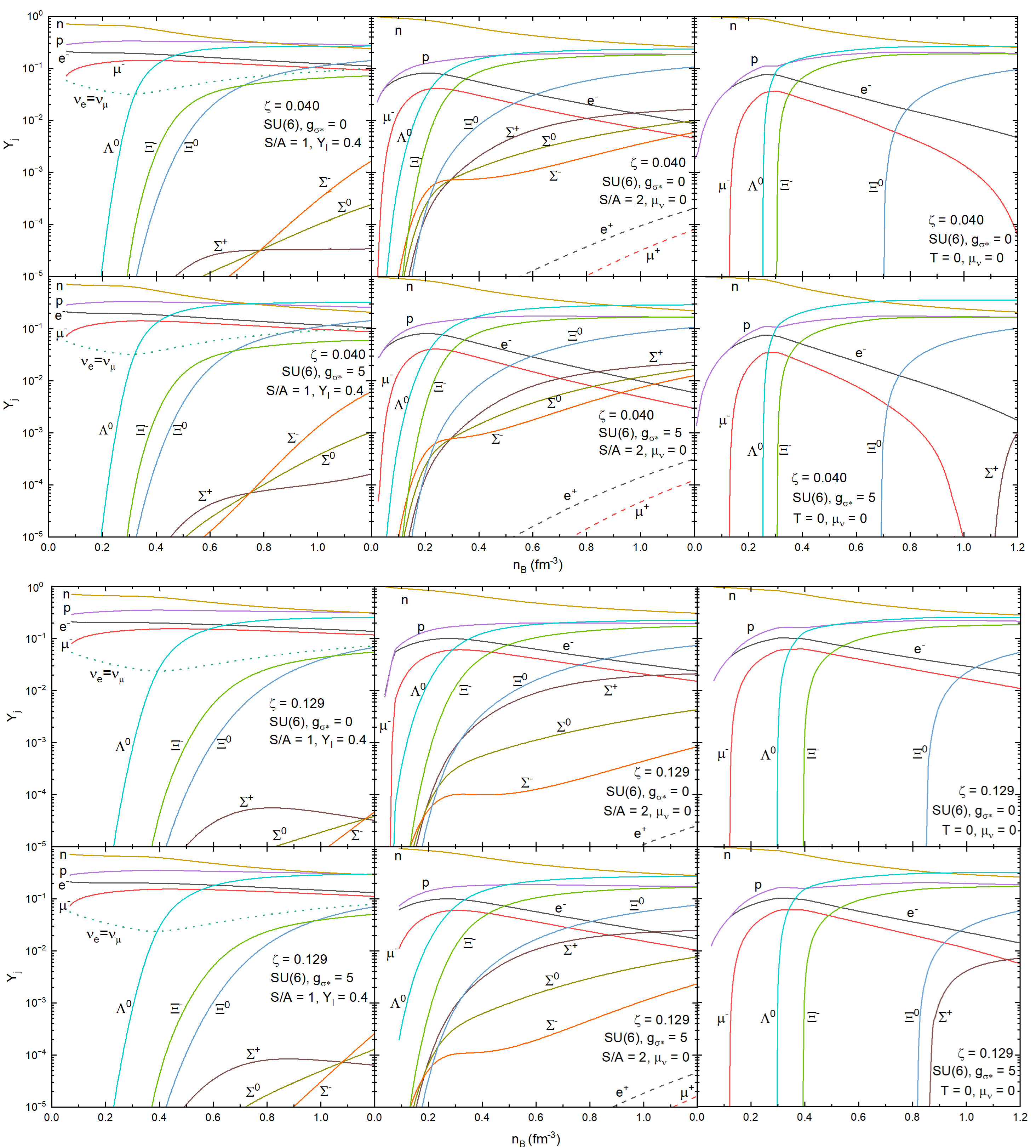}
\caption{\label{POPSU6} {\bf{Populations of particles -- SU(6):}} Same as Fig. \ref{POPU} but for the SU(6) coupling scheme for hyperons.}
\end{figure*}

Figs. \ref{FSgeral} and \ref{FSsequencia} show the strangeness fraction $Y_S$ as a function of the baryon density for different evolutionary conditions and hyperon-meson coupling schemes. Particle fractions are shown in Figs. \ref{POPnucleons}--\ref{POPSU6}. In all cases, $Y_S$ is negligible at low densities, increases rapidly at the onset of hyperons (particularly for $T=0$), and then grows monotonically as additional strange baryons are populated. The onset density and growth rate depend on the evolutionary conditions, the parameter $\zeta$, and the adopted hyperon coupling scheme. Finite temperature lowers the hyperon threshold, leading to an earlier but more gradual increase of $Y_S$ than in cold, neutrino-free matter. The effect of $\zeta$ depends strongly on the adopted hyperon coupling scheme. 

For $SU(6)$ couplings, the stiffer parameterization ($\zeta=0.040$) produces larger strangeness fractions, whereas the opposite trend is found for the universal coupling ($U$). In contrast, the Moszkowski prescription ($M$) is almost insensitive to $\zeta$. This coupling-dependent behavior persists at finite temperature and entropy, indicating that the relation between EoS stiffness and strangeness production is not universal. At finite temperature (or fixed entropy), thermal smearing largely suppresses the dependence of the hyperon onset on the coupling scheme. For cold matter, however, the $SU(6)$ and Moszkowski ($M$) prescriptions predict slightly earlier hyperon formation than the universal coupling, for both values of $\zeta$.

The additional attraction mediated by the strange scalar meson $\sigma^*$ lowers the hyperon thresholds, leading to a modest increase in the hyperon population. The effect is most pronounced for the $SU(6)$ prescription, which favors the appearance of $\Lambda$ hyperons through the attractive potential of Eq.~(\ref{hyperonpotentials}). Nevertheless, $\sigma^*$ produces only quantitative changes, preserving the ordering and overall shape of the $Y_S(n_B)$ curves for all values of $\zeta$ and all coupling schemes.

Fig. \ref{FSsequencia} compares the effects of the hyperon coupling schemes and the parameter $\zeta$ for fixed evolutionary conditions. For $S/A=1$, neutrino trapping suppresses negatively charged hyperons, resulting in the smallest strangeness fractions. For $S/A=2$, thermal effects lower the hyperon thresholds while preserving the ordering $SU(6)>M>U$ in the growth of $Y_S$. The dependence on $\zeta$ follows the same coupling-dependent pattern discussed previously.

\onecolumngrid

\twocolumngrid

For cold, neutrino-free matter, the hyperon thresholds become much sharper, making the differences among the coupling schemes more evident. The influence of $\zeta$ remains coupling dependent, while the hyperon-meson couplings primarily determine the onset of strangeness. As in Fig.~\ref{FSgeral}, the inclusion of the strange scalar meson $\sigma^*$ produces only a modest enhancement of the strangeness fraction at high densities, without changing the qualitative behavior of the curves. When compared to other approaches, different hyperon coupling schemes in our case give rise to somewhat larger spread of $Y_S$ than different RMF models in Ref.~\cite{Stone:2019blq}.

Figs. \ref{POPnucleons}--\ref{POPSU6} summarize the particle populations for the different evolutionary  conditions and hyperon coupling schemes. Except for antiparticles, all populations correspond to net particle fractions. For purely nucleonic matter (Fig.~\ref{POPnucleons}), particle fractions evolve smoothly and rapidly approach their equilibrium values because only nucleons and leptons participate in the chemical equilibrium.
When hyperons are included (Figs.~\ref{POPU}--\ref{POPSU6}), the dominant source of variation is the hyperon-meson coupling scheme. By contrast, changing $\zeta$ or including the strange scalar meson $\sigma^*$ produces only quantitative modifications, without altering the sequence in which particle species appear.

The universal ($U$) and Moszkowski ($M$) prescriptions lead to very similar particle populations, reflecting the identical meson-hyperon couplings assigned to all hyperons by Eqs.~(\ref{Ucouplings}) and (\ref{Mcouplings}).
By contrast, the particle-dependent couplings of the $SU(6)$ prescription given by (\ref{SU6couplings}) produce a markedly different hyperon population. In particular, the repulsive $\Sigma$ and attractive $\Xi$ optical potentials adopted in Eq.~(\ref{hyperonpotentials}) suppress the former while favoring the latter.

Consistent with the behavior of the total strangeness fraction, the hyperon populations increase systematically from the universal ($U$) to the Moszkowski ($M$) and finally to the $SU(6)$ prescription. This behavior reflects the progressively stronger attractive interaction in the strange sector, which modifies the chemical potentials and the competition among hyperonic species.

As discussed previously, the strange scalar meson $\sigma^*$ has only a limited impact on the particle populations. Its most visible effect occurs for antiparticles, which become abundant only in the $S/A=2$ configuration owing to thermal pair production and are slightly enhanced by the additional attraction mediated by $\sigma^*$. A second, more subtle effect appears for the universal coupling at $T=0$, where the inclusion of $\sigma^*$ reverses the onset order of the $\Lambda^0$ and $\Sigma^-$ hyperons, illustrating the sensitivity of threshold competition to relatively weak strange-sector interactions.

The particle populations evolve markedly with the thermodynamic state. For $S/A=1$, neutrino trapping suppresses the growth of the hyperon population and limits the restructuring of the nucleonic and leptonic sectors despite the lower thermal thresholds. For $S/A=2$, thermal effects dominate the composition, leading to the early appearance of all hyperon species largely independently of $\zeta$ and the coupling prescription. 

\begin{figure*}
\includegraphics[width=0.8\textwidth]{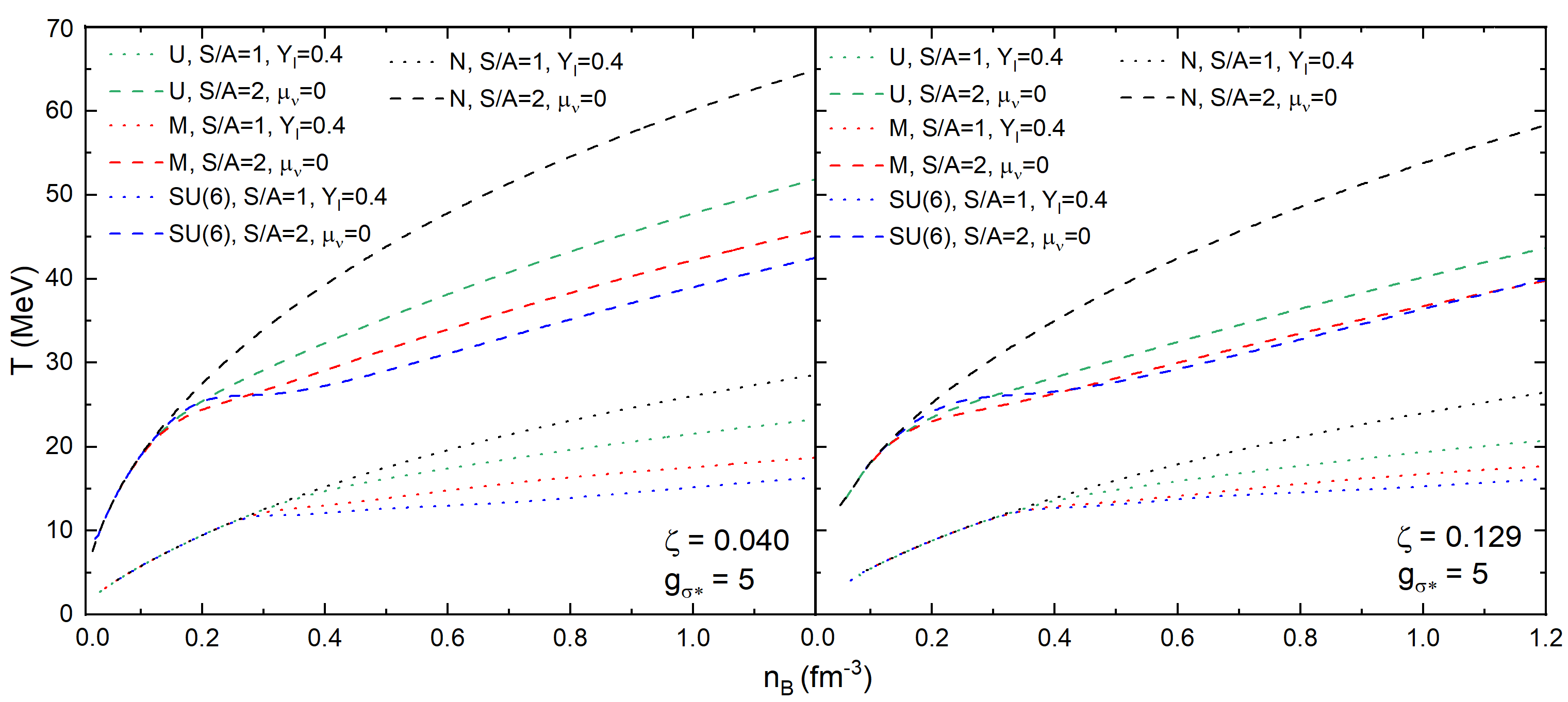}
\caption{\label{Tgeral} {\bf{SET 1 (Temperature profiles):}} Temperature ($T$) as a function of baryon density ($n_B$) for different proto-neutron star evolution snapshots, particle composition, and hyperon couplings. The different panels show different MBF parameterizations.}
\end{figure*}

For cold, neutrino-free matter, hyperon thresholds become sharper and more strongly dependent on the hyperon-meson coupling scheme. In particular, the $SU(6)$ prescription favors the earliest hyperon onset, whereas the universal coupling delays the appearance of strange baryons. At $T=0$, the appearance of negatively charged hyperons ($\Sigma^-$ and $\Xi^-$) rapidly reduces the electron and muon fractions in the absence of neutrino trapping and thermal smearing. This reflects the progressive replacement of leptons by hyperons in maintaining charge neutrality. Consequently, the proton and neutron fractions evolve more gradually than in the finite-temperature cases. In addition, the absence of thermal pair production suppresses antiparticles, leaving hyperons as the only strange degrees of freedom.

Overall, the strangeness fraction $Y_S$ and the particle populations $Y_j$ demonstrate that the microscopic composition of dense matter is governed by the interplay between temperature, neutrino content, and the hyperon-meson coupling scheme in addition to the $\zeta$ parameter that controls the many-body forces.

\subsection{\label{sec:temperature}Temperature profiles}

\begin{figure*}
\includegraphics[width=0.8\textwidth]{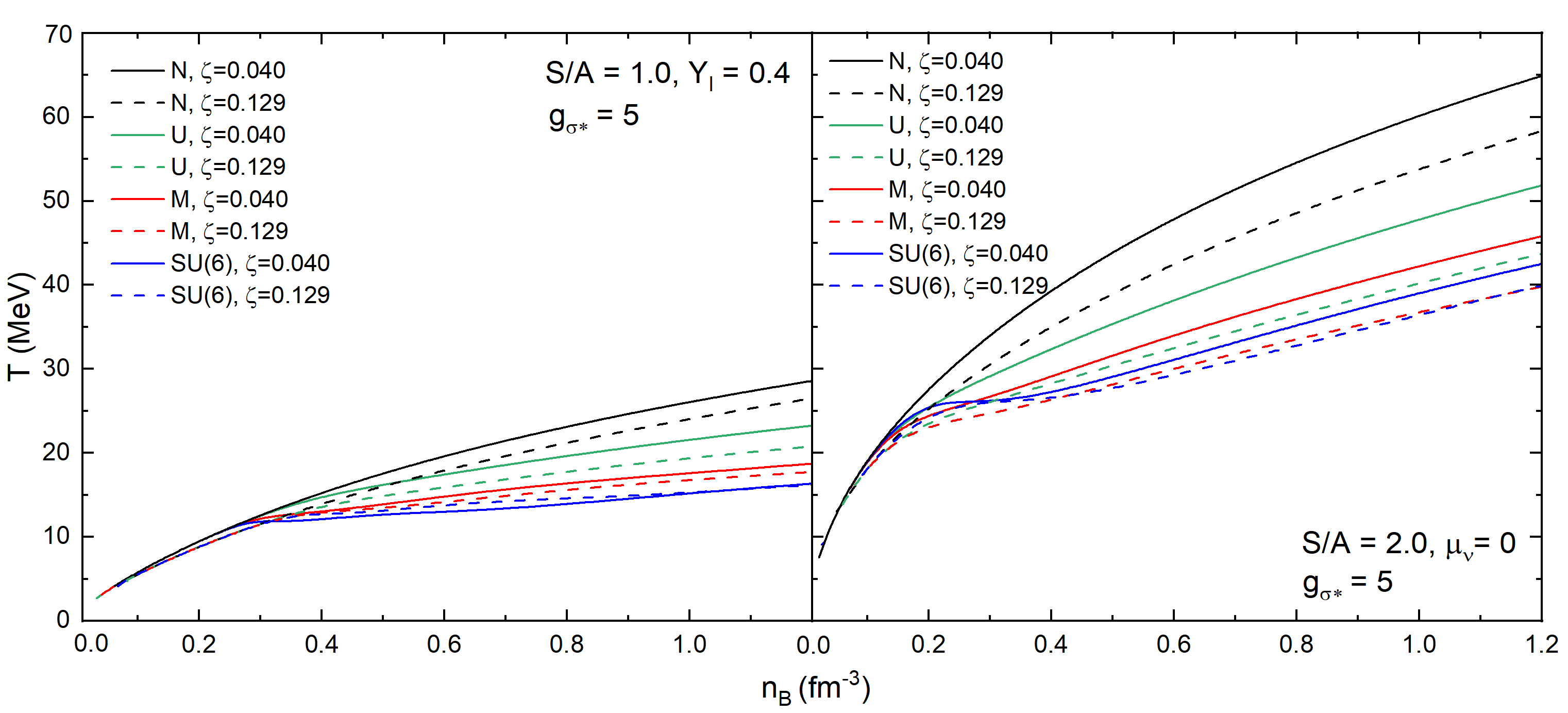}
\caption{\label{Tsequencia} {\bf{SET 2 (Temperature profiles):}} Same as Fig. \ref{Tgeral} but now the different panels show different evolution snapshots.}
\end{figure*}

Figs. \ref{Tgeral} and \ref{Tsequencia} show the temperature profiles obtained for the two MBF parameterizations ($\zeta=0.040$ and $\zeta=0.129$) and the different hyperon coupling schemes, considering the evolutionary  conditions $S/A=1$ with trapped neutrinos ($Y_l=0.4$) and $S/A=2$ with $\mu_\nu=0$, representative of successive stages of proto-neutron-star evolution~\cite{Prakash:1996xs,Pons:1998mm}.

For both values of $\zeta$, purely nucleonic matter exhibits the highest temperatures at a given density. The inclusion of hyperons softens the EoS \cite{glendenning1985,Schaffner:1995th}
by introducing additional baryonic degrees of freedom. At fixed entropy, the thermal energy is therefore distributed among more particle species, resulting in lower temperatures. Accordingly, the temperature profiles follow the ordering $N>U>M>SU(6)$. Because the $SU(6)$ prescription provides the weakest vector repulsion among hyperons~\cite{Moszkowski:1974gj,Schaffner:1995th}, it favors the earliest hyperon onset and consequently the lowest temperatures.

A comparison between Figs.~\ref{EOSgeral} and \ref{Tgeral} illustrates that similar equations of state do not necessarily imply similar thermal behavior. Although the nucleonic ($N$) and universal-coupling ($U$) EoSs nearly overlap in Fig.~\ref{EOSgeral}, their temperature profiles differ substantially, particularly at high densities. The difference between the $S/A=1$ and $S/A=2$ profiles reflects the effect of neutrino trapping: the larger lepton fraction delays the onset of negatively charged hyperons and produces smoother temperature profiles~\cite{Prakash:1996xs,Pons:1998mm}. This behavior is consistent with the particle populations discussed in Sec.~\ref{sec:populations}.

The influence of the many-body parameter $\zeta$ is also evident. As expected, the stiffer parameterization ($\zeta=0.040$) produces steeper temperature profiles, whereas the softer one ($\zeta=0.129$) leads to flatter curves and favors an earlier onset of hyperons owing to the reduced vector repulsion~\cite{furnstahl1997}. At sub-saturation densities, all temperature profiles nearly coincide because hyperons are strongly suppressed~\cite{Prakash:1996xs}. Differences among coupling schemes and MBF parameterizations become appreciable only after the onset of hyperons. At the highest density shown ($n_B=1.2~\rm{fm}^{-3}$), temperatures span $16$--$23~\rm{MeV}$ ($S/A=1$) and $40$--$52~\rm{MeV}$ ($S/A=2$) for hyperonic matter, compared with about $29$ and $65~\rm{MeV}$, respectively, for purely nucleonic matter. The corresponding numerical values are listed in Table~\ref{quadro_temperatura}.

\begin{table}   
\caption{\label{quadro_temperatura}
Temperature (in $\rm{MeV}$) reached for baryon density $n_B = 1.2~{\rm{fm}^{-3}} \simeq 8~n_0$ in different configurations of the MBF Model and two scenarios of finite entropy per baryon ($S/A$). Here, $N$ stands for pure nucleonic matter. These results consider $g_{\sigma^*}=5$.
}
\begin{ruledtabular}
\resizebox{1.0\linewidth}{!}{
\begin{tabular}{cccccc}
S/A & $\zeta$ & $SU(6)$ & $M$ & $U$ & $N$ \\ 
\hline 
$1$ & $0.040$ & $16$ & $19$ & $23$ & $29$ \tabularnewline  
$1$ & $0.129$  & $16$ & $18$ & $21$ & $26$ \tabularnewline 
\hline
$2$ & $0.040$ & $43$ & $46$ & $52$ & $65$ \tabularnewline  
$2$ & $0.129$  & $40$ & $40$ & $44$ & $58$ \tabularnewline 
\end{tabular}}
\end{ruledtabular}
\end{table}

Fig.~\ref{Tsequencia} complements Fig.~\ref{Tgeral} by directly comparing the two MBF parameterizations under fixed evolutionary conditions. The same ordering observed previously is recovered: for every coupling prescription, the stiffer parameterization ($\zeta=0.040$) produces higher temperatures than the softer one ($\zeta=0.129$), while the coupling schemes follow the sequence $N>U>M>SU(6)$. Neutrino trapping reduces the separation among the temperature profiles by delaying hyperon formation, whereas the neutrino-free configuration enhances the differences between the various parameterizations. Overall, the temperature profiles are governed by three competing effects: hyperon softening, which lowers the temperature at fixed entropy; the MBF parameter $\zeta$, which controls the stiffness of the EoS; and neutrino trapping, which delays hyperon formation and modifies the thermal evolution. Our results are qualitatively consistent with previous finite-temperature RMF studies including hyperons~\cite{Prakash:1996xs,Schaffner:1995th,fattoyev2010} and therefore support their application to PNSs and other hot, dense hadronic systems. Nevertheless, when compared to~\cite{Stone:2019blq} for the $S/B=2$ case, there is a much larger spread for different hyperon coupling schemes in our case than in Ref.~\cite{Stone:2019blq} for different RMF models. In particular, Figs.~\ref{Tgeral} and \ref{Tsequencia} are in good agreement with recent effective-field calculations~\cite{Kunkel:2024otq,Stone:2019blq,Kumar:2025dlc}. 

The temperature profiles also reflect the thermal stratification expected in compact stars, with temperature increasing toward the high-density inner core. The low-density outer regions are more efficient at radiating energy and therefore remain comparatively cool, whereas the dense inner layers are thermally opaque and retain heat more effectively, resulting in substantially higher temperatures. While the outer layers cool efficiently, the dense stellar core is thermally more opaque and therefore remains significantly hotter. Consequently, the dependence of $T$ on $n_B$ provides a direct thermodynamic picture of the stellar interior.

\subsection{\label{sec:speedsound}Speed of sound}

The speed of sound is one of the main diagnostics of the stiffness of the EoS, since it quantifies the response of pressure to changes in energy density. It is therefore particularly sensitive to changes in the microscopic composition of dense matter. While $c_s$ generally increases with density, causality requires it to remain below the speed of light~\cite{Pasqualotto:2023hho}.

Realistic EoSs may also exhibit non-monotonic behavior associated with the onset of new degrees of freedom, including hyperons, exotic hadrons, or deconfined quarks. Such transitions may generate characteristic features (e.g., kinks or plateaus) in the sound-speed profile~\cite{Dutra2015,McLerran:2018hbz,Stone:2019blq,PhysRevLett.114.031103,Annala:2019puf,Ferreira:2020kvu,Jakobus:2020nxw,Zacchi:2019ayh,Zhao:2020dvu,Minamikawa:2020jfj,Hippert:2021gfs,Pisarski:2021aoz,Kapusta:2021ney,Somasundaram:2021ljr,Malfatti:2020onm,Sen:2020qcd} and have been associated with ultra-massive neutron stars~\cite{Tan:2021ahl}.

The speed of sound $c_s$ for a fixed parameter $\alpha$, in units of the speed of light, can be obtained from the expression:
\begin{equation}\label{eq:cs2}
c_s^2 \big|_{\alpha} = \frac{d p}{d \varepsilon}\Big|_{\alpha} \textrm.
\end{equation}
A variety of choices for $\alpha$ may be found in literature, such as $\alpha=s/n_B$, $s$, $n_B$, $T$, $\mu_B$, etc. In neutron star matter, the quantity $c_s^2 |_T$ has been studied extensively with considerable experimental importance~\cite{Parui:2025zlb}; however, in the present work, for the reasons presented in the introduction of Sec.~\ref{sec:resultsNM}, we study the speed of sound with the fixed parameter $\alpha=S/A=s/n_B$, i.e., along the isentropic curve. 

\begin{figure*}
\includegraphics[width=0.75\textwidth]{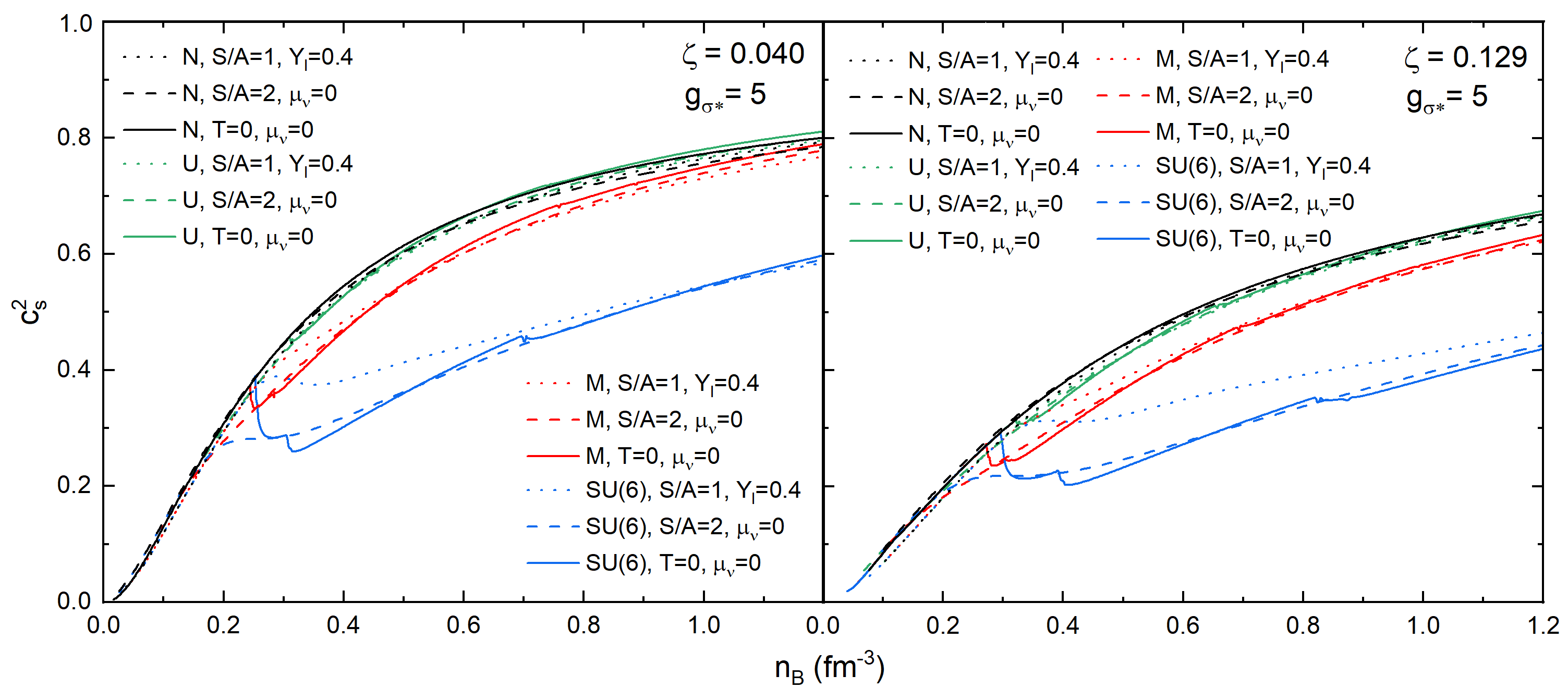}
\caption{\label{CSgeral} {\bf{SET 1 (Speed of sound):}} Speed of sound squared ($c_s^2$) as a function of baryon density ($n_B$) for different proto-neutron star evolution snapshots, particle composition, and hyperon couplings. The different panels show different MBF parameterizations.}
\end{figure*}

\begin{figure*}
\includegraphics[width=1.0\textwidth]{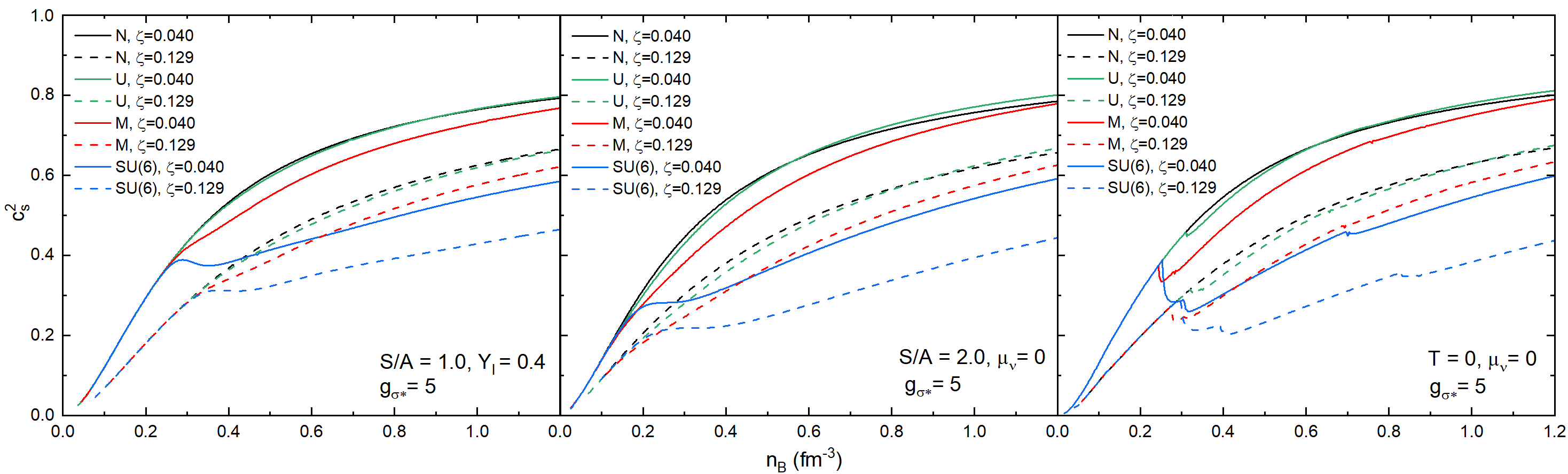}
\caption{\label{CSsequencia} {\bf{SET 2 (Speed of sound):}} Same as Fig. \ref{CSgeral} but now the different panels show different evolution snapshots.}
\end{figure*}

Figs. \ref{CSgeral} and \ref{CSsequencia} show the sound speed for the different MBF parameterizations, hyperon coupling schemes, and evolutionary conditions. In all cases, $c_s^2 \rightarrow 0$ at low density, consistent with the corresponding EoS, and increases rapidly around nuclear saturation density. For a fixed evolutionary condition, all coupling schemes coincide below the hyperon threshold, since matter is composed exclusively of nucleons and leptons in this density range. At higher densities, $c_s^2$ gradually approaches an approximately constant value.

For a given model configuration, the asymptotic value is nearly independent of temperature, indicating that high-density effects dominate over thermal contributions (see Fig.~\ref{CSgeral}). The asymptotic values depend mainly on $\zeta$ and the hyperon coupling prescription: for $\zeta=0.040$, $c_s^2$ approaches $\sim0.6$ for the $SU(6)$ scheme and $\sim0.8$ for the remaining cases, whereas for $\zeta=0.129$ the corresponding values are $\sim0.45$ and $\sim0.65$. Therefore, all MBF parameterizations satisfy the causality constraint throughout the explored density range, both at zero and finite temperature ($0\leq c_s^2\leq1$).

At asymptotically high densities, perturbative QCD predicts that $c_s^2$ approaches the conformal limit $1/3$ from below. However, this regime is reached only at densities of order $\sim40\,n_0$~\cite{Brown:2024gqu,Gorda:2018gpy,Gorda:2021znl}, well above those expected in neutron-star interiors. The MBF Model exceeds the conformal limit. Nevertheless, at the intermediate densities relevant for NSs, the EoS cannot yet be determined from first principles, and the behavior of $c_s$ remains uncertain. Accordingly, both sub-conformal and super-conformal behaviors remain viable in this density range~\cite{Altiparmak:2022bke}.

\begin{figure*}
\includegraphics[width=0.75\textwidth]{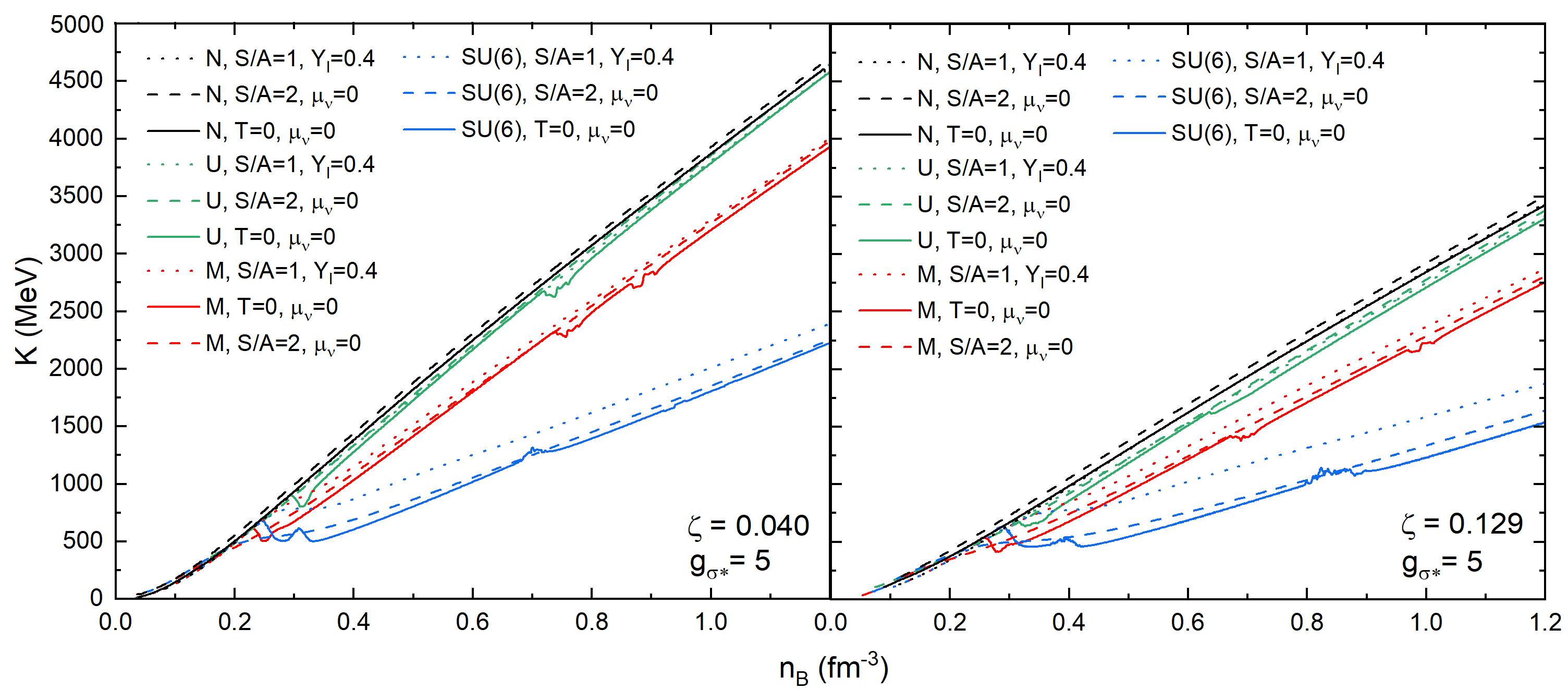}
\caption{\label{Kgeral} {\bf{SET 1 (Compressibility):}} Compressibility ($K$) as a function of baryon density ($n_B$) for different proto-neutron star evolution snapshots, particle composition, and hyperon couplings. The different panels show different MBF parameterizations.}
\end{figure*}

\begin{figure*}
\includegraphics[width=1.0\textwidth]{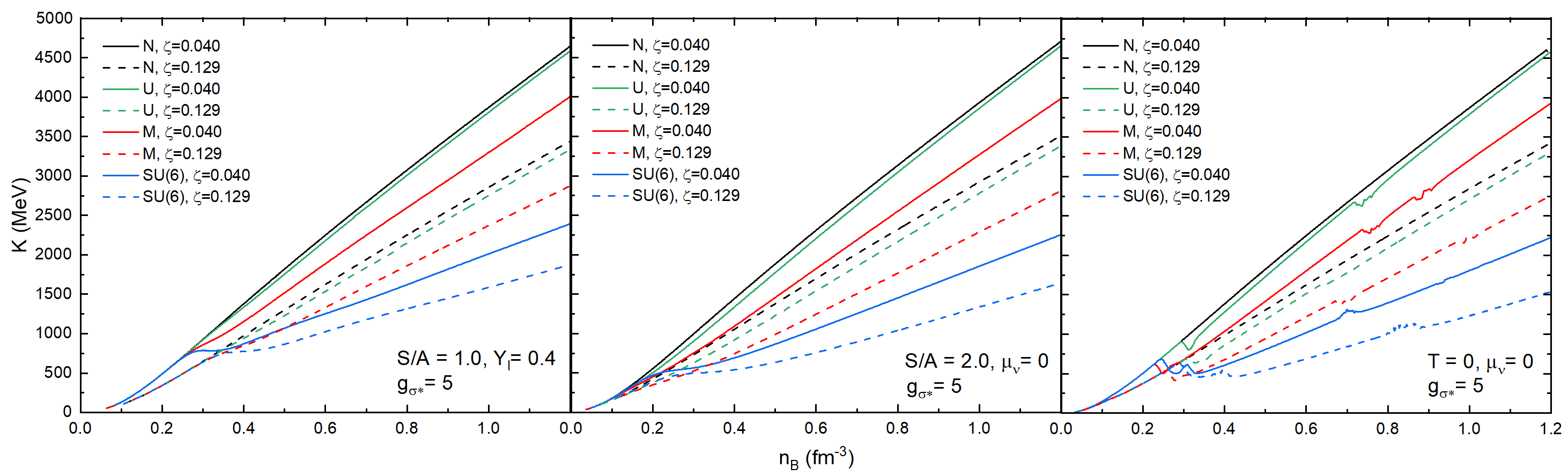}
\caption{\label{Ksequencia} {\bf{SET 2 (Compressibility):}} Same as Fig. \ref{Kgeral} but now the different panels show different evolution snapshots.}
\end{figure*}

Although the conformal bound was once regarded as a natural expectation, accumulating evidence from QCD and astrophysical observations indicates that it need not hold at neutron-star densities. Such violations have been reported in a wide variety of QCD-inspired approaches, including large-isospin QCD, two-color QCD, quarkyonic matter, effective models, and gauge/gravity duality~\cite{Carignano:2016lxe,459l-qqd1,AYALA2025139396,PhysRevD.107.074027,Hands:2006ve,Pasqualotto:2025kpo,sym18020220,McLerran:2018hbz,Margueron:2021dtx,Duarte:2021tsx,Ecker:2017fyh,Demircik:2021zll,Kovensky:2021kzl}. Likewise, observations of massive NSs and theoretical studies of their maximum masses favor sufficiently stiff EoSs, often requiring $c_s^2\gtrsim1/3$ over part of the relevant density range~\cite{Antoniadis:2013pzd,NANOGrav:2019jur,Fonseca:2021wxt,Margalit:2017dij,Rezzolla:2017aly,Ruiz:2017due,Shibata:2019ctb,Nathanail:2021tay,PhysRevLett.114.031103}. Current evidence therefore favors a non-monotonic, causal behavior with $c_s^2>1/3$ over part of the neutron-star density range, while purely sub-conformal EoSs appear to be statistically disfavored~\cite{Altiparmak:2022bke}.

Figs.~\ref{CSgeral} and \ref{CSsequencia} show that, in purely nucleonic matter, $c_s^2$ increases monotonically with density. In contrast, hyperonic matter exhibits characteristic bumps associated with the onset of successive hyperon species, whose appearance temporarily softens the EoS. This behavior directly reflects the sensitivity of the sound speed to the stiffness of the EoS~\cite{Tews:2018kmu}. The effect is most pronounced at $T=0$, where each peak can be directly associated with the onset of a different hyperon species (see Sec.~\ref{sec:populations}). The zero-temperature results agree qualitatively well with previous density-dependent RMF calculations~\cite{Shrivastava:2025ags}
and with other effective approaches~\cite{Stone:2019blq}.

At finite temperature ($S/A=1$ and $S/A=2$), the onset of hyperons is thermally broadened, smoothing the sound-speed profiles and largely suppressing the sharp peaks observed at $T=0$. The finite-temperature behavior is likewise qualitatively consistent with previous effective-model calculations~\cite{Stone:2019blq,Raduta:2021coc,Miyatsu:2025rzn}. In all evolutionary conditions we studied, there is a similar spread for different hyperon coupling schemes in our case to the one in Ref.~\cite{Stone:2019blq} for different RMF models.

\subsection{\label{sec:compressibility}Compressibility}

The compressibility quantifies the response of nuclear matter to density variations and is another important measure of the stiffness of the EoS. It is closely related to the pressure around saturation density~\cite{Blaizot:1980tw} and therefore influences neutron-star structure and dynamics~\cite{PhysRevC.68.034324,Perego:2021mkd}.

In the present work, the compressibility is defined as
\begin{equation}\label{compress}
K=9 \frac{\partial p}{\partial n_B}\, .
\end{equation}

The corresponding results for the different MBF parameterizations, hyperon coupling schemes, and evolutionary conditions are shown in Figs.~\ref{Kgeral} and \ref{Ksequencia}. In all cases, $K\rightarrow0$ as $n_B\rightarrow0$ and increases with density, without approaching an asymptotic value. For purely nucleonic matter, this increase is smooth, whereas the onset of hyperons produces pronounced non-monotonic structures, particularly at $T=0$. The compressibility is even more sensitive than $c_s^2$ to the details of the MBF parameterization. For example, the universal coupling ($U$) already deviates noticeably from the purely nucleonic case, and the positions of the peaks correlate closely with the successive appearance of hyperon species discussed in Sec.~\ref{sec:populations}.

This physical quantity also confirms the stiffness hierarchy established in the previous sections: for a fixed hyperon coupling, $\zeta=0.040$ always yields a more incompressible EoS than $\zeta=0.129$, whereas, for fixed $\zeta$, the incompressibility increases from $SU(6)$ to $M$ and finally to the universal ($U$) coupling. For example, at $n_B=1.2~\rm{fm}^{-3}$ ($\simeq8\,n_0$) with $\zeta=0.040$, the compressibility increases from about $2300~\rm{MeV}$ for $SU(6)$ to $3900~\rm{MeV}$ for $M$ and $4600~\rm{MeV}$ for the universal coupling, remaining slightly larger for purely nucleonic matter.

The compressibility also reveals how temperature affects the stiffness of the EoS. For every model configuration, the finite-temperature curves ($S/A=1$ and $S/A=2$) lie above the corresponding $T=0$ curve, indicating that thermal effects generally increase the incompressibility. This thermal stiffening is strongest for the softer hyperonic configurations ($M$ and $SU(6)$) and less pronounced for the intrinsically stiffer ones (pure nucleonic matter and the universal coupling). The effect is particularly evident for the softer parameterization ($\zeta=0.129$), where the separation between the finite-temperature and zero-temperature curves is systematically larger than for $\zeta=0.040$.

This behavior reflects the reduced sensitivity of already stiff EoSs to thermal effects. In purely nucleonic matter and under the universal coupling, the $S/A=1$ curves are almost indistinguishable from the $T=0$ ones, whereas the hotter $S/A=2$ configurations exhibit a measurable increase in compressibility. The behavior changes for the $M$ and $SU(6)$ coupling schemes. Because these parameterizations are intrinsically softer, even the moderate temperatures corresponding to $S/A=1$ produce a noticeable increase in $K$, despite only a small change in the hyperon population. At $S/A=2$, however, the much earlier and more abundant production of hyperons outweighs the direct stiffening effect of temperature, leading instead to a softer EoS. Accordingly, the $S/A=2$ curves for the $M$ and $SU(6)$ prescriptions lie below the corresponding $S/A=1$ curves in Fig.~\ref{Kgeral}.

Therefore, the compressibility is governed by two competing mechanisms: thermal effects, which tend to increase $K$, and hyperon production, which softens the EoS and reduces it. The observed behavior results from the balance between these two effects.

\subsection{\label{sec:gamma}Adiabatic index}

The adiabatic index is another quantity that characterizes the stiffness of the EoS. In the case of constant entropy, it is directly related to the speed of sound through
\begin{equation}
  \Gamma =\frac{\partial \ln p}{\partial \ln n_B}=\frac{n_B}{p} \frac{\partial p}{\partial n_B}=\frac{(p+\varepsilon)}{p}  \frac{dp}{d \varepsilon} = \frac{(p+\varepsilon)}{p}\, c_s^2 \, .
   \label{eq:GS}
\end{equation}

In neutron-star matter, $\Gamma$ is particularly useful because it is highly sensitive to changes in the microscopic composition of matter and to the stability of stellar oscillations~\cite{PhysRevC.58.1804,Chamel:2008ca,Casali2010AdiabaticIO,Haensel:2002qw}. Recently, structure in $\Gamma$ has been associated with longer gravitational-wave signals from neutron-star mergers, whereas in that case they were associated with higher-order phase transitions to a deconfined quark phase~\cite{Hammond:2025kki}.

The $\Gamma$ density dependence reflects changes in the microscopic composition of matter, making it a valuable diagnostic of phase transitions and other structural changes in the EoS.
\begin{figure*}
\includegraphics[width=0.75\textwidth]{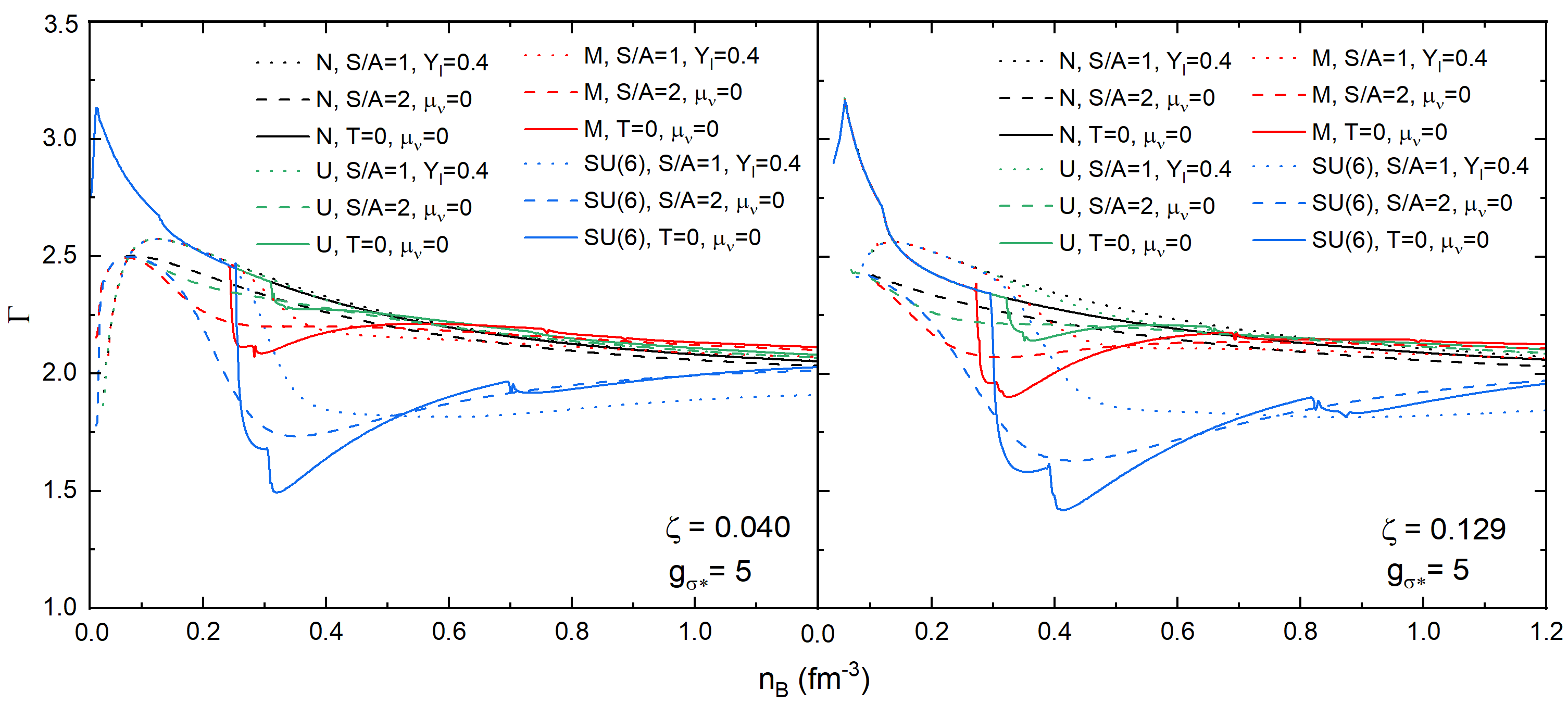}
\caption{\label{GAMAgeral} {\bf{SET 1 (Adiabatic index):}} Adiabatic index ($\Gamma$) as a function of baryon density ($n_B$) for different proto-neutron star evolution snapshots, particle composition, and hyperon couplings. The different panels show different MBF parameterizations.}
\end{figure*}
\begin{figure*}
\includegraphics[width=1.0\textwidth]{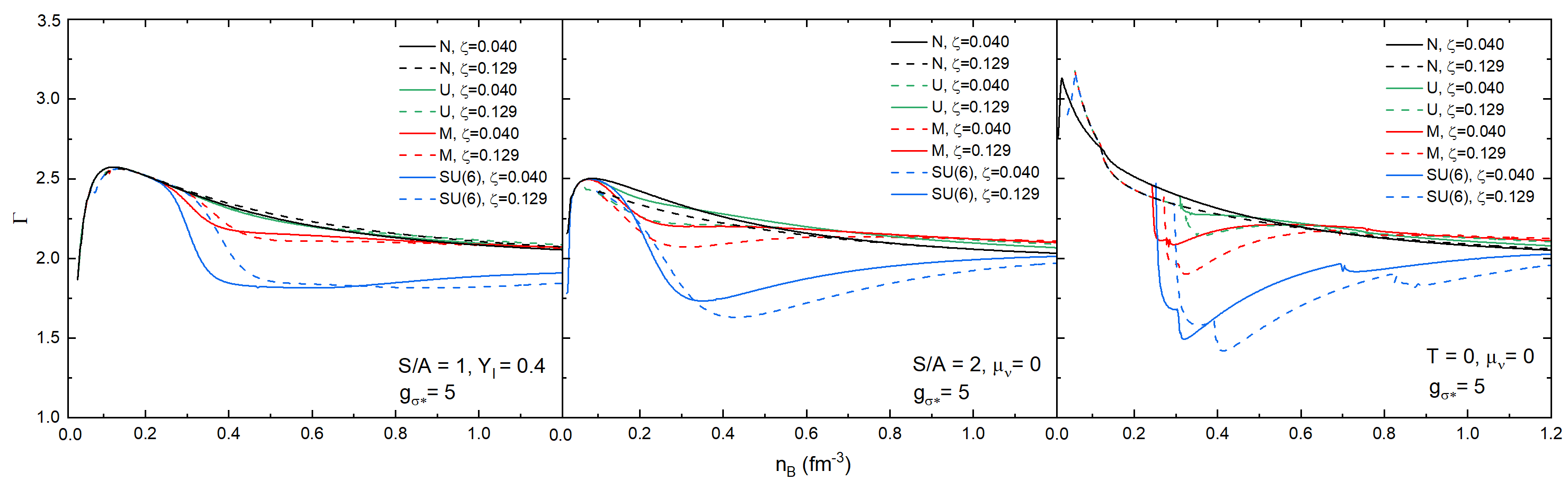}
\caption{\label{GAMAsequencia} {\bf{SET 2 (Adiabatic index):}} Same as Fig. \ref{Kgeral} but now the different panels show different evolution snapshots.}
\end{figure*}
In particular, values of $\Gamma$ below approximately $4/3$ signal instability against gravitational collapse~\cite{osti_5063967,PhysRevC.50.460}. Only a few classes of EoSs, such as those based on Skyrme energy-density functionals, exhibit a weak density dependence of $\Gamma$~\cite{PhysRevC.50.460}. Relativistic many-body models generally predict a much richer behavior owing to the interplay among the different interaction channels.

The adiabatic index also depends on temperature, particularly in hyperonic matter, making constant-$\Gamma$ approximations inadequate~\cite{Kochankovski:2022ygd}. The MBF Model naturally incorporates both density and temperature effects, whose consequences are shown in Figs.~\ref{GAMAgeral} and \ref{GAMAsequencia}. For both zero and finite temperature, $\Gamma$ rises rapidly above its low-density value in the vicinity of saturation density. This follows directly from Eq.~(\ref{eq:GS}): although $c_s^2$ is small at low densities, the large ratio $\varepsilon/p$ dominates the product $(1+\varepsilon/p)c_s^2$, leading to the observed increase in $\Gamma$.

At high densities, where the EoS becomes nearly linear and $c_s^2$ approaches a constant value, $\Gamma$ also reaches a plateau around $\Gamma\simeq2$ for $n_B\gtrsim1.2~\rm{fm}^{-3}$ ($\simeq8\,n_0$). The intermediate-density region contains the most distinctive features. Each hyperon threshold produces a marked decrease in $\Gamma$. The magnitude of these drops is governed primarily by the nucleon-hyperon interaction. This effect is most evident at $T=0$ for both MBF parameterizations (right panel of Fig.~\ref{GAMAsequencia}).

As expected, the purely nucleonic and universal-coupling ($U$) cases display comparatively smooth profiles, whereas the $M$ and $SU(6)$ prescriptions generate pronounced discontinuity-like drops associated with the successive appearance of hyperons. The strongest drops occur for the $SU(6)$ prescription, reflecting its more realistic treatment of nucleon-hyperon interactions through symmetry relations and hyperon optical potentials (Eqs.~\ref{SU6couplings}--\ref{hyperonpotentials}). The correspondence between these drops and the hyperon thresholds is confirmed by comparison with the particle populations discussed in Sec.~\ref{sec:populations}. The $T=0$ results are also qualitatively consistent with those reported for other effective models~\cite{Stone:2019blq}

As in the previous observables, finite temperature smooths the $\Gamma$ profiles by promoting the gradual appearance of hyperons over a broader density range. The finite-temperature results are qualitatively consistent with those obtained in other effective models~\cite{Raduta:2021coc,Stone:2019blq}.

Independently of the evolutionary conditions, when compared with Ref.~\cite{Stone:2019blq}, our results show a much smaller spread at larger densities than for different hyperon coupling schemes than in Ref.~\cite{Stone:2019blq} for different RMF models. In all configurations, the MBF Model predicts $\Gamma>4/3$, thus satisfying the canonical stability criterion.

\section{\label{sec:MRrelations}Mass-radius relation\\ of hot and cold neutron stars}

The equilibrium structure of nonrotating NSs is obtained by solving the Tolman-Oppenheimer-Volkoff (TOV) Eqs.~\cite{Tolman,OppenheimerVolkoff}, which relate the microscopic EoS to the stellar mass and radius. Using natural units ($G=c=1$), the TOV equations read
\begin{equation}
\label{tovp} \frac{dp(r)}{dr} = -\frac{\varepsilon(r)M(r)}{r^2} \frac{
\left[1+ {p(r)}/{\varepsilon(r)} \right] \left[1+{4\pi
r^3p(r)}/{M(r)}\right] } {\left[1-{2M(r)}/{r}\right]}\, ,
\end{equation}
\begin{equation}
\label{tovm} M(r)=\int_0^r 4\pi {r'}^2 \varepsilon(r')dr'\, .
\end{equation}
Here $p(r)$, $\varepsilon(r)$ and $M(r)$ denote the pressure, energy density and enclosed gravitational mass in radius $r$, respectively. For each central energy density $\varepsilon_c$, the equations are integrated outward until $p(R)=0$, yielding one equilibrium configuration, where $R$ is the stellar radius. Repeating the procedure generates the mass-radius relation for the chosen EoS.

\begin{figure*}
\includegraphics[width=0.75\textwidth]{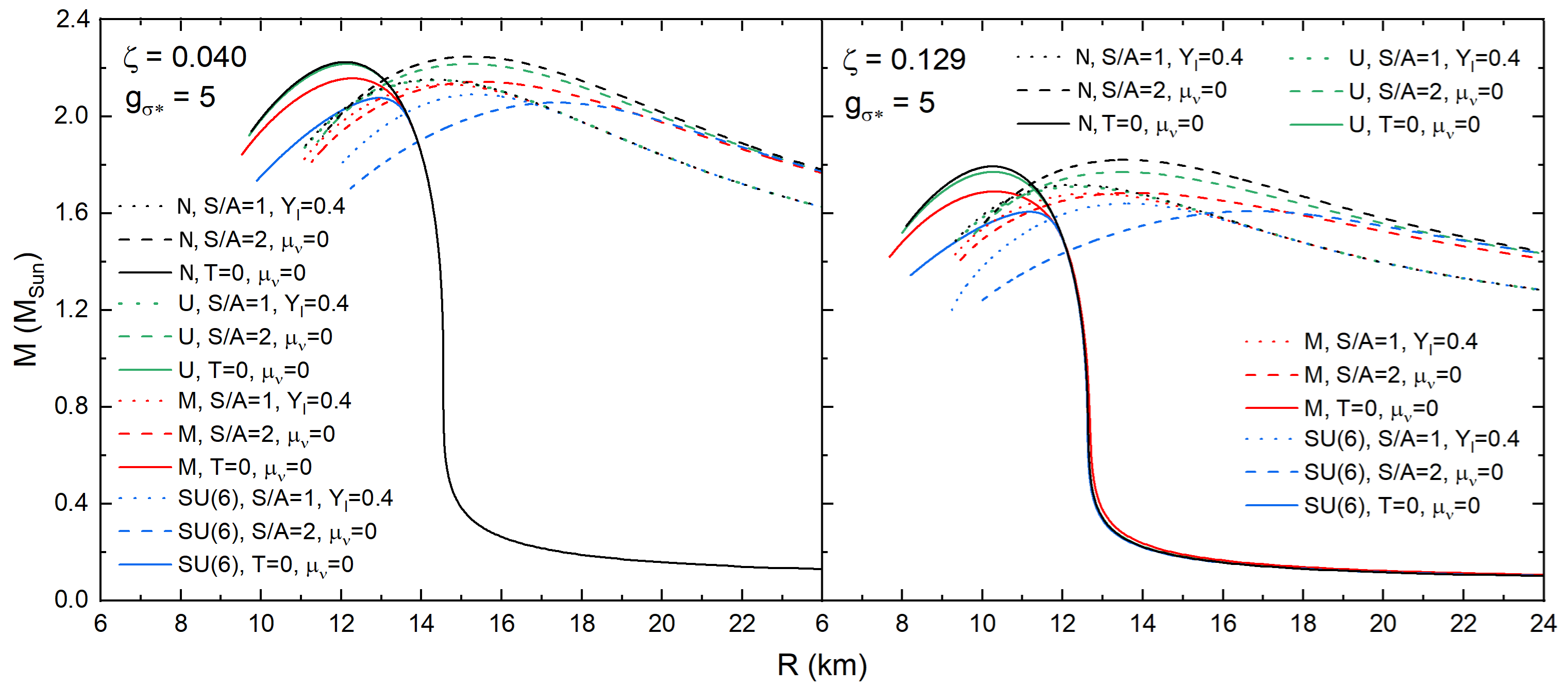}
\caption{\label{TOVgeral} {\bf{SET 1 (Mass-radius diagram):}} Gravitational mass ($M$) as a function of radius ($R$) for different proto-neutron star evolution snapshots, particle composition, and hyperon couplings. The different panels show different MBF parameterizations.}
\end{figure*}

\begin{figure*}
\includegraphics[width=1.0\textwidth]{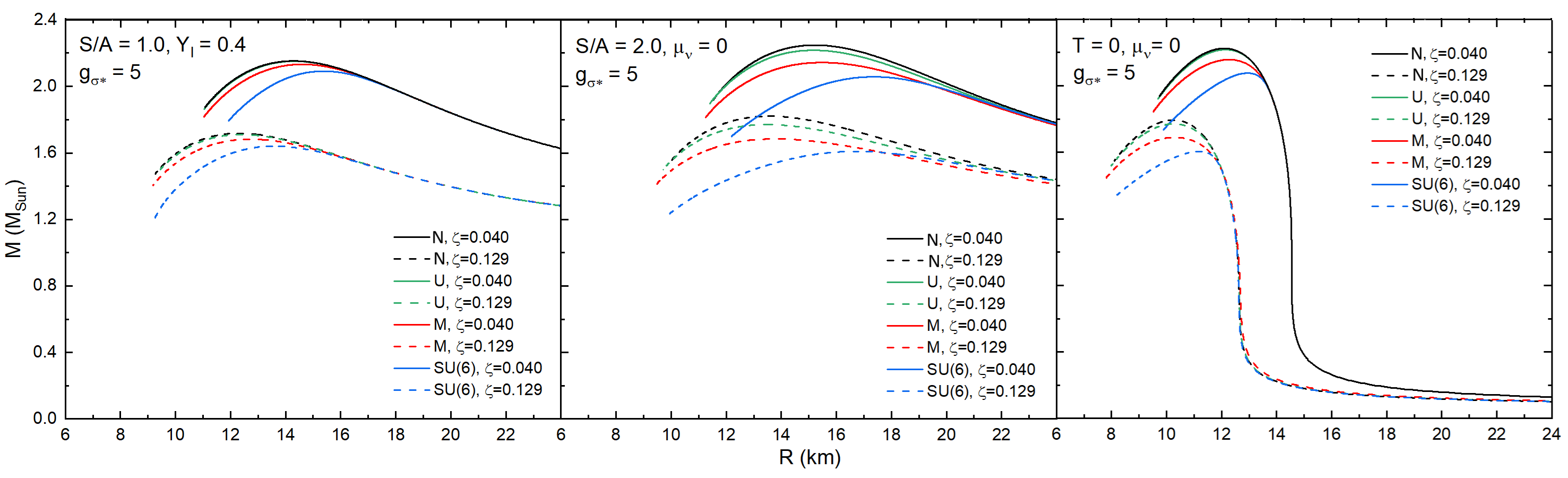}
\caption{\label{TOVsequencia} {\bf{SET 2 (Mass-radius diagram):}} Same as Fig. \ref{TOVgeral} but now the different panels show different evolution snapshots.}
\end{figure*}

Different EoSs produce different families of TOV solutions. Consequently, changes in composition, hyperon couplings, temperature, or entropy per baryon directly translate into different mass-radius relations and maximum masses, allowing astrophysical observations to constrain the high-density EoS.

In the present work, different evolutionary  conditions, neutrino contents and hyperonic compositions generate distinct EoSs and, consequently, different families of TOV solutions, allowing us to follow the evolution from a hot PNS to a cold NS. For cold neutron stars ($T=0$), the MBF EoSs were matched to the BPS (Baym-Pethick-Sutherland) crust~\cite{osti_4718088}. For finite-temperature configurations, the Lattimer-Swesty thermal mantle was adopted~\cite{LATTIMER1991331}.

Figs.~\ref{TOVgeral} and \ref{TOVsequencia} summarize the mass-radius relations obtained for the different evolutionary conditions, hyperon coupling schemes and MBF parameterizations. Finite entropy mainly affects stellar radii rather than maximum masses. For a given gravitational mass, hotter configurations are systematically larger, particularly for $S/A=1$, where neutrino trapping delays hyperon formation through the larger lepton fraction.

The influence of the many-body parameter $\zeta$ is equally robust. As expected, the stiffer parameterization ($\zeta=0.040$) systematically yields larger radii and higher maximum masses than the softer one ($\zeta=0.129$), with the differences being more pronounced for hyperonic matter. This behavior is particularly evident in Fig.~\ref{TOVsequencia}. The corresponding maximum masses and radii are listed in Table~\ref{quadro_massradius}.

The impact of hyperons on the mass-radius relation is evident in Figs.~\ref{TOVgeral} and \ref{TOVsequencia}. As expected from previous RMF studies~\cite{Colucci:2014wda,Oertel:2014qza,Zhao:2015luv,Zhao:2015tra,Zhao:2017pim,Banik:2017zia}, pure nucleonic matter always yields the largest maximum masses and radii. From the universal to the Moszkowski and finally the $SU(6)$ prescription, the progressive softening of the EoS shifts the mass-radius curves downward and reduces the maximum mass, an effect that is most evident for cold matter. Neutrino trapping mitigates this effect by delaying hyperon formation, whereas the differences among coupling schemes become progressively more pronounced for $S/A=2$ and, especially, for $T=0$.

\begin{table}   
\caption{\label{quadro_massradius}
Maximum star mass and the corresponding radius for different configurations of the MBF Model and three scenarios of entropy per baryon ($S/A$). Here, $N$ stands for pure nucleonic matter. These results consider $g_{\sigma^*}=5$.
}
\begin{ruledtabular}
\resizebox{1.0\linewidth}{!}{
\begin{tabular}{cccccc}
S/A & $\zeta$ & $SU(6)$ & $M$ & $U$ & $N$ 
\tabularnewline
\hline 
$1$ & $0.040$ & $2.10\,\rm{M_{Sun}}$ & $2.13\,\rm{M_{Sun}}$ & $2.15\,\rm{M_{Sun}}$ & $2.15\,\rm{M_{Sun}}$ \tabularnewline  
 & & $15.29 $\,km & $14.59 $\,km & $14.33 $\,km & $14.31 $\,km 
\tabularnewline
$1$ & $0.129$  & $1.65\,\rm{M_{Sun}}$ & $1.68\,\rm{M_{Sun}}$ & $1.71\,\rm{M_{Sun}}$ & $1.72\,\rm{M_{Sun}}$ \tabularnewline 
 & & $13.46 $\,km & $12.71 $\,km & $12.38 $\,km & $12.32 $\,km 
\tabularnewline
\hline
$2$ & $0.040$ & $2.07\,\rm{M_{Sun}}$ & $2.15\,\rm{M_{Sun}}$ & $2.22\,\rm{M_{Sun}}$ & $2.25\,\rm{M_{Sun}}$ \tabularnewline 
 & & $17.15 $\,km & $15.52 $\,km & $15.17 $\,km & $15.22 $\,km 
\tabularnewline
$2$ & $0.129$  & $1.62\,\rm{M_{Sun}}$ & $1.69\,\rm{M_{Sun}}$ & $1.77\,\rm{M_{Sun}}$ & $1.82\,\rm{M_{Sun}}$ \tabularnewline 
 & & $16.31 $\,km & $13.81 $\,km & $13.50 $\,km & $13.55 $\,km 
\tabularnewline
\hline 
$0$ & $0.040$ & $2.10\,\rm{M_{Sun}}$ & $2.16\,\rm{M_{Sun}}$ & $2.22\,\rm{M_{Sun}}$ & $2.22\,\rm{M_{Sun}}$ \tabularnewline  
 & & $12.87 $\,km & $12.27 $\,km & $12.14 $\,km & $12.10 $\,km 
\tabularnewline
$0$ & $0.129$ & $1.62\,\rm{M_{Sun}}$  & $1.70\,\rm{M_{Sun}}$ & $1.77\,\rm{M_{Sun}}$ & $1.79\,\rm{M_{Sun}}$ \tabularnewline 
 & & $11.08 $\,km & $10.30 $\,km & $10.26 $\,km & $10.25 $\,km 
\tabularnewline
\end{tabular}}
\end{ruledtabular}
\end{table}

Beyond the maximum-mass point, the steep decline of each curve marks the onset of radial instability, whose location depends on both $\zeta$ and the hyperon coupling scheme.

The strongest softening occurs for the combination of $SU(6)$ couplings and $\zeta=0.129$, which becomes unstable at comparatively low central densities and therefore predicts the lowest maximum masses. This result reinforces the well-known tendency of hyperons to reduce the maximum mass of neutron stars in the absence of additional repulsive interactions.

The three thermodynamic snapshots illustrate the evolution from a hot, neutrino-rich PNS to a cold NS. While the maximum mass changes only modestly during this evolution, the stellar radius decreases from approximately $14$--$17\,\rm{km}$ in the hot configurations to $10$--$13\,\rm{km}$ for the cold star. Overall, the mass-radius relations demonstrate how thermal effects, neutrino trapping, hyperons and the many-body parameter $\zeta$ jointly determine the global properties of neutron stars within the MBF Model.

\begin{figure}
\includegraphics[width=0.48\textwidth]{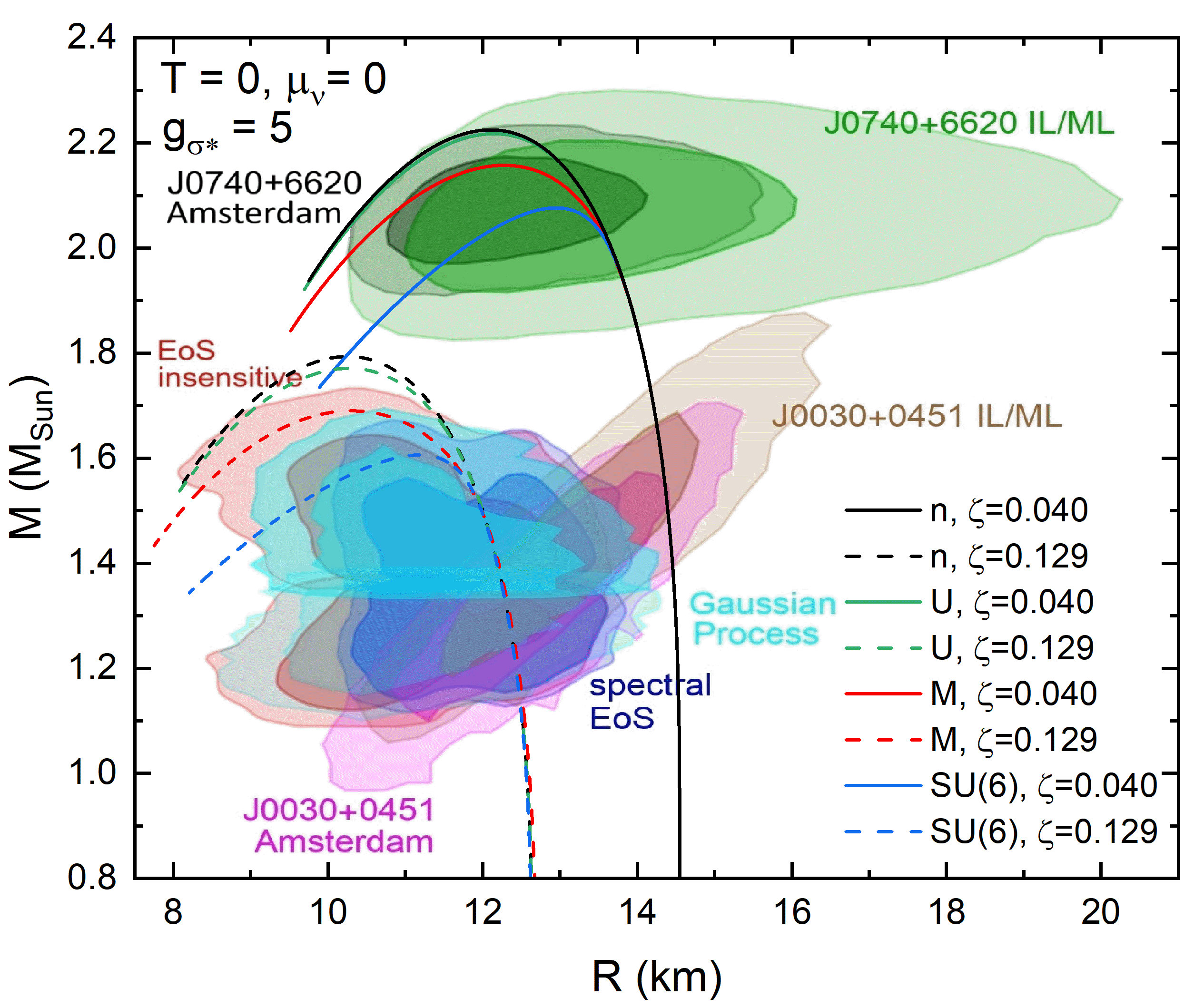}
\caption{\label{TOVconstraints} Mass-radius diagram predicted by the MBF Model for fully evolved NSs, compared with multimessenger observational constraints. Solid and dashed curves represent, respectively, the
$\zeta=0.040$ and $\zeta=0.129$ MBF parameterizations for different configurations (pure nucleonic matter $N$, and the $U$, $M$, and $SU(6)$ coupling schemes for hyperons). The shaded contours in the background, extracted from Ref.~\cite{MUSES:2023hyz}, summarize current astrophysical inferences: NICER measurements for PSR J0030+0451 (brown) and PSR J0740+6620 (green) from both the Amsterdam and Illinois/Maryland groups; the EoS-insensitive constraints from GW170817 (old pink); the spectral-EoS constraints (dark blue); and the Gaussian Process reconstruction combining GW170817 and heavy-pulsar mass measurements (light blue). All contours show the $68\%$ and $95\%$ confidence regions.}
\end{figure}

Fig.~\ref{TOVconstraints} compares the $T=0$ mass-radius relations with the astrophysical constraints inferred from NICER and LIGO/Virgo observations.

Since 2010, several NSs with masses exceeding $2\,\rm{M_{Sun}}$ have been observed. Two pioneering observations are those of PSR J1614-2230, with a mass of $1.97 \pm 0.04\,\rm{M_{Sun}}$~\cite{PMID:20981094}and an estimated radius of $10.06^{+1.25}_{-0.87}$ km~\cite{mauviard2026nicerviewpsrj16142230}, and PSR J0348+0432, with a mass of $2.01 \pm 0.04\,\rm{M_{Sun}}$~\cite{Antoniadis:2013pzd} and an estimated radius of $12.07 $\,km~\cite{Zhao:2015tra}. A significant milestone was the observation of PSR J0740+6620, the heaviest NICER pulsar, with a mass of $2.08 \pm 0.07\,\rm{M_{Sun}}$~\cite{NANOGrav:2019jur} and an estimated radius of $13^{+1.30}_{-0.98} $\,km~\cite{Riley:2021pdl}, which was the record holder for the heaviest NS until 2022, when the title was taken by PSR J0952–0607 with a reported mass of $2.35 \pm 0.17\,\rm{M_{Sun}}$~\cite{Romani:2022jhd} and an estimated radius of $12.39^{+1.30}_{-0.98} $\,km~\cite{Kumar:2023mlp}. It is also important to record, at the other extreme, the lightest pulsar observed by the NICER telescope: PSR J0030+0451, with a mass of $1.44 \pm 0.14\,\rm{M_{Sun}}$ and an estimated radius of $13.02^{+1.24}_{-1.06} $\,km~\cite{Riley:2019yda,Miller:2019cac}. 

Another relevant source of data for the mass-radius relation is the a gravitational wave event GW170817 observed by the LIGO and Virgo detectors in 2017~\cite{LIGOScientific:2017vwq}. The wave was produced by the last moments of the inspiral of a binary pair of neutron stars, ending with their merger. Combining these observations with a quasi-universal relation between the maximum mass of nonrotating stellar models and the maximum mass that can be supported through uniform rotation, the neutron-star mass $M$ was estimated to be in the range $2.01^{+0.04}_{-0.04} M_{Sun}\leq M \lesssim 2.16^{+0.17}_{-0.15}\,\rm{M_{Sun}}$~\cite{Rezzolla:2017aly}. In Ref.~\cite{PhysRevLett.121.091102}, tidal deformabilities and masses of GW170817 were used to infer the radii of the objects in the binary system, adopting a spectral parameterization for the EoS, leading to $11.9 \pm 1.4 $\,km. An alternative approach follows the procedure of Ref.~\cite{YAGI20171} and uses an EoS that is insensitive to the dependence between the tidal deformability and the mass-radius relation, leading to inferred radii of $10.8^{+2.0}_{-1.7} $\,km and $10.7^{+2.1}_{-1.5} $\,km. In this calculation, uncertainty in the EoS-insensitive relation was modeled as Gaussian white noise in the relation. A different EoS-insensitive relation was used in Ref.~\cite{PhysRevLett.121.091102} to determine the common neutron-star radius, which was reported as $10.7^{+2.1}_{-1.6} $\,km.   

The regions (contours) in the background of Fig. \ref{TOVconstraints}, extracted from Ref.~\cite{MUSES:2023hyz}, summarize some of these astrophysical data. The universal-relation constraint (EoS-insensitive, colored in old pink) and the spectral-EoS constraint (colored in dark blue) are obtained using GW170817 data. The Gaussian Process constraint (colored in light blue) is obtained by combining GW170817, PSR 1614-2230, PSR 0348+0432, and PSR 0740+6620 (mass only) data. NICER constraints from both the Amsterdam and Illinois/Maryland groups are show for both J0030+0451 (colored in brown) and J0740+6620 (colored in green). Shaded regions show the $68\%$ and $95\%$ confidence region.

Although these different methods produce posterior distributions for the mass and radius of GW170817 that look different, they are statistically consistent. Importantly, one cannot simply overlay an EoS mass–radius curve on these contours and decide by eye whether the EoS is ruled out, because every point on the plane (even outside the $68\%$ or $95\%$ regions) still carries nonzero posterior weight; therefore, a proper Bayesian comparison would be required. Nevertheless, even a less rigorous comparison between the curves describing the mass-radius relationship of the MBF model and these contours can serve as a first and strictly qualitative evaluation of our results.

\begin{figure}
\includegraphics[width=0.48\textwidth]{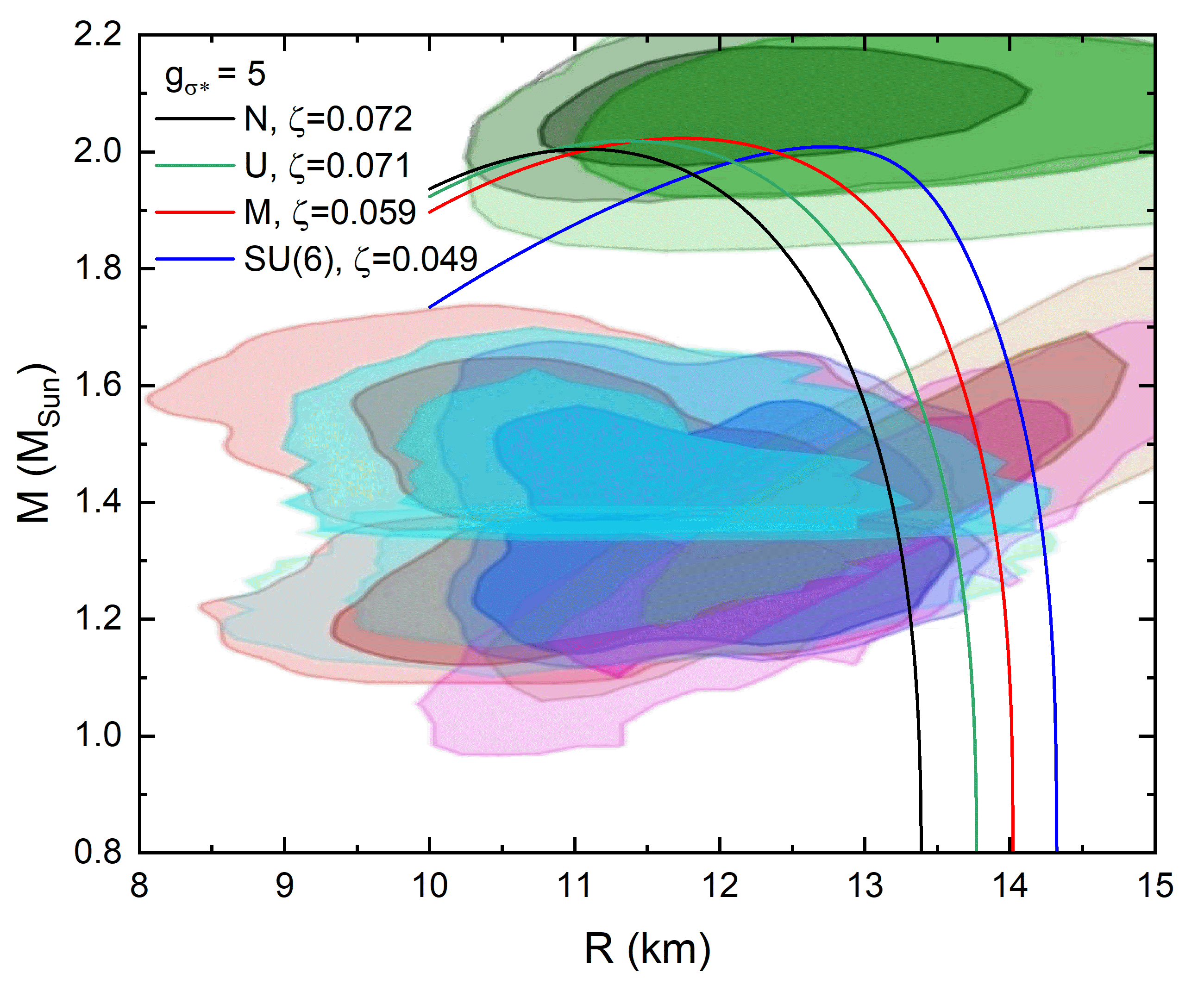}
\caption{\label{TOVgold}{\bf Representative observationally compatible parameterizations of the MBF Model.}
Mass-radius relations obtained at $T=0$ for the four particle compositions considered in this work: purely nucleonic matter ($N$), universal ($U$), Moszkowski ($M$), and $SU(6)$ hyperon-meson coupling schemes. For each composition, the value of the many-body parameter $\zeta$ was chosen to yield a mass-radius relation consistent with current astrophysical constraints. The shaded regions represent the observational constraints discussed in Fig.~\ref{TOVconstraints}.}
\end{figure}

Fig.~\ref{TOVconstraints} shows that the $\zeta=0.040$ parameterization is broadly consistent with NICER observations of neutron-star masses and radii. All $\zeta=0.040$ sequences exceed the $2\,\rm{M_{Sun}}$ threshold and are compatible with the NICER constraints from both PSR~J0740+6620 and PSR~J0030+0451. Although the $M$ and $SU(6)$ prescriptions reduce the maximum mass relative to pure nucleonic matter, the effect remains moderate for $\zeta=0.040$, and all sequences stay within the observationally favored region. By contrast, the softer $\zeta=0.129$ parameterization remains compatible with GW170817 and PSR~J0030+0451 but fails to reproduce the mass of PSR~J0740+6620, placing it in tension with current observations. These results highlight the strong sensitivity of the hyperonic sector to both the coupling prescription and the many-body parameter $\zeta$.

To further assess the astrophysical performance of the MBF Model, Fig.~\ref{TOVgold} presents a set of representative observationally compatible parameterizations. Instead of considering only the extreme values of the many-body parameter $\zeta$, we selected, for each hyperon-meson coupling prescription, the value of $\zeta$ that simultaneously reproduces the two most important observational requirements currently available, namely a maximum stellar mass close to $2\,\rm{M_{Sun}}$ and radii consistent with the constraints inferred from NICER and the GW170817 event. The resulting optimal parameterizations are $\zeta=0.072$ for purely nucleonic matter, $\zeta=0.071$ for the universal ($U$) coupling, $\zeta=0.059$ for the Moszkowski ($M$) prescription, and $\zeta=0.049$ for the $SU(6)$ coupling. In these configurations, the corresponding radii for a star with $\simeq 2\,\rm{M_{Sun}}$ are $11.07 $\,km for purely nucleonic matter, $11.36 $\,km for the universal ($U$) coupling, $11.75 $\,km for the Moszkowski ($M$) prescription, and $12.73$\,km for the $SU(6)$ coupling. These are the $4$ parameterizations that we provide in the CompOSE repository~\cite{Typel:2013rza,refId0}. For each parameterization, we generate 1-Dimensional charge neutral, $\beta$-equilibrated, $T=0$ tables as a function of $n_B$ and 3-Dimensional general purpose tables as a function of $n_B$, $T$ , and charge fraction $Y_Q$.

A clear physical trend emerges from these solutions. As the hyperonic interaction becomes progressively more favorable -- moving from purely nucleonic matter to the universal, Moszkowski, and finally the $SU(6)$ prescription -- the corresponding EoS becomes increasingly softer owing to the earlier onset and larger abundance of hyperons. To compensate for this additional softening while preserving agreement with astrophysical observations, progressively smaller values of $\zeta$ are required. Since decreasing $\zeta$ enhances the repulsive many-body contribution and stiffens the equation of state, the monotonic sequence $\zeta_{\rm N}=0.072$,
$\zeta_{U}=0.071$, $\zeta_{M}=0.059$, $\zeta_{SU(6)}=0.049$, reveals a compensation mechanism between hyperon-induced softening and many-body interactions. In other words, the stronger the softening introduced by the hyperonic sector, the weaker the many-body effects required to recover agreement with present observations.

Rather than being an observational determination of $\zeta$, these solutions should be regarded as illustrating the parameter dependence of the model and the existence of parameterizations that remain compatible with observational constraints on neutron-star masses and radii while including hyperonic degrees of freedom.

\section{\label{sec:conclusions}Summary and outlook}

In this work we have extended the MBF Model to finite-temperature matter and systematically investigated different hyperon-meson coupling prescriptions within a unified RMF framework. This framework allowed us to assess the combined effects of many-body interactions, temperature and hyperons on the microscopic and macroscopic properties of dense matter relevant to neutron stars.

We constructed finite-temperature EoSs (with fixed entropy per baryon and lepton fraction) for both nucleonic and hyperonic matter and analyzed their thermodynamic properties. From these EoSs we derived the speed of sound, compressibility and adiabatic index, which characterize the stiffness of dense matter.

The MBF Model was shown to generate a broad family of equations of state spanning the phenomenologically relevant range between the $\zeta=0.040$ and $\zeta=0.129$ parameterizations. The hyperon coupling prescription strongly affects the stiffness of the EoS, with the universal scheme producing the stiffest hyperonic configurations and the $SU(6)$ prescription the softest. All configurations satisfy both the stability criterion ($\Gamma>4/3$) and the causality constraint on the speed of sound.

The appearance of hyperons produces characteristic non-monotonic features in the speed of sound, compressibility, and adiabatic index, although thermal effects progressively smooth them by shifting hyperon populations to lower densities. Moreover, despite the fact that some parameterizations produce nearly overlapping EoSs, their temperature profiles differ substantially, particularly at high densities. The coupling-dependent behavior noticed for these thermodynamical quantities  persists at finite temperature and entropy, indicating that the  relation between EoS stiffness and strangeness production is not universal.

We also determined temperature profiles at fixed entropy per baryon, showing how the stellar thermal structure depends on the hyperon coupling prescription and the many-body parameter $\zeta$. The results show that many-body effects can partially compensate for the hyperon-induced softening, particularly for the $\zeta=0.040$ parameterization. The astrophysical implications were investigated by solving the Tolman-Oppenheimer-Volkoff equations for both cold and hot beta-equilibrated matter. For cold NSs, the softer parameterization ($\zeta=0.129$) reproduces the general radius trends but fails to support the largest observed stellar masses, particularly in the presence of hyperons. By contrast, the $\zeta=0.040$ parameterization satisfies both the mass and radius constraints for nucleonic and hyperonic stars, but fails to reproduce gravitational-wave inferences for neutron-star radii.

More generally, these two values delimit the phenomenological range of the scalar MBF parameterization, indicating that intermediate choices of $\zeta$ can generate a continuous family of equations of state compatible with different astrophysical constraints. Intermediate values allow one to cover a broad range of physics. We give as an example selected values $\zeta$ that reproduce standard observational constraints, referred to as ``representative observationally compatible parameterizations''. In this case, a different value of $\zeta$ is chosen for each hyperon coupling scheme, with stronger hyperonic softening requiring smaller values of $\zeta$ and, consequently, weaker many-body effects to recover agreement with present observations. We make tables available for these equations of state in the CompOSE repository.

The finite-temperature TOV solutions reproduce the expected evolutionary sequence from hot proto-neutron stars to cold neutron stars. Neutrino trapping delays hyperonization and partially compensates the hyperon-induced softening, leading to larger stellar radii during the early evolutionary stages. Throughout this evolution, cooling has only a modest impact on the maximum mass, whereas the stellar radius decreases by approximately $25\%$.

These results highlight the importance of a self-consistent thermal description when connecting proto-neutron-star evolution to the final cold configuration. Overall, our results clarify the interplay between thermal effects, many-body forces, and hyperonic degrees of freedom in determining both the microscopic properties of dense matter and the global structure of compact stars.

Several natural extensions of the present work deserve future investigation. First, a Bayesian inference framework combining nuclear-matter properties with current astrophysical observations could be employed to place quantitative constraints on the MBF many-body parameter $\zeta$ and assess its correlations with the hyperonic sector. Moreover, the inclusion of the full spin-$3/2$ baryon decuplet may modify the onset of hyperons at high densities and temperatures. Another important extension is the self-consistent treatment of thermal mesons through Bose-Einstein distributions, which is expected to influence the EoS in hot environments such as binary mergers. Finally, the inclusion of strong magnetic fields will allow the MBF Model to describe magnetars and post-merger remnants, including anisotropic pressures and magnetic oscillatory effects.

In summary, this work establishes the finite-temperature, hyperon-inclusive MBF Model as a robust framework for describing dense baryonic matter over a wide range of conditions relevant to neutron stars and proto-neutron stars. The extensions outlined above should further broaden its applicability to the diverse environments probed by current and future multimessenger observations.

The data presented in this work are available upon request and can be obtained by contacting the corresponding author.

\begin{acknowledgments}
This work was partially supported by the Department of Energy under grant DE-SC0024700; National Science Foundation under grants MUSES OAC2103680 and NP3M PHY2116686;
Coordena\c{c}\~ao de Aperfei\c{c}oamento de Pessoal de N\'{\i}vel Superior (CAPES) Finance Code 001; Conselho Nacional de Desenvolvimento Cient\'{\i}fico e Tecnol\'ogico (CNPq) under grants 312032/2023-4, 402963/2024-5, and 445182/2024-5; Funda\c{c}\~ao de Amparo \`a Pesquisa do Estado do Rio Grande do Sul (FAPERGS) under grant 24/2551-0001285-0; Instituto Nacional de Ci\^encia e Tecnologia - F\'isica Nuclear e Aplica\c{c}\~oes (INCT - FNA) under grants 464898/2014-5 and 408419/2024-5; and Ser\-ra\-pi\-lhei\-ra Institute under grant Serra2211-42230. 
We wish to thank the following researchers, for offering comments, suggestions and encouragement throughout the making of this work: Germán Lugones, Sidney dos Santos Avancini, Dyana Cristine Duarte, Marcus Emmanuel Benghi Pinto, Carlos Conde Ocazionez, André da Silva Schneider, Adriana Raduta, and Yuhan Wang. R.L.S.F. and R.B.J. acknowledge the kind hospitality of the Center for Nuclear Research at Kent State University, where part of this work was done.
\end{acknowledgments}

\appendix*

\section{Calculation of the Coupling Constants}
\label{app:couplings}

In this appendix, we present the method for determining the coupling constants for the nucleons $g_{\sigma}$, $g_{\omega}$, $g_{\varrho}$, and $g_{\delta}$ in the MBF Model. This calculation produces the values shown in Table \ref{quadro_acoplamentos}. After choosing one specific value for the adjustable parameter $\zeta$, the calculation proceeds as follows:

{\bf{STEP 1}}: At saturation, the cold isopin-symmetric nuclear matter has vanishing pressure~\cite{Hugenholtz:1958zz}, and is populated only by nucleons ($N$), not by leptons nor hyperons. Also, due to isospin symmetry, the mean values of the isovector mesons $\varrho$ and $\delta$ are zero. Moreover, at $T=0$, the integral in (\ref{PressaoGasFermiTocteto}) can be performed analytically, resulting:
\begin{equation}
\label{pre}
\begin{split}
p &=  -\frac{1}{2} m_{\sigma}^2\sigma_0^2+ \frac{1}{2}m_{\omega}
^2\omega_0^2 \\ & +
\frac{2}{3\pi^2} \Bigg[ \left(\frac{1}{4}(k_{F})_0^3-\frac{3}{8}(m_N^*)_0^2(k_{F})_0\right)
\sqrt{(k_{F})_0^2+(m_N^*)_0^2}
\\ & +\frac{3}{8}(m_N^*)_0^4\ln \left( {\frac{(k_{F})_0+\sqrt{(k_{F})_0^2+(m_N^*)_0^2
}} {(m_N^*)_0}} \right) \Bigg]=0\, ,
\end{split}
\end{equation}
where ${(k_{F})}_0$ and ${(m_N^*)}_0$ represent, respectively, the Fermi momentum of the nucleons and the effective mass of the nucleons, both at saturation. In this equation, the value of the effective mass at saturation ${(m_N^*)}_0$ is written in terms of the constant $g_\sigma$ and the field $\sigma_0$, according to expression (\ref{meffMBFtcm}):
\begin{equation}
\label{mefetivacoupl}
{(m_N^*)}_0 = m_N - \Pi_\zeta\, g_\sigma\, \sigma_0  ,
\end{equation}
where $m_N$ is the rest mass of the nucleon and where the parametric term $\Pi_\zeta$, given by (\ref{mstarTCM}), has reduced to
\begin{equation}
\label{mstarreduced} 
\Pi_{ \zeta} = \left(1+\frac{g_{\sigma }\sigma_0}{\zeta\,m_N} \right)^{-\zeta}\, .
\end{equation}
The Fermi momentum $(k_{F})_0$ can be obtained from the experimental value of the saturation density $n_0$, after evaluating the integral in expression (\ref{rhoBMBF}):
\begin{equation}
\label{rho0eq}n_0=\frac{2{(k_{F})}_0^3}{3\pi^2} \rightarrow {(k_{F})}_0 = { \left ( \frac{3 \pi^2 n_0}{2} \right) }^{\frac{1}{3}}\, . \end{equation}
Furthermore, $\omega_0$ can be written in terms of the coupling constant $g_\omega$ using the second equation of motion in (\ref{tcm}) and the experimental value of the saturation density $n_0$:
\begin{equation}
\label{omegaSAT}
\omega_0 = \frac{1}{m_\omega^2} g_\omega n_0\, .
\end{equation}
The value $\varepsilon_0$ of energy density at saturation at $T=0$ can be obtained from expressions (\ref{EnergiaMBF}) and (\ref{EnergiaGasFermiTocteto}):
\begin{equation}
\label{ener}
\begin{split}
\varepsilon_0 &=  \frac{1}{2} m_{\sigma}^2\sigma_0^2 +
\frac{1}{2}m_{\omega}^2\omega_0^2 \\ &
+\frac{2}{\pi^2}\Bigg[\left(\frac{1}{4}(k_{F})_0^3+\frac{1}{8}(m_N^*)_0^2(k_{F})_0\right)
\sqrt{(k_{F})_0^2+(m_N^*)_0^2}
\\ & -\frac{1}{8}(m_N^*)_0^4\ln \left ( {\frac{(k_{F})_0+\sqrt{(k_{F})_0^2+(m_N^*)_0^2}} {(m_N^*)_0}} \right) \Bigg]\, .
\end{split}
\end{equation}
In this formula, once again, the substitutions indicated by Eqs.~(\ref{rho0eq}) and (\ref{omegaSAT}) are made, in order to exchange the variables ${(k_{F})}_0$ and $\omega_0$ for $n_0$ and $g_\omega$; furthermore, the value of $\varepsilon_0$ can be determined by knowing the experimental value of the binding energy per baryon ($B/A$), the density of nuclear matter at saturation ($n_0$) and the rest mass of the nucleon ($m_N$), according to:
\begin{equation}
\frac{B}{A} = \frac{\varepsilon_0}{n_0} - m_N \rightarrow \varepsilon_0 = n_0 (m_N + B/A)\, .
\end{equation}

Finally, the third equation of this system is the equation of motion of the scalar field, according to (\ref{tcm}):
\begin{equation}
\label{sigmacoupl}
\sigma_{0}=\frac{1}{m_{\sigma}^{2}}
\left[g_{\sigma}\left(\Pi_{\zeta}\right) -\frac{g_{\sigma}^2}{m_N}
\left(\Pi_{\zeta}\right)^{\frac{\zeta+1}{\zeta}}
\sigma_{0} \right]n_{s}\, ,
\end{equation}
where the parametric term $\Pi_\zeta$ also assumes the simplified form presented in (\ref{mstarreduced}), and the scalar density is determined according to (\ref{rhosMBF}):
\begin{equation}
\begin{split}
\label{rhoss}n_s &=\frac{1}{\pi^2} \Bigg[ (k_{F})_0(m_N^*)_0 \sqrt{(k_{F})_0^2+(m_N^*)_0^2} \\& -(m_N^*)_0^3
\ln \left( {\frac{(k_{F})_0+\sqrt{(k_{F})_0^2+(m_N^*)_0^2}}{(m_N^*)_0}} \right) \Bigg]
\, ,
\end{split}
\end{equation}
where, again, the momentum $(k_F)_0$ is written in terms of the density $n_0$ using relation (\ref{rho0eq}).

Thus, (\ref{pre}), (\ref{ener}) and (\ref{sigmacoupl}) make up a system of three equations and three variables ($\sigma_0$, ${g_\sigma}$ and
${g_\omega}$) that must be solved numerically for each chosen value of the adjustable parameter $\zeta$ of the scalar (S) version of MBF Model. The values obtained for $\sigma_0$, ${g_\sigma}$ and ${g_\omega}$ then allow the calculation of the effective mass of the nucleon ${(m^*_N)}_0$, using Eq.~(\ref{mefetivacoupl}), and also the compressibility for symmetric nuclear matter at saturation $K_0$, which (at $T=0$) is related to the curvature of the EoS by:
\begin{equation}
\label{compressibility}
K_0= 9 n_B^2 \left[ \frac{d^2(\varepsilon/n_B)}{d n_B^2} \right]_{n_B=n_0}\, .
\end{equation}

In the present work, for these calculations, we consider a saturation density $n_0= 0.15~ \rm{fm}^{-3}$ (in agreement with~\cite{Blaizot:1980tw,Gross-Boelting:1998qhi,PREX:2021umo}) and a binding energy per baryon $B/A = -15.75~\rm{MeV}$ (a value which lies in the range usually adopted in literature, e.g.,~\cite{livro:Glendenning,Weber:2006ep}, and not far from the result presented in Ref.~\cite{Myers:1966zz}). 

{\bf{STEP 2:}} In order to determine the coupling constants of the nucleon with respect to the isovector mesons, it is necessary to consider the EoS of cold asymmetric nuclear matter (still in the absence of leptons) and solve the system of equations of the symmetry energy
\begin{equation}
\label{symmetry}
a_{sym}^0= \frac{1}{2} \left[ \frac{d^2(\varepsilon/n_B)}{dt^2} \right]_{t=0} \, ,
\end{equation}
and its slope
\begin{equation}
\label{slope}
  L_0= 3 n_0 \left( \frac{d a_{sym}}{d n_B} \right)_{n_B=n_0},  
\end{equation}
to find the corresponding values of $(g_\varrho/m_\varrho)$ and $(g_\delta/m_\delta)$. In Eq.~(\ref{symmetry}), the asymmetry between protons ($p$) and neutrons ($n$) is quantified by $t=(n_{B_p}-n_{B_n})/n_B$. 

In this context, it is necessary to employ the appropriate relation to describe the asymmetric nuclear matter at saturation, according to (\ref{EnergiaMBF}) and (\ref{EnergiaGasFermiTocteto}):
\begin{equation}
\begin{split}
\varepsilon&=  \frac{1}{2} m_{\sigma}^2\sigma_0^2+
\frac{1}{2}m_{\omega}^2\omega_0^2+\frac{1}{2}m_{\varrho}^2{\varrho_{0}^3}^2+\frac{1}{2}m_{\delta}^2{\delta_{0}^3}^2
\\ & +\frac{1}{\pi^2}\Bigg[\int_0^{k_{F_n}}dk ~ k^2
\sqrt{k^2+(m_n^*)^2} \\ & +\int_0^{k_{F_p}}dk ~ k^2
\sqrt{k^2+(m_p^*)^2}\Bigg]\, ,
\end{split}
\end{equation}
where $k_{F_n}$ and $k_{F_p}$ represent, respectively, the Fermi momentum for neutrons and the protons, respectively.

\setlength{\parskip}{0pt}

Thus, Eqs.~(\ref{symmetry}) and (\ref{slope}), together with the equations of motion of the fields $\varrho_0^3$ and $\delta_0^3$, given by the equations of motion (\ref{tcm}), combine into a system of four equations to be solved for the variables $g_\varrho$, $g_\delta$, $\varrho_0^3$ and $\delta_0^3$. Note that, within this procedure, the effective masses of the nucleons and the fields $\sigma_0$ and $\omega_0$ must be recalculated in this new scenario of asymmetric matter, according to Eqs.~(\ref{tcm}), taking their interdependence into account. In fact, for each choice of the parameter $\zeta$, there will be a new solution, setting the coupling constants, the effective mass of the nucleons and the compressibility as functions of the adjustable parameter in the MBF Model.

Considering the phenomenologically acceptable values for ${(m_N^*)}_0$ (\cite{Johnson:1987zza,Jaminon:1989wj}), $K_0$ (\cite{Blaizot:1980tw,Krivine:1980kzz,Stone:2014wza,Colo:2013yta,Colo:2004mj,Todd-Rutel:2005yzo,Agrawal:2003xb}), $a_{sym}^0$ (\cite{Tsang:2012se,Lattimer:2012xj,Horowitz:2014bja}) and $L_0$ (\cite{Tsang:2012se,Reed:2021nqk,Li:2019xxz}), these two steps allow us to build a map of the meson-nucleon coupling constants. Additionally, due to the rapid convergence of the adjustable parameter (\ref{mstar}) to the exponential form shown in (\ref{exponential coupling}), the range of values for the adjustable parameter $\zeta$ is quite narrow, and Table \ref{quadro_acoplamentos} presents the coupling constants for different values of $\zeta$ within the phenomenological range.



\bibliography{referencias}

\end{document}